\pdfoutput=1

\documentclass[10pt]{article}

\usepackage{listings}    % for snippets of code using \begin{lstlisting}.
\usepackage{graphicx}    % for figures, eg, \includegraphics
\usepackage{subcaption}  % for subfigures, ie figures with internal figures
\usepackage{algorithm}   % for displaying algorithm in pseudo-code, using \begin{algorithm}
\usepackage{algorithmic} % and \begin{algorithmic}
\usepackage{booktabs}    % for better tables, eg using \toprule
\usepackage{hyperref}    % to handle URL links with \url
\usepackage{xurl}        % to handle very long URLs inside \url{} (done automatically, nothing to set manually)
\usepackage[htt]{hyphenat} % for line breaks in texttt, using \-
\usepackage{microtype} % ligatures are extremely annoying, especially when copy&paste text from PDF
\DisableLigatures{}    % https://tex.stackexchange.com/questions/439651/how-do-i-disable-ligatures
\usepackage{adjustbox}
\usepackage{colortbl}

\usepackage{caption}
\usepackage{setspace}
\usepackage{multirow}
\usepackage{enumerate}
\usepackage{pdflscape}
\usepackage{pgfplots} % bar lib

\usepackage{xcolor}
\definecolor{codegreen}{rgb}{0.25,0.5,0.35}
\definecolor{codegray}{rgb}{0.5,0.5,0.5}
\definecolor{codepurple}{rgb}{0.6,0,0}
\definecolor{backcolour}{rgb}{0.95,0.95,0.92}
\definecolor{colorstring}{rgb}{0.5,0,0.35}
\definecolor{rltred}{rgb}{0.5,0,0}
\definecolor{rltgreen}{rgb}{0,0.5,0}
\definecolor{rltblue}{rgb}{0,0,0.5}
\definecolor{DarkGreen}{rgb}{0.00,0.60,0.00}
\definecolor{ScarletRed}{rgb}{0.80,0.00,0.00}
\definecolor{blizzardblue}{rgb}{0.67, 0.9, 0.93}
\definecolor{green-yellow}{rgb}{0.68, 1.0, 0.18}
\definecolor{dkgreen}{rgb}{0,0.6,0}
\definecolor{gray}{rgb}{0.5,0.5,0.5}
\definecolor{mauve}{rgb}{0.58,0,0.82}
\definecolor{lightgrey}{rgb}{0.90,0.90,0.90}
\definecolor{grey}{gray}{0.75}
\definecolor{light-gray}{gray}{0.80}

\lstdefinestyle{mystyle}{
    escapechar=©, %  use ©\label{}© when needing \label pointing to line numbers
	backgroundcolor=\color{backcolour},
    basicstyle=\footnotesize\ttfamily,
   	identifierstyle=\footnotesize\ttfamily,
	commentstyle=\color{codegreen},
	keywordstyle=\color{colorstring}\bfseries,
	numberstyle=\ttfamily\color{codegray},
	stringstyle=\ttfamily\color{DarkGreen},
	breakatwhitespace=false,
	breaklines=true,
	captionpos=b,
	keepspaces=true,
	numbers=left, % possible values are (none, left, right)
	numbersep=2pt,
	showspaces=false,
	showstringspaces=false,
	showtabs=false,
	tabsize=2
}
\lstdefinelanguage{yaml}{
	morekeywords={true,false,null,y,n},
	sensitive=false,
	morecomment=[l]{\#},
	morestring=[b]",
	morestring=[b]',
}

\usepackage{xspace}
\newcommand{\evo}{{\sc EvoMaster}\xspace}
\newcommand{\etal}{{\emph{et al.}}\xspace}

\newcommand{\base}{Base\xspace}
\newcommand{\aratrl}{ARAT-RL\xspace}
\newcommand{\emrest}{EmRest\xspace}

\newcommand{\evomaster}{EvoMaster\xspace}
\newcommand{\evomasterbbnoauth}{EvoMaster\xspace}
\newcommand{\llamaresttest}{LLamaRestTest\xspace}
\newcommand{\restler}{RESTler\xspace}
\newcommand{\schemathesis}{Schemathesis\xspace}
\newcommand{\deeprest}{DeepRest\xspace}
\newcommand{\apirl}{APIRL\xspace}
\newcommand{\cats}{CATS\xspace}
\newcommand{\wuppie}{WuppieFuzz\xspace}
\newcommand{\wuppiefuzz}{WuppieFuzz\xspace}

\usepackage{boxedminipage}
\newenvironment{result}%
{\smallskip
	\noindent
	\let\emph=\textbf
	\begin{boxedminipage}{\columnwidth}\begin{center}\em}%
		{\end{center}\end{boxedminipage}%
}

\usepackage{ifthen}
\newboolean{showcomments}
\setboolean{showcomments}{true} % comment this line to deactivate comments

\ifthenelse{\boolean{showcomments}}{
	\newcommand{\nbc}[3]{
		{\colorbox{#3}{\bfseries\sffamily\scriptsize\textcolor{white}{#1}}}
		{\textcolor{#3}{\sf\small$\langle$\textit{#2}$\rangle$}}}
	
}{
	\newcommand{\nbc}[3]{}

}

\newcommand{\nsuts}{56\xspace}
\newcommand{\nsutsOld}{36\xspace}
\newcommand{\ntools}{24\xspace}
\newcommand{\noracles}{52\xspace}

\lstdefinestyle{yaml}{
     basicstyle=\color{blue}\footnotesize,
     rulecolor=\color{black},
     string=[s]{'}{'},
     stringstyle=\color{blue},
     comment=[l]{:},
     commentstyle=\color{black},
     morecomment=[l]{-}
 }

\usepackage{authblk}                % for \affil
\usepackage[square,numbers]{natbib} % needed for the references, when using ACM-Reference-Format outside an
\usepackage{amsmath}                % for math operations
\usepackage{amssymb}                % for symbols like \checkmark

\usepackage[nomessages]{fp}% http://ctan.org/pkg/fp

\newcommand{\maxnum}{100.0}
\newlength{\maxlen}
\newlength{\barht}

\newcommand{\databar}[3][blue!20]{%
	\settowidth{\maxlen}{\maxnum}%
	\addtolength{\maxlen}{\tabcolsep}%
	\setlength{\barht}{\dimexpr\ht\strutbox+\dp\strutbox\relax}%
	\FPeval\result{max(0.0,min(1.0,round(#2/\maxnum:4)))}%
	\makebox[\maxlen][l]{%
		\rlap{\color{gray!20}\rule[-\dp\strutbox]{\maxlen}{\barht}}%
		\rlap{\color{#1}\rule[-\dp\strutbox]{\result\maxlen}{\barht}}%
		\makebox[\maxlen][c]{#3}%
	}%
}

\newcommand{\rankbest}[1]{\cellcolor{green!20}\textbf{#1}}
\newcommand{\ranksecond}[1]{\cellcolor{yellow!20}#1}
\newcommand{\rankworst}[1]{\textcolor{red!55!black}{#1}}

\usepackage{geometry}
\title{
{
Web Fuzzing Commons, Dataset, Guidelines and Evaluations to Support Experimentation in REST API Fuzzing
}
}

\author[1,2]{Omur Sahin}
\author[3]{Man Zhang}
\author[4]{Dmitry Dygalo}
\author[5]{Madalin Ilie}
\author[6]{Thomas Rooijakkers}
\author[6]{Erieke Weitenberg}
\author[2,7]{Andrea Arcuri}

\affil[1]{Erciyes University, Türkiye}
\affil[2]{Kristiania University of Applied Sciences, Norway}
\affil[3]{Beihang University, China}
\affil[4]{Schemathesis.io, Czechia}
\affil[5]{Endava, Romania}
\affil[6]{The Netherlands Organisation for Applied Scientific Research (TNO), The Netherlands}
\affil[7]{Oslo Metropolitan University, Norway}

\date{}

\begin{document}

\maketitle

\begin{abstract}
Fuzzing REST APIs is an important research problem, with practical applications and impact in industry.
As such, lot of research work has been carried out on this topic in the last few years.
However, there are three major issues that hinder further progress:
(1) how to deal with API authentication;
(2) how to catalog and compare different fault types found by different fuzzers;
(3) what to use as case study to facilitate fair comparisons among fuzzers.

In this paper, to address these important challenges, we present Web Fuzzing Commons (WFC), Web Fuzzing Dataset (WFD) {and Web Fuzzing Evaluations (WFE)}.
WFC is a set of open-source libraries and schema definitions to declaratively specify authentication info and catalog
different types of faults that fuzzers can automatically detect.
WFD is a collection of \nsuts open-source APIs, { configured to run with Docker-Compose}, supported by WFC.
{
WFE provides scaffolding scripts to automatically run experiments and collecting/analysing their results, employing different REST API fuzzers configured via Docker, using WFD as case study.
}

{
This is a community driven work,
currently stewarded by the authors of the most popular open-source REST API fuzzers, including
\cats,
\evo,
\schemathesis
and
\wuppie.
}
To show the usefulness of WFC/WFD{/WFE},
{
a set of experiments is carried out with these fuzzers on WFD, using WFE.
In our empirical analyses, we also include other state-of-the-art fuzzers such as
\aratrl,
\emrest,
\llamaresttest,
and
\restler.
}

We {also} discuss common pitfalls in tool comparisons, as well as providing guidelines with support of WFC/WFD to avoid them.

\end{abstract}


%%%%%%%%%%%%%%%%%%%%%%%%%%%%%%%%%%%%%%%%%%%%%%%%%%%%%%%%%%%%%%%%%%%%%%%%%%%%
\section{Introduction}

As of 2025, cloud applications are widely used in industry.\footnote{https://edgedelta.com/company/blog/how-many-companies-use-cloud-computing}
There are different kinds of web services, where REST APIs are the most common~\cite{newman2021building,rajesh2016spring}.
Not only are they widely used to build the backends of web applications, but also they are widely used to provide all different services and functionalities over the
internet.\footnote{\label{foot:apisguru}https://apis.guru/}$^{,}$\footnote{https://rapidapi.com/}

Due to their importance and widespread use in industry, a lot of research has been carried out in academia on how to automatically
test REST APIs~\cite{golmohammadi2023testing}.
Several fuzzers have been proposed in the literature, like for example:
APIF~\cite{wang2024beyond},
APIRL~\cite{foley2025apirl},
\aratrl~\cite{kim2023adaptive},
ASTRA~\cite{sondhi2025utilizing},
AutoRestTest~\cite{kim2025autoresttest},
bBOXRT~\cite{laranjeiro2021black},
DeepREST~\cite{corradini2024deeprest},
\emrest~\cite{xu2025effective},
\evo~\cite{arcuri2025tool},
KAT~\cite{le2024kat},
\llamaresttest~\cite{kim2025llamaresttest},
LogiAgent~\cite{zhang2025logiagent},
MINER~\cite{lyu2023miner},
Morest~\cite{liu2022icse},
Nautilus~\cite{deng2023nautilus},
OpenAPI-Fuzzer~\cite{ferech2023efficient},
RAFT~\cite{saha2025rest},
RestCT~\cite{wu2022icse},
RESTest~\cite{martinLopez2021Restest},
\restler~\cite{restlerICSE2019},
RestTestGen~\cite{viglianisi2020resttestgen},
\schemathesis~\cite{hatfield2022deriving},
VoAPI2~\cite{du2024vulnerability}
and
WuppieFuzz~\cite{rooijakkers2025wuppiefuzz}.
Furthermore, there are also open-source fuzzers developed in industry with no published research paper (as far as we know),
like for example
CATS,\footnote{\url{https://github.com/Endava/cats}}
Dredd,\footnote{\url{https://github.com/apiaryio/dredd}}
and
fuzz-lightyear.\footnote{\url{https://github.com/Yelp/fuzz-lightyear}}

Few studies have also been carried out to compare several fuzzers on large, diverse sets of APIs~\cite{Kim2022Rest,sartaj2024restapitestingdevops}, including our own previous work~\cite{zhang2023open}.
When developing a novel technique, it is important to empirically compare it with the state-of-the-art to assess its strengths and limitations.
However, there are some major challenges that complicate and hinder the running and analysis of large case studies.
These challenges also reduce the impact of academic results in industry.
In particular, these challenges are related to how to deal with \emph{authentication}, how to catalog and compare detected \emph{faults}, and how the \emph{case study} is selected.

Most real-world APIs require \emph{authentication} information on each API call.
This can be login/password information sent at each request, or based on dynamically generated authentication tokens.
Such information would be missing in (OpenAPI) schemas, as it depends on which users exist in the API's test environment (e.g., pre-filled user information in databases).
Somehow, this information needs to be provided as input to the fuzzers.
However, there are major challenges when dealing with short-lived authentication tokens.
Furthermore, there is no standard way to provide authentication information to a fuzzer.
Each fuzzer provides its own custom way, or no support for authentication at all.
The problem is at least twofold:
\begin{enumerate}
\item for \emph{academics}, comparing different fuzzers in academic experiments become cumbersome, because different manual settings would be required for \emph{each} fuzzer on \emph{each} employed API in the experiments.
    This hinders the ability of running large empirical studies on several fuzzers on many real-world APIs.
\item for \emph{practitioners}, these custom configurations create a sort of ``vendor-lock-in'', which prevents trying out different fuzzers without major time investment in learning and setting up authentication configurations for each new fuzzer.
\end{enumerate}

When comparing the performance of different fuzzers, this is usually based on code coverage metrics and \emph{fault} detection.
For many programming languages, there is support for collecting coverage metrics of a running API,
like for example JaCoCo\footnote{https://www.eclemma.org/jacoco/}
for the JVM.
Therefore, comparing REST API fuzzers based on code coverage metric is usually not problematic.
Unfortunately, though, this is not the case for fault detection.
To automatically detect a fault with a fuzzer, you need an automated oracle.
The most common automated oracle is to check if an application does crash.
However, this is not the case for REST APIs, as they are typically run inside an HTTP server, where internal crashes
just result in returning HTTP responses with status code 500 (Server Error).
This can be used as an automated oracle~\cite{marculescu2022faults}, but there are many other automated oracles that can be used to detect faults,
like for
{
verifying HTTP compliance~\cite{decrop2026detecting,sahin2026validating}, as well as for
}
robustness~\cite{viglianisi2020resttestgen,laranjeiro2021black}  and security~\cite{deng2023nautilus,du2024vulnerability,arcuri2025fuzzing} testing.
To detect a fault, not only do you need a test case that triggers the fault, but also some information stating which fault was found.
As each fuzzer has its own way to report which and how many faults were found in its generated test cases, comparisons of fault detections among fuzzers are cumbersome.
Typically, in the literature, only 500 HTTP status codes are checked in fuzzer comparisons, based on for example HTTP reverse-proxies (e.g., using tools such as
\emph{mitmproxy}\footnote{https://www.mitmproxy.org/})
or on analyzing server logs.
Unfortunately, this provides only a limited view of what these fuzzers are capable of.
For example, many 500-status related faults have limited severity (e.g., many are just related to incomplete input validation~\cite{marculescu2022faults}), especially compared to potentially disastrous, security-related faults.

In this paper, we address and propose working solutions to these two major issues in fuzzing REST APIs.
First, we have designed a novel \emph{declarative} approach to define authentication information.
Such information can be provided in configuration files such as YAML and TOML.
We have created a JSON Schema to capture all this information, which then can be used to automatically generate parser libraries for different programming languages (e.g., Java).

For fault detection, we have cataloged existing automated oracles in the literature of fuzzing REST APIs, and defined a JSON schema to represent fuzzer's results, including fault detection.
A schema with documentation is useful to resolve ambiguities, and to enable fuzzers to easily compare their results in a standardized way.
Furthermore, a standardized output format enables the possibility of providing separated, professional-looking, usable HTML-based test report files.
As a proof-of-concept, we have created an HTML web report template to visualize test results from our new test result outputs.

To enable different fuzzers to easily use these schemas and tool support, all this work is released in a new library, called \emph{Web Fuzzing Commons} (WFC).
This new WFC library is open-source on GitHub,\footnote{\label{foot:wfc}\url{https://github.com/WebFuzzing/Commons}}
with each new release automatically uploaded on Zenodo (e.g., version 0.1.0~\cite{wfc010}).

Originally, the work on simplifying the configuration of authentication information, and on comparing and visualizing fault detection, was started as part of the work on the state-of-the-art fuzzer \evo~\cite{arcuri2025tool}.
This work was based on the needs of fuzzing tens of open-source APIs,
direct industry collaborations with Fortune 500 enterprises such as
Meituan~\cite{zhang2023rpc,zhang2024seeding,zhang2025fuzzing}
and
Volkswagen~\cite{poth2025technology,icst2025vw},
and informal feedback from tens of engineers using \evo in industry at different companies.
However, what presented in this paper is not specific to \evo, and would apply to any fuzzer that aims at being applicable on real-world APIs in industry.
This motivates the creation of candidate standardized schema definitions and tool support in a library such as WFC, independent from \evo.
{
Due to its importance,  this is now a community-driven project, stewarded by the authors of the most popular REST API fuzzers in industry, namely
\cats,
\schemathesis,
and \wuppie (besides \evo).
}
A democratic process has been established {
 with a formal governance charter in place,\footnote{\url{https://github.com/WebFuzzing/Commons/blob/master/GOVERNANCE.md}}
}
  to steer changes and updates to our proposed candidate standards.
   This is a type of effort that is unfortunately rare in the literature in a research field prone to fragmentation.

To show the usefulness of WFC, we have collected a set of REST APIs in which, for each API requiring authentication, we provide authentication information in WFC format.
To enable and simplify the running of academic experiments, we provide all scaffolding needed to run experiments, including Docker files (e.g., to start, stop and collect coverage results from the APIs) and database test configurations.
This selection of APIs are collected in an open-source repository called \emph{Web Fuzzing Dataset} (WFD),\footnote{\url{https://github.com/WebFuzzing/Dataset}} which is the new updated version from the existing EMB benchmark~\cite{icst2023emb}.
{
The scaffolding for running experiments with different fuzzers on WFD, and analyzing their results, is provided in a separated repository called \emph{Web Fuzzing Evaluations (WFE)}.\footnote{\url{https://github.com/WebFuzzing/Evaluations}}
}
Considering \nsuts REST APIs, this is by far the largest selection of open-source REST APIs used in academic experiments, to date~\cite{golmohammadi2023testing}.

{
In this work, besides the contributions of WFC, WFD and WFE,
we aim at
}
pinpointing and highlighting issues in how fuzzers and techniques are compared.
We provide clear guidelines on how to avoid such issues, with an in-depth discussion on the practical and research consequences of ignoring these issues.
In particular, we discuss issues about tool usability, API selection for the empirical studies, the impact of authentication configurations, and the critical consequences of how test suites are generated.
Our goal is not to study which fuzzer technique is best (to avoid biases, the interested reader is referred to \emph{independent} studies where the authors are not involved in the development of any compared fuzzer~\cite{Kim2022Rest,sartaj2024restapitestingdevops}).
Our aim is to highlight current issues in existing empirical studies and provide actionable guidelines for researchers to avoid these issues in future studies, as well as for reviewers to be able to spot and recognize these issues.
The final goal is to ``raise-the-bar'' in the empirical research on this field, pinpointing important research problems of practical consequence that are currently ignored.
Ultimately, this  will enable a better technology transfer from academic research to industrial practice, as {for example with \evo} we have experienced with large enterprises such as Meituan~\cite{zhang2023rpc,zhang2024seeding,zhang2025fuzzing}
and  Volkswagen~\cite{poth2025technology,icst2025vw}.

{
To provide more evidence to the usefulness of WFC, WFD and WFE, we carried out two different sets of experiments, first in 2025, and then in the late 2026.

In the first set of experiments, we used our guidelines to identify a set of six representatives for the state-of-the-art  in fuzzing REST APIs in 2025, namely
\aratrl~\cite{kim2023adaptive},
\emrest~\cite{xu2025effective},
\evo~\cite{arcuri2025tool},
\llamaresttest~\cite{kim2025llamaresttest},
\restler~\cite{restlerICSE2019}
and
\schemathesis~\cite{hatfield2022deriving}.
Experiments were run on an earlier version of WFD, including \nsutsOld APIs.
At that time, only \evo supported WFC, and so it was not used (e.g., experiments were run without authentication support for any fuzzer).

Due to the importance of what is provided by WFC, all the most used REST API fuzzers in industry (i.e., at least 100 stars on GitHub, and a new release within the last 12 months) have already joined the WFC project as of 2026.
We hence carried out a second set of experiments with
\cats,
\evo,
\schemathesis,
and \wuppie
on the latest version of WFD containing \nsuts APIs.
The authors of all these four fuzzers were directly involved in this study.
Authentication information for all fuzzers was provided in WFC format.
Fault detection comparisons were not limited to just 500 status codes, but also they included \emph{all} the different types of automated oracles these fuzzers support (\noracles in total).
}

The practical engineering contribution of creating Docker Compose files for all \nsuts APIs with JaCoCo integration, mitmproxy logging, configurable ports, and database initialization required substantial effort.
Although this was mainly engineering effort, it is a practical contribution that will hopefully help the research community in carrying out new empirical studies that push forward the state-of-the-art in this research field.

This paper provides the following research and engineering contributions:

\begin{itemize}
	\item A candidate standardization in the \emph{Web Fuzzing Commons} (WFC)  by introducing:
		\begin{itemize}
			\item A declarative specification and schema for defining authentication details required to fuzz REST APIs.
			\item A declarative specification and schema for defining fuzzer result reports, using the presented fault catalog.
		\end{itemize}
	\item Proposing the new Web Fuzzing Dataset (WFD), Evaluations (WFE) and guidelines by including:
		\begin{itemize}
			\item A review and analysis of issues in current empirical studies.
			\item A collection of \nsuts open-source REST APIs that adopt WFC.
			\item A set of guidelines for fuzzer usability and fuzzer selection in web fuzzing evaluation.
		\end{itemize}
	\item Carrying out empirical studies with {eight} state-of-the-art fuzzers by using WFD, {WFE} and WFC.
	\item Developing open-source tool support in the new WFC library, and implementing a Web HTML viewer for the fuzzer result reports in our proposed candidate standard format.
	\item Building a public repository for WFD to support continuous addition of new APIs.
    \item {Building a public repository for WFE to support continuous addition of new fuzzers and types of  analyses.}
	\item The establishment of a democratic process to steer changes in our proposed candidate standards, already involving the authors of {all} the most used {REST API} fuzzers in industry.

\end{itemize}

This paper provides several  novel contributions.
% No specific contribution is ``primary'', and significantly more important than the others.
% Showing WFC being an effective common specification is as important as showing that WFD is a meaningful benchmark resource, and as important as empirically demonstrating that current REST API fuzzing evaluations are methodologically limited, where our proposed guidelines can shape future studies.
{
Current REST API fuzzing research lacks reproducible, authentication-aware, and comparable infrastructure.
 WFC/WFD/WFE are a solution to these problems, and our experiments show methodological problems in existing evaluations.
}

The paper is organized as follows.
%Section~\ref{sec:background} provides background information to better understand the rest of the paper.
Related work is discussed in Section~\ref{sec:relatedwork}.
Web Fuzzing Commons (WFC) is presented in Section~\ref{sec:all_wfc}, encompassing our new candidate standard for authentication information and the candidate standard for reporting detected faults.
{
Section~\ref{sec:all_wfd} introduces our new corpus Web Fuzzing Dataset (WFD), together with an in-depth analysis on API selection bias.
Sound guidelines to select fuzzers for empirical comparisons are presented in Section~\ref{sec:selection-criteria}.
Technical scaffolding for running and analyzing experiments on WFD is discussed in Section~\ref{sub:wfe}, where Web Fuzzing Evaluations (WFE) is presented.
}
Our empirical study, including experiment settings, result analysis, and a detailed discussion of test generation, is provided in Section~\ref{sec:experiments}.
Summarizing discussions are presented in Section~\ref{sec:discussion}, followed by threats to validity in Section~\ref{sec:threats}.
Finally, Section~\ref{sec:conclusions} concludes the paper.

\section{Related Work}
\label{sec:relatedwork}

In software testing, conducting experiments to evaluate the effectiveness of new testing approaches requires a collection of programs or applications.
Existing research has explored the development of such collections to support experimentation and analysis, thereby easing and standardizing the assessment process.

Siemens Benchmark Suite~\cite{hutchins1994experiments} is one of the first benchmarks built for evaluating the effectiveness of testing approaches for C programs.
The Software-Artifact Infrastructure Repository (SIR)~\cite{DER05b} is a repository that contains various programs developed in Java, PHP, C\#, C, and C++ across multiple versions, along with supporting information such as test suites and fault data, facilitating controlled experiments for evaluating testing approaches.
SF110~\cite{fraser2014large}\footnote{\url{https://www.evosuite.org/experimental-data/sf110/}} is a corpus of classes developed for studying unit testing in Java programs, representing a sample of 100 Java projects sourced from SourceForge.\footnote{\url{https://sourceforge.net/}}
Specifically in the context of the faults, Defects4J~\cite{just2014defects4j}\footnote{\url{https://github.com/rjust/defects4j}} is a widely used dataset for assessing and comparing testing approaches for Java programs.
The initial version of Defects4J contains 357 real bugs from 5 projects, and it has since been extended to include 835 bugs from 17 projects~\cite{gay2020defects4j}.
BugsJS~\cite{gyimesi2019bugsjs}\footnote{\url{https://bugsjs.github.io/}} is a bug benchmark for JavaScript programs that contains 453 bugs from 10 server-side programs.

Recently, several benchmarks have been introduced for evaluating  fuzzers~\cite{hazimeh2020magma,li2021unifuzz,metzman2021fuzzbench,bohme2022reliability,ounjai2023green,miao2025program}.
For example, Magma~\cite{hazimeh2020magma}\footnote{\url{https://hexhive.epfl.ch/magma/}} focuses on the assessment of bug findings.
FuzzBench~\cite{metzman2021fuzzbench} enables a platform for comparing fuzzers using real-world software in terms of various metrics, such as code and bug coverage, independent code coverage, and differential coverage.
FeatureBench~\cite{miao2025program} consists of 153 programs generated based on seven program features related to control flow and data flow.
By incorporating these features, the benchmark enables comparison and analysis of fuzzer performance across programs with varying characteristics.
Benchmarks have also been developed in other domains, such as video games~\cite{li2022gbgallery} and microservices~\cite{amoroso2024dataset}.

In terms of REST API testing,
Di Meglio~\etal~\cite{di2023starting} proposed a benchmark for REST API frameworks and execution environments (e.g., Spring Boot with OpenJDK \emph{vs.} Express with Node.js) to facilitate comparisons of the different technologies in performance (e.g., response time) through load and stress testing.
The benchmark can be applied for the selection of suitable technologies when building new REST APIs.
% PRAB
The Public REST API Benchmark (PRAB)~\cite{decrop2025public}\footnote{~\url{https://github.com/alixdecr/PRAB}} was created by collecting public REST APIs that have been included in the evaluation of REST API fuzzing approaches from the literature.
PRAB includes schemas defined using OpenAPI/Postman, along with various structural characteristics of the collected 60 REST APIs.
% EMB
\evo Benchmark (EMB)~\cite{icst2023emb}\footnote{\url{https://github.com/WebFuzzing/EMB}}, open-source since 2017, includes a collection of open-source web APIs, comprising not only REST APIs but also GraphQL and RPC APIs.
In addition to the API source code, it provides libraries that facilitate database setup and the starting, resetting, and stopping of the APIs.
These features enable experimentation for both black-box and white-box testing approaches.
Several studies in the literature provide ``replication-packages'', where not only the APIs but also the compared tools and experiment scripts are provided, like for example in the case of the work for \aratrl.\footnote{https://github.com/codingsoo/ARAT-RL}

%-------------------------------------------------
\begin{table}[!t]
\centering
\caption{\label{tab:restgym}
{Comparison of features and details between RestGym, EMB and WFD.}
}
\begin{tabular}{lrrr}
\toprule
Features/Details  &      RestGym &           EMB &          WFD \\
\midrule
Inception Year    &         2024 &          2017 &         2017 \\
Publication Year  &         2025 &          2023 &            - \\
API Count         &           11 &            20 &       \nsuts \\
Multi-OS          &              &  $\checkmark$ & $\checkmark$ \\
Docker Support    & $\checkmark$ &               & $\checkmark$ \\
White-Box Support &              &  $\checkmark$ & $\checkmark$ \\
Authentication    &              &               & $\checkmark$ \\
Multi-Faults      &              &               & $\checkmark$ \\
R-Analyses        &              &               & $\checkmark$ \\
Multi-Authors     &              &               & $\checkmark$ \\
\bottomrule
\end{tabular}
\end{table}
%-------------------------------------------------

The authors of RestTestGen/DeepREST put their experiment infrastructure in a repository called RestGym~\cite{corradini2025restgym},
{
to enable comparisons of black-box REST API fuzzers.
RestGym came years later than EMB, and provides less functionalities and features compared to first EMB, and now its rebranding in WFD, as shown in more details in Table~\ref{tab:restgym}.
In particular, the provided number of available APIs for experiments was much smaller than what the state-of-the-art had already been providing for years before (e.g., 11 vs.~20).
Also, RestGym, due to technical limitations, does not work on Windows (and this still seems the case at the time of writing).
And it only supports black-box testing, and not white-box testing.
However, to simplify the installation of APIs and compared tools, RestGym introduced the use of Docker, which was not previously done in EMB.
Now, compared to EMB and RestGym,  WFD introduces a standardized way to provide authentication information to the fuzzers, as well as the ability to report and compare all different kinds of faults a fuzzer can find (and not limited to just the HTTP 500 status code), besides a much larger selection of APIs.
This is together with a set of R-analysis scripts to fully automate the creation of tables/figures with all the needed analyses to automatically evaluate the results of the run experiments.
Finally, as WFD is a community driven project, co-authored by the authors of four of the most used fuzzers in industry, there is less bias towards any specific fuzzer.
}

In the literature, there is a lack of a standardized approach for experimental setup (e.g., authentication configuration), result reporting (e.g., fuzzing reports), and analysis (e.g., fault detection), which would simplify and standardize the comparison of fuzzers across various APIs.

Regarding the status of evaluating fuzzing techniques,
Klees \etal~\cite{klees2018evaluating} reports on an analysis of 32 existing studies on fuzzing of parsers and utility libraries.
Several issues were found, related for example to low statistical power (e.g., experiments repeated only once), metrics used for comparisons and selection of artifacts for the studies.

%%%%%%%%%%%%%%%%%%%%%%%%%%%%%%%%%%%%%%%%%%%%%%%%%%%%%%%%%%%%%%%%%%%%%%%%%%%%
\section{Web Fuzzing Commons}
\label{sec:all_wfc}

%---------------------------------------------------------------------------
\subsection{Candidate Standard for Declarative Authentication}
\label{sec:auth}

Some APIs are ``read-only'', providing data without the option of modifying it.
If the data is in the public domain, then anyone can easily access it by calling the API directly (e.g., with \texttt{GET} endpoints).
This is for example the case for APIs related to weather forecast or news-feeds.
Still, even in these cases, if the API is a commercial one, it might still require some way to identify the users.
For example, they might provide a ``free-tier'' version with a limited number of API calls per-hour (\emph{rate-limiter}).
More usage would require payment for a subscription, which needs to be identified and verified on each API call.

When dealing with APIs that enable modifying data, like for example users' information for commercial products, identifying users is essential (\emph{authentication}), especially needed to verify access policies (\emph{authorization}).
Otherwise, users could delete/modify all data created by other users.

Due to its importance, apart from example/toy APIs and not-rate-limited, read-only APIs, most real-world APIs require to be able to identify the users that use the API.
As HTTP is a stateless protocol, authentication information needs to be sent on \emph{each} API call.

Typically, a user would be identified with a unique \emph{ID}, and a shared secret that only the user and the API know, i.e., a \emph{password}.
Sending the user ID and password on each HTTP call is possible.
An example is by using the ``Basic'' HTTP Authentication Scheme (RFC 7617\footnote{https://www.rfc-editor.org/rfc/rfc7617}), where ID and password are Base64 encoded and sent via the \texttt{Authorization} HTTP header.
In these cases, the authentication information is \emph{static}.
In other words, for a given user the authentication information sent as part of an HTTP call would always be the same, until they change their password or user ID.

This approach has limitations.
If for any reason a hacker would intercept one such message, then the entire account would be compromised.
Furthermore, there are many cases in which there is the need to temporarily grant access to some resources with limited rights.
To handle these common cases, it is a widespread practice to use \emph{dynamic}, token-based authentication mechanisms.
A user would first make a request to a ``login'' endpoint, by providing their user id and password.
Then, the API would return a \emph{token} that can be used for authentication (e.g., sent via the \texttt{Authorization} HTTP header).
These tokens are usually short-lived.
Even if a hacker manages to steal one, they would be able to use it only for a limited amount of time before it expires.

To be able to test real-world APIs, a fuzzer needs to be able to send valid authentication information, on each API call.
Randomly generating user ids and passwords that match what present in the databases of the APIs would be extremely unlikely.
Even if an API enables creating new users with its endpoints, creating high-permission users (e.g., administrators) would typically not be possible.
This is not a problem just for black-box fuzzers, as also grey and white-box fuzzers would have the same issues.
Even if a white-box fuzzer would directly access the content of the API's databases, typically for security reasons passwords are ``hashed'', i.e., they are not stored in plain-text.
Automatically reverse-engineering passwords would not be a viable approach.

Somehow, testers using a fuzzer need to provide it with authentication information for some users that are recognized by the tested APIs.
How this is done can significantly impact how fuzzers are used and can be compared.
This strongly depends on whether the authentication information is provided \emph{statically} or \emph{dynamically}.
Then, the fuzzer can use this provided authentication information when making calls toward the tested APIs.

\begin{table*}[!t]
\centering
\caption{Authentication support info (\emph{Static} and \emph{Dynamic}) for existing REST API fuzzers in the literature.}
\label{tab:auth}
% \resizebox{.95\textwidth}{!}{\protect
\begin{adjustbox}{width=.95\textwidth,center}
\begin{tabular}{l r cc rr l}
\toprule
Name & Ref. & Static & Dynamic & Year & Update & Availability \\
\midrule
APIF         & \cite{wang2024beyond}            & \checkmark &            & 2024 &      & (broken) \url{https://github.com/apif-tool/APIF_tool_2024}\\
APIRL        & \cite{foley2025apirl}            & \checkmark &            & 2025 & 2025 & \url{https://github.com/ICL-ml4csec/APIRL}\\
ARAT-RL      & \cite{kim2023adaptive}           &            &            & 2023 & 2024 & \url{https://github.com/codingsoo/ARAT-RL}\\
ASTRA        & \cite{sondhi2025utilizing}       &            &            & 2025 &      & \\
AutoRestTest & \cite{kim2025autoresttest}       & \checkmark &            & 2024 & 2025 & \url{https://github.com/selab-gatech/AutoRestTest/}\\
bBOXRT       & \cite{laranjeiro2021black}       &            &            & 2021 &      & \url{https://eden.dei.uc.pt/%7Ecnl/papers/2020-access.zip}\\
DeepREST     & \cite{corradini2024deeprest}     & \checkmark & \checkmark & 2024 & 2025 & \url{https://github.com/SeUniVr/DeepREST-Docker}\\
EmRest       & \cite{xu2025effective}           &            &            & 2025 & 2025 & \url{https://github.com/GIST-NJU/EmRest} \\
\evo         & \cite{arcuri2025tool}            & \checkmark & \checkmark & 2016 & 2025 & \url{https://github.com/WebFuzzing/EvoMaster}\\
KAT          & \cite{le2024kat}                 &            &            & 2024 &      & \\
LlamaRestTest & \cite{kim2025llamaresttest}     &            &            & 2024 & 2025 & \url{https://github.com/codingsoo/LlamaRestTest}\\
LogiAgent   & \cite{zhang2025logiagent}        &            &            & 2025 &      & \\
MINER        & \cite{lyu2023miner}              & \checkmark & \checkmark & 2023 & 2023 & \url{https://github.com/puppet-meteor/MINER}\\
Morest       & \cite{liu2022icse}               &            &            & 2023 & 2024 & \url{https://github.com/sumleo/morest}\\
Nautilus     & \cite{deng2023nautilus}          &            &            & 2023 &      & \\
OpenAPI-Fuzzer & \cite{ferech2023efficient}     & \checkmark &            & 2020 & 2023 & \url{https://github.com/matusf/openapi-fuzzer} \\
RAFT         & \cite{saha2025rest}              &            &            & 2025 &      & \\
RestCT       & \cite{wu2022icse}                & \checkmark &            & 2022 & 2025 & \url{https://github.com/GIST-NJU/RestCT}\\
RESTest      & \cite{martinLopez2021Restest}    & \checkmark &            & 2018 & 2024 & \url{https://github.com/isa-group/RESTest}\\
RESTler      & \cite{restlerICSE2019}           & \checkmark & \checkmark & 2019 & 2025 & \url{https://github.com/microsoft/restler-fuzzer}\\
RestTestGen  & \cite{viglianisi2020resttestgen} & \checkmark & \checkmark & 2021 & 2025 & \url{https://github.com/SeUniVr/RestTestGen}\\
Schemathesis & \cite{hatfield2022deriving}      & \checkmark & \checkmark & 2019 & 2025 & \url{https://github.com/schemathesis/schemathesis}\\
VoAPI2       & \cite{du2024vulnerability}       & \checkmark & \checkmark & 2024 & 2024 & \url{https://github.com/NSSL-SJTU/VoAPI2}\\
WuppieFuzz   & \cite{rooijakkers2025wuppiefuzz} & \checkmark & \checkmark & 2024 & 2025 & \url{https://github.com/TNO-S3/WuppieFuzz}\\
\bottomrule
\end{tabular}
%}
\end{adjustbox}
\end{table*}

According to their documentation {in late 2025}, several fuzzers in the literature are unable to work on real-world APIs, as they do not support authentication.
Table~\ref{tab:auth} shows a summary of all the \ntools existing fuzzers in the literature that we are aware of.
In particular, we show whether they support \emph{static} authentication, and/or \emph{dynamic} authentication.
Note: this information is collected from what is stated in their respective research articles and tool documentation, if any is available.
For each tool, we also report if it is open-source (with a URL), and its inception year.
In these cases, we also report when was the last commit in their repository (this information was collected in December 2025).
If a tool is not open-source, we simply report the publication year of the articles in which they were presented.
{
Table~\ref{tab:auth} only includes fuzzers presented in research articles.
Open-source fuzzers developed in industry without an associated research article, such as CATS (first released in 2020 and still currently maintained, supporting both static and dynamic authentication), Dredd and fuzz-lightyear, are not listed.
}

Out of \ntools fuzzers, only eight can be effectively applied on real-world APIs, as they support \emph{dynamic} authentication.
Note that MINER and VoAPI2 are built on top of RESTler, and DeepREST is built on top of RestTestGen.
As such, we can safely assume they share the same authentication capabilities.

{In 2025}, besides \evo, the other three tools support the same kind of dynamic authentication based on
manually written Python scripts.
This is specified in their documentation, for
RESTler,\footnote{\url{https://github.com/microsoft/restler-fuzzer/blob/main/docs/user-guide/Authentication.md}}
RestTestGen\footnote{\url{https://github.com/SeUniVr/RestTestGen?tab=readme-ov-file\#auth}}
and
Schemathesis.\footnote{\label{link:schemathesis:auth}\url{https://schemathesis.readthedocs.io/en/stable/guides/auth/}}
This means that the users of fuzzers need to manually write scripts to make HTTP calls with authentication credentials, and then extract tokens from the received responses in these scripts.
Then, these fuzzers can call these scripts to collect tokens each time they need a new one.

This approach  with manually written Python scripts works.
After all, RESTler and Schemathesis are the two most popular fuzzers in  industry, according to the number of ``stars'' on GitHub (both having more than 2900, as of December 2025).
However, as mentioned in the introduction, this poses a few issues.
In particular:
\begin{itemize}
\item  Not all testers in industry would have the necessary coding skills to write code scripts, possibly in programming languages
        they are not familiar with (e.g., Java or Python).
\item  Calling a custom script from a fuzzer requires to address few technical challenges and limitations, based on the programming
       language the fuzzer is written in and the programming language of the target language of the scripts (the two might not be necessarily the same). This might limit the applicability of the approach.
\item  A tester that wants to try out a new fuzzer would first need to recreate all the authentication scripts, based on the programming language and libraries supported by this new fuzzer. Based on the number of tested APIs, this might become a major effort.
\item  When fuzzers are compared in academic experiments, significant time would need to be spent to prepare scripts for each different fuzzer for each different API in the study. This is a major endeavor, which possibly explains why in many studies in the literature no real-world API requiring authentication is used for the experiments (recall Table~\ref{tab:auth}).
\end{itemize}

{
Script-based approaches are not necessarily bound to a specific programming language.
For example, since 2022 the fuzzer CATS can run an arbitrary user-provided executable (e.g., a Bash or Python script) at a configurable time interval, and use its output as the new value of the authentication headers.
This removes the dependency on the programming language of the fuzzer, but it does not remove the need to write and maintain custom code for each tested API.
}

The most recent fuzzer we were aware of in 2025, i.e., WuppieFuzz~\cite{rooijakkers2025wuppiefuzz}, published on arXiv the same month of this writing (i.e., December 2025), provides authentication mechanisms similar to \evo, i.e., based on YAML configuration files.\footnote{https://github.com/TNO-S3/WuppieFuzz/blob/main/authentication.md}

%------------------------------------------------------
\begin{figure}[!t]
\begin{lstlisting}[style=yaml,basicstyle=\footnotesize]
auth:
  - name: ADMIN  ©\label{line:ocvn:admin}©
    loginEndpointAuth:
      payloadRaw: "username=admin&password=admin"
  - name: user1  ©\label{line:ocvn:user1}©
    loginEndpointAuth:
      payloadRaw: "username=user1&password=password"

authTemplate: ©\label{line:ocvn:authTemplate}©
    loginEndpointAuth:
        endpoint: /login
        verb: POST
        contentType: application/x-www-form-urlencoded
        expectCookies: true ©\label{line:ocvn:expectCookies}©
\end{lstlisting}
\caption{\label{fig:auth:ocvn}
Authentication configuration in YAML for the API called \emph{ocvn}.
}
\end{figure}
%------------------------------------------------------

To ease its use in industry, especially among our industrial partners, \evo has used a \emph{declarative} approach for defining dynamic authentication~\cite{icst2025vw}.
This means providing configuration files (e.g., in YAML format) to tell \emph{where} and \emph{how} tokens can be obtained.
Then, based on this information, the fuzzer (\evo in this case) makes the calls to the login endpoints and extracts the needed tokens, without the user needing to write any code.
Figure~\ref{fig:auth:ocvn} shows an example of authentication configuration for the API called \emph{ocvn} (more details on the evaluated APIs will be discussed in Section~\ref{sec:wfd}).
In this example, authentication information for two users are specified: \texttt{ADMIN} (Line~\ref{line:ocvn:admin}) and \texttt{user1} (Line~\ref{line:ocvn:user1}).
Each one has its own username and password.
As for these two users the authentication info is obtained from the same endpoint, where the only difference is in the payload containing the username and password, all shared settings are specified in an \texttt{authTemplate} declaration (Line~\ref{line:ocvn:authTemplate}).
This includes the path of the login endpoint (i.e., \texttt{/login}), the HTTP verb to use when sending the payload (i.e., \texttt{POST}) and the \texttt{contentType} type of the payload.
Then, in this particular API, the authentication tokens are sent via cookies.
This information is specified  with \texttt{expectCookies} (Line~\ref{line:ocvn:expectCookies}).

%------------------------------------------------------
\begin{figure}[!t]
\begin{lstlisting}[language=java,basicstyle=\footnotesize]
/**
* Calls:
* (500) GET:/api/percentTotalProjectsFlaggedByYear
* Found 2 potential faults. Type-codes: 100, 101
*/
@Test(timeout = 60000)
public void test_38_getOnPercentTotalProjectsFlaggedByYearShowsFaults_100_101() throws Exception {

   final Map<String,String> cookies_ADMIN = given()   ©\label{line:ocvn:test:extract}©
         .contentType("application/x-www-form-urlencoded")
         .body("username=admin&password=admin")
         .post(baseUrlOfSut + "/login")
         .then().extract().cookies();


   // Fault100. HTTP Status 500. org/devgateway/ocds/web/rest/controller/GenericOCDSController_353_lambda$getContrMethodFilterCriteria$1 ©\label{line:ocvn:test:context}©
   // Fault101. Received A Response From API That Is Not Valid According To Its Schema. Response status 500 not defined for path '/api/percentTotalProjectsFlaggedByYear'.
   given().accept("application/json")
         .header("x-EMextraHeader123", "_EM_50571_XYZ_")
         .cookies(cookies_ADMIN) ©\label{line:ocvn:test:use}©
         .get(baseUrlOfSut + "/api/percentTotalProjectsFlaggedByYear?" +
   ...
\end{lstlisting}
\caption{\label{fig:test:ocvn}
Beginning of a test case generated by \evo in Java, in which authentication information has been exploited to make dynamic calls.
}
\end{figure}
%------------------------------------------------------

Such information is enough for a fuzzer to automatically extract dynamic tokens whenever it needs.
For example, Figure~\ref{fig:test:ocvn} shows the beginning of a test case automatically generated by \evo on \emph{ocvn} when exploiting the authentication information provided in the configuration file shown in Figure~\ref{fig:auth:ocvn}.
On Line~\ref{line:ocvn:test:extract} a call is made toward the \texttt{/login} endpoint, using authentication information for the \texttt{ADMIN} user.
From this call, the resulting cookies are saved in the \texttt{cookies\_ADMIN} variable, which is then used in the following API calls for testing the API (e.g., Line~\ref{line:ocvn:test:use}).
Note that this approach of extracting tokens dynamically in the generated tests is essential for their use in industry.
If the cookies were hardcoded in the generated tests, then the test would fail as soon as these cookies expire, making it unusable for regression testing and debugging purposes.

%------------------------------------------------------
\begin{figure}[!t]
\begin{lstlisting}[style=yaml,basicstyle=\footnotesize]
auth:
  - name: admin
    loginEndpointAuth:
      payloadRaw: "{\"usernameOrEmail\": \"admin\", \"password\": \"bar123\"}"
  - name: user
    loginEndpointAuth:
      payloadRaw: "{\"usernameOrEmail\": \"user\", \"password\": \"bar123\"}"

authTemplate:
    loginEndpointAuth:
        endpoint: /api/auth/signin
        verb: POST
        contentType: application/json
        token:   ©\label{line:blogapi:auth:token}©
            extractFromField: /accessToken ©\label{line:blogapi:auth:extract}©
            httpHeaderName: Authorization
            headerPrefix: "Bearer "
\end{lstlisting}
\caption{\label{fig:auth:blogapi}
Authentication configuration in YAML for the API called \emph{blogapi}.
}
\end{figure}
%------------------------------------------------------

In this example, tokens were provided via cookies.
However, what if the tokens are present in the body payloads of the responses?
Figure~\ref{fig:auth:blogapi} shows an authentication configuration for the API called \emph{blogapi}.
Similar to the other case, two users are configured, one being an administrator.
However, in this case the payload is expected to be sent as JSON (Line~\ref{line:ocvn:test:use}),
and no cookie is expected in the response.
The \texttt{token} configuration (Line~\ref{line:blogapi:auth:token}) is used to specify how to extract and use the token.
In particular, a JSON Pointer (RFC 6901\footnote{https://datatracker.ietf.org/doc/html/rfc6901})
is used in the variable \texttt{extractFromField} (Line~\ref{line:blogapi:auth:extract}) to specify where in the HTTP response the token is located.
Then, the configurations for \texttt{httpHeaderName} and \texttt{headerPrefix} are used to instruct the fuzzer on how such token should be used when making API calls.
In this particular case, using the \texttt{Authorization} header with the token prefixed with the ``\texttt{Bearer }'' text.

%------------------------------------------------------
\begin{figure}[!t]
\begin{lstlisting}[language=python,basicstyle=\footnotesize]
# Calls:
# (500) GET:/api/tags/{id}
# Found 1 potential fault of type-code 100
@timeout_decorator.timeout(60)
def test_0_with500(self):

     token_admin = "Bearer "
     headers = {}
     headers["content-type"] = "application/json"
     body = " { " + \
            " \"usernameOrEmail\": \"admin\", " + \
            " \"password\": \"bar123\" " + \
            " } "
     res_admin = requests \
                .post(self.baseUrlOfSut + "/api/auth/signin",  ©\label{line:blogapi:test:login}©
                    headers=headers, data=body)
     token_admin = token_admin + res_admin.json()["accessToken"] ©\label{line:blogapi:test:token}©


     # Fault100. HTTP Status 500. GET:/api/tags/{id}
     headers = {}
     headers["Authorization"] = token_admin # admin  ©\label{line:blogapi:test:header}©
     headers['Accept'] = "*/*"
     res_0 = requests \
                .get(self.baseUrlOfSut + "/api/tags/2",
                    headers=headers)

     assert res_0.status_code == 500
     assert "application/json" in res_0.headers["content-type"]
     assert res_0.json()["status"] == 500.0
     assert res_0.json()["error"] == "Internal Server Error"
     assert res_0.json()["message"] == "could not extract ResultSet; SQL [n/a]; nested exception is org.hibernate.exception.SQLGrammarException: could not extract ResultSet"
     assert res_0.json()["path"] == "/api/tags/2"
\end{lstlisting}
\caption{\label{fig:test:blogapi}
Test case generated by \evo in Python, in which authentication information has been exploited to make dynamic calls.
}
\end{figure}
%------------------------------------------------------

Figure~\ref{fig:test:blogapi} shows an example of a test generated by \evo using the authentication configuration for \emph{blogapi} in Figure~\ref{fig:auth:blogapi}.
The call to the login endpoint \texttt{/api/auth/signin} is done on Line~\ref{line:blogapi:test:login},
where the token is extracted from the field \texttt{accessToken} and combined with the ``\texttt{Bearer }'' prefix
on Line~\ref{line:blogapi:test:token}.
Then, this dynamically obtained token is used in the \texttt{Authorization} header in the following API calls towards the API, like for example done on Line~\ref{line:blogapi:test:header}.

Industry-strengths fuzzers can support outputs in different formats.
As of version 3.4.0~\cite{zenodo340evomaster}, \evo supports test outputs in Java, Kotlin, JavaScript and Python~\cite{arcuri2025widening}.
For example, the test shown in Figure~\ref{fig:test:ocvn} uses Java, whereas the one in Figure~\ref{fig:test:blogapi} uses Python.
This is needed to make the use of Web API fuzzers more widespread in industry.
However, this impacts how authentication configuration mechanisms can be designed.
For example, if using Python scripts to define authentication configurations (like in the case of RESTler and Schemathesis), then calling those scripts from different programming languages (e.g., JavaScript) might be cumbersome.
This could limit the number of potential programming language outputs that can be supported in a fuzzer, possibly harming its widespread usage in industry.
Using a declarative approach, e.g., based on YAML configuration files, where the HTTP calls are delegated to the fuzzer, solves this issue.

There can be other kinds of authentication mechanisms, besides the examples shown in this section.
An example is based on static information sent via headers, or when authentication tokens are obtained from a different service instead of the tested API.
Since its inception year in 2016, in \evo we have handled every single authentication mechanism we have encountered in any open-source API we have run experiments on, and all industrial APIs used by all of our industrial partners now and in the past (e.g., currently Fortune 500 enterprises such as Meituan and Volkswagen).
Still, there might be other less common ways to do authentication that we are not aware of, and that, as such, we  do not support yet.
For these edge cases, enabling a fuzzer to have custom scripts to do authentication is a reasonable fallback approach.

%------------------------------------------------------
\begin{figure}[!t]
\begin{lstlisting}[style=yaml,basicstyle=\footnotesize]
$schema: "https://json-schema.org/draft/2020-12/schema"
$id: "https://github.com/WebFuzzing/Commons/blob/master/src/main/resources/wfc/schemas/auth.yaml"
title: "Web Fuzzing Commons Authentication"
description: "Schema Definition for Web Fuzzing Commons Authentication"
type: object
properties:
  schema_version:
    type: string
    description: "The schema version of WFC needed to use to validate and process this document."
  auth:
    description: "List of authentication information for different users."
    type: array
    items:
      $ref: "#/$defs/AuthenticationInfo"
  authTemplate:
    description: "Optional authentication information template. This is used to avoid duplication in the auth list. \
                 Entries defined in the template will be applied to all elements in the auth list that do not specify them."
    allOf:
    - $ref: "#/$defs/AuthenticationInfo"
    - type: object
  configs:
    description: "Optional map of configuration parameters, in the form key:value strings. \
                  This can be useful to provide extra custom information in the same configuration file, \
                  independently of the defined authentication information."
    type: object
    additionalProperties:
      type: string
required: ["auth"]
$defs:
...
\end{lstlisting}
\caption{\label{fig:auth:schema}
Snippet of JSON Schema definition for our authentication configurations.
}
\end{figure}
%------------------------------------------------------

Having each fuzzer defining their own declarative approach would be detrimental, as these configuration files would not be reusable between fuzzers.
Furthermore, using different configuration approaches would increase the cognitive load for testers in industry.
To solve this major problem, in this paper we extracted the authentication configuration approach of \evo into its own separated project, independently of \evo.
To formally define how authentication configuration files should be written, we have designed a JSON Schema,\footnote{https://json-schema.org/} with extensive documentation to clarify the semantic of each different fields.
For reason of space, we do not show the full schema here in this paper, or discuss every single of its entries and fields, but a snippet can be seen in Figure~\ref{fig:auth:schema}.
This schema is now part of a new open-source project, called \emph{Web Fuzzing Commons} (WFC).
%, which is independent of \evo.
More details on WFC will be provided in Section~\ref{sec:wfc}.

%------------------------------------------------------
\begin{figure}[!t]
\begin{lstlisting}[language=Python,basicstyle=\footnotesize]
# auth.py
import requests
import schemathesis

@schemathesis.auth()
class TokenAuth:
    def get(self, case, ctx):
        response = requests.post(
            "http://localhost:8000/auth/token",
            json={"username": "demo", "password": "test"}
        )
        return response.json()["access_token"]

    def set(self, case, data, ctx):
        case.headers = case.headers or {}
        case.headers["Authorization"] = f"Bearer {data}"
\end{lstlisting}
\caption{
\label{fig:auth:schemathesis:python}
Example of Python authentication script from Schemathesis's documentation.\footref{link:schemathesis:auth}
}
\end{figure}

%------------------------------------------------------
\begin{figure}[!t]
\begin{lstlisting}[style=yaml,basicstyle=\footnotesize]
schemaVersion: "0.2.0"
auth:
  - name: demo_user
    loginEndpointAuth:
      verb: POST
      endpoint: /auth/token
      contentType: application/json
      payloadRaw: '{"username": "demo", "password": "test"}'
      token:
        extractFrom: body
        extractSelector: /access_token
        sendIn: header
        sendName: Authorization
        sendTemplate: "Bearer {token}"
\end{lstlisting}
\caption{\label{fig:auth:schemathesis:wfc}
WFC version of the Python authentication script from Schemathesis's documentation shown in Figure~\ref{fig:auth:schemathesis:python}.
}
\end{figure}

By using WFC, it is currently possible to define the following types of authentication mechanisms:
\begin{itemize}

\item static information that is sent via HTTP headers at each call (e.g., for the ``Basic'' HTTP Authentication Scheme).

\item calls to a login endpoint (which might be on an external server), where the authentication token is returned via a cookie.

\item calls to a login endpoint (which might be on an external server), where the authentication token needs to be extracted either from a header or a field in the returned body payload. The schema enables then to specify how the token needs to be sent in the following calls (e.g., if in headers or query parameters).

\end{itemize}

For a fuzzer, to implement WFC authentication it would require at least the following:

\begin{itemize}
\item parse the YAML/JSON configuration files written in WFC format.
\item make HTTP calls based on the provided information extracted from those files (e.g., endpoints, header values and body payloads).
\item read the JSON responses, parse them, and extract info from fields, specified for example with a JSON Pointer
(RFC 6901\footnote{\url{https://datatracker.ietf.org/doc/html/rfc6901}}).
\item add the obtained tokens in the fuzzer's following HTTP calls toward the API, in the location (e.g., HTTP headers) specified in the WFC configuration files.
\end{itemize}

As \evo fully implements WFC, its source code can be used as reference, or used it directly as it is, as long as the terms of its open-source license (i.e., LGPL) are not violated.

Furthermore, the complexity of writing the WFC configuration files does not seem so particularly different from writing code scripts directly.
As an example, Figure~\ref{fig:auth:schemathesis:python} shows an example of Python configuration script, from \schemathesis's documentation, to obtain an authentication token via a login endpoint.
Figure~\ref{fig:auth:schemathesis:wfc} shows the equivalent script in WFC configuration format.
In terms of complexity, we argue they are similar.

Defining a formal schema has also many advantages.
For example, the schema can be used to automatically validate the correctness of configuration files.
Furthermore, many programming languages have tool and library support to automatically generate code from these schemas,
like the \emph{jsonschema2pojo}\footnote{https://www.jsonschema2pojo.org/}
tool we use to generate POJO classes to parse these WFC auth configuration files.

WFC is an attempt to define \emph{standards} for common problems faced in fuzzing Web APIs.
How to define authentication information is one of such critical problems.
To facilitate its use among the different fuzzer developers and vendors,
we invite them all to join us and be part of the open-source WFC project. %, as WFC is independent of \evo.

{
Since its original definition coming from \evo, the authentication mechanism of WFC is now currently fully supported by other fuzzers such as \cats, \schemathesis and \wuppie.
After a democratic process among the authors of these fuzzers, several improvements have been carried out on the specifications of WFC (e.g., including renaming variables and adding more documentation and clarifications).

Implementing WFC in different fuzzers also raised semantic questions that the schema does not answer yet.
For example, when an authentication file defines several users (e.g., an administrator and a regular user), and the fuzzer is not told which one to use, CATS starts with the first entry, and switches to the next one if its baseline requests are rejected with 401 or 403 status codes.
Whether such behavior should be standardized in WFC is left to the democratic process among the participating fuzzers.
}

%%%%%%%%%%%%%%%%%%%%%%%%%%%%%%%%%%%%%%%%%%%%%%%%%%%%%%%%%%%%%%%%%%%%%%%%%%%%
\subsection{Candidate Standard for Fuzzer Report}
\label{sec:report}

When generating test cases, to assess their quality, there is a need to know what has been covered by these test cases.
Test cases can be run with the tested API instrumented for code coverage (e.g., JaCoCo for Java APIs).
However, there are many properties of REST APIs which would not result in these code coverage reports.
These would include which endpoints were called, what status codes were returned per endpoint, and which types of faults were detected.
This is important information for the final users.
As such, besides generating test cases, fuzzers often create some report files with this information as part of their outputs.

As each fuzzer would have its own, non-standardized way to report this information, comparing results between fuzzers on the same API is cumbersome.
To deal with this issue, there have been tools such as Restats~\cite{corradini2021restats} that takes as input an OpenAPI schema and HTTP logs of requests and responses (e.g., collected with an HTTP proxy, like the commercial Burp Proxy).
From this data, it can compute black-box, REST API coverage metrics, including faults related to the HTTP 500 status code.

A limitation here is on what kind of detected faults can be measured with this data.
Looking at HTTP 500 status codes is only one type of fault~\cite{marculescu2022faults}.
In the literature, different \emph{automated oracles} have been evaluated to detect different kinds of faults in REST APIs.
A common oracle is to check if, for each API response, such response is valid according to the API schema.
Other cases include \emph{robustness} testing~\cite{viglianisi2020resttestgen,laranjeiro2021black}, where invalid data (e.g., not matching what specified in the schema) is sent towards the API on purpose, and then the oracle checks if the responses are correctly classifying the requests as user errors (e.g., returning an HTTP status code in the 4xx family).
If the API returns a response in the 2xx success range, then it is a fault.

Security properties can  also be verified during fuzzing, and used as automated oracles for detecting faults.
These include for example attacks with SQL Injection and XSS payloads~\cite{deng2023nautilus,du2024vulnerability}, as well as checking for Mass Assignment vulnerabilities~\cite{corradini2023automated},
 access policy violations~\cite{arcuri2025fuzzing} and other security-related properties~\cite{atlidakis2020checking} (e.g., a deleted resource should no longer be accessible).

{
A REST API is based on HTTP.
To avoid misleading clients, APIs should follow the semantics of HTTP.
Compliance with HTTP is something that can be automatically verified during fuzzing.
For example, this could be based on checking for the proper use of status codes~\cite{decrop2026detecting} (e.g., a \texttt{DELETE} operation should never return a 201 ``Created'' response),
as well as for example creating specific sequences of HTTP calls to check different properties, such as idempotency for \texttt{PUT} endpoints~\cite{sahin2026validating}.

Not all automated oracles have been introduced in academic work.
Fuzzers developed in industry have also designed oracles, driven by practical needs.
For example, CATS checks
whether \texttt{GET} operations that succeeded before fuzzing start failing with server errors after write operations on the same path have been fuzzed (i.e., corrupted data was stored),
whether successful updates are reflected in subsequent reads, and
whether write operations carrying a stale If-Match ETag are accepted.
It also checks security properties such as command injection, NoSQL injection, excessive data exposure (e.g., writeOnly fields or credentials returned in responses) and missing security headers.
All these oracles have been added to the WFC fault catalog as well.
}

When a fault is detected, a fuzzer needs to explicitly inform the user of it.
There are different ways to do it.
In \evo, for example, we generate descriptive comments directly in the code of the generated tests cases where the faults are revealed (e.g., recall the example tests in Figure~\ref{fig:test:ocvn} and Figure~\ref{fig:test:blogapi}).

To be able to easily compare fuzzers for the different kinds of faults that they can detect, there is the need of a standard format to represent results.
Furthermore, for each different kind of automated oracles proposed in the literature, there is a need of a unique way to identify them in these reports.
To address this problem, in this paper we have designed a JSON Schema to represent fuzzer reports, where all the different oracles used in the literature (not just the ones implemented in \evo) have been assigned a unique code id.

%------------------------------------------------------
\begin{figure}[!t]
\begin{lstlisting}[style=yaml,basicstyle=\footnotesize]
$schema: "https://json-schema.org/draft/2020-12/schema"
$id: "https://github.com/WebFuzzing/Commons/blob/master/src/main/resources/wfc/schemas/report.yaml"
title: "Web Fuzzing Commons Report"
description: "Schema Definition for Web Fuzzing Commons Reports"
type: object
properties:
  # REQUIRED
  schema_version:
    type: string
    description: "The schema version of WFC needed to use to validate and process this document."
  tool_name:
    type: string
    description: "The name of the tool used to create the test cases reported in this document."
  tool_version:
    type: string
    description: "The version number of the used tool, e.g., 1.0.0."
  creation_time:
    type: string
    format: date-time
    description: "The timestamp of when this report file was created."
  faults:
    $ref: "#/$defs/Faults"
  problem_details:
    type: object
    properties:
      rest:
        $ref: "#/$defs/RESTReport"
  total_tests:
    type: integer
    minimum: 0
    description: "The total number of test cases generated by the tool."
  test_file_paths:
    type: array
    items:
      $ref: "#/$defs/TestFilePath"
    uniqueItems: true
    description: "The list of relative paths (compared to this document) of all the generated test suite files."
  test_cases:
    description: "Information on each generated test case."
    type: array
    items:
      $ref: "#/$defs/TestCase"
...
\end{lstlisting}
\caption{\label{fig:report:schema}
Snippet of JSON Schema definition for our test reports.
}
\end{figure}
%------------------------------------------------------

As already discussed for authentication configurations in Section~\ref{sec:auth}, using a formal schema has many advantages.
Different fuzzers can use the schema to make sure their reports are in the same format.
A formal schemas can resolve ambiguities among different implementations.
A schema can also be used to automatically generate code (e.g., POJO for Java) to parse and output these reports (e.g., in JSON files).
This schema, together with our code classification for the different automated oracles presented in the literature, is now part of the new open-source project WFC (more details in Section~\ref{sec:wfc}).
For reasons of space, we cannot show here the full schema, but Figure~\ref{fig:report:schema} shows an excerpt of it.

A fault would be uniquely identified by three properties:
its fault category unique code (e.g., 100), the endpoint that was called, and an optional ``context''.
For example, given the fault category ``Schema Validation'', for a response in a call for an endpoint, such a response might be invalid according to the schema for several reasons (e.g., a field was not supposed to be returned as null, whereas another one was supposed to be a date formatted in a specific way).
Each of these reasons would end up defining a different fault having a different ``context''.
Also, in the case of HTTP 500 status codes, for the same endpoint there could be more than one fault.
This could be determined based on content of server logs (if those are accessible to the fuzzer).
In white-box testing in \evo, we use the last executed line in the business logic of the API to define different contexts (e.g., see Line~\ref{line:ocvn:test:context} in Figure~\ref{fig:test:ocvn}).
{
In black-box testing, where no execution trace is available, other heuristics can be used.
For example, CATS derives the context of a 500 fault from a short root-cause summary extracted from the response body (e.g., exception type and message), and groups similar error responses based on string similarity metrics.
}

There might be several strategies to define a context.
At least for the time being, defining a standardized way for them is premature, as it requires more research.
{
As such, contexts should not be used when comparing the fault detection capability of different fuzzers.
Still, they are useful when presenting results to users, as they can provide extra information to the users to help them in debugging, and finally fixing, any detected fault.
}

A fault categorization needs to be open to deal with unknown fault category codes.
As part of research, new automated oracles can be defined and evaluated.
Until they have been published in a peer-reviewed venue, their authors might not be willing to share their descriptions, and ask to add a new unique code for them in WFC.
However, this potential issue can be easily solved by reserving a special range of codes (e.g., 900--999) dedicated for these experimental oracles not in WFC yet (i.e., no defined code in WFC would ever reserve a value in such 900--999 range).
The validity of these ranges is then enforced in the schema via number constraints.
{
As an example, CATS currently uses the code 907 to report unexpected API behaviors that do not match any of the categories defined and formalized in WFC yet (the process is ongoing at the time of this writing).
}

On the one hand,
one limitation of using test reports in fuzzer comparisons is that the reporting of detected faults is ``self-reported''.
Even excluding malicious cases, by mistake a fuzzer could wrongly report the number of faults it finds.
Fuzzer comparisons would hence be affected negatively.
As such, external ways to measure performance (e.g., code coverage, and HTTP 500 faults checked with an HTTP proxy) are still necessary.
On the other hand, ignoring all the different kinds of faults that can be automatically detected would only provide a limited picture of the compared fuzzers.
These reports can provide the needed extra information on fault detection, but care needs to be taken when using them to draw conclusions in scientific experiments.
{
This threat can be partially mitigated by cross-checking self-reported results against independent sources.
For example, the endpoints and the 500 status codes listed in a WFC report can be compared with the HTTP logs collected by the mitmproxy instances used in WFD (Section~\ref{sec:wfd}).
Reported faults that are inconsistent with these logs can then be flagged.
}

A further benefit of generating test reports (e.g., in JSON format), following a proposed standardized schema, is that it provides a clear contract for visualization.
For example, an HTML-based application can be used to read these report files and show a GUI to navigate and visualize these test results.
Such an effort to improve the usability of fuzzers in industry would then be detached from the fuzzer itself.
As discussed in more detail in the next section, we have built such a HTML-based application, and published it as part of WFC.
Any fuzzer that outputs report in WFC format would then be able to use such HTML-based application to visualize the test results in a professional manner.

Compared to supporting WFC authentication, supporting WFC reports should be significantly easier to implement for a fuzzer.
All data already collected about the generated test cases (e.g., their names and what status codes were returned in the HTTP calls) just need to be put in a JSON file that complies to our WFC schema.
Such file only needs to be written out: no need to read it and act upon its content.

% \begin{table}[t]
% 	\centering
% 	\caption{\label{table:wfc:faults} Mapping existing tool outputs to WFC fault categories}
% 	\resizebox{.95\textwidth}{!}{
% % 	\begin{tabular}{|l|p{6cm}|l|}
% 	\begin{tabular}{l|p{6cm}|l}
% % 		\hline
%         \toprule
% 		\textbf{Tool} & \textbf{Example Output} & \textbf{Mapped WFC Category} \\
% 		\midrule
% % 		\hline
%
% 		Schemathesis & API accepted schema-violating request & F102 -- Schema validation bypass \\
% % 		\hline
%
% 		Schemathesis & API rejected schema-compliant request & F101 -- Schema invalid response \\
% % 		\hline
%
% 		Schemathesis & 500 Internal Server Error / unsupported method returns 500 & F100 -- HTTP 500 \\
% % 		\hline
%
% 		RESTler & 5xx server error (main\_driver) & F100 -- HTTP 500 \\
% % 		\hline
%
% 		RESTler & Use-after-delete resource access & F103 -- Delete not working \\
% % 		\hline
%
% 		RESTler & Unauthorized access (Namespace rule violation) & F206 -- Wrong authorization \\
% 		%\hline
% 		\bottomrule
% 	\end{tabular}
% 	}
% \end{table}

If a fuzzer uses different type of oracles to detect faults, then such fault types need to be mapped to the right codes in our WFC fault categorization.
If there is no valid code representing some of these fault types, new temporary 9xx codes can be used, until such new fault types are added to WFC.

% To illustrate the applicability of our fault catalog, Table~\ref{table:wfc:faults} shows how some types of faults detected by popular fuzzers such as \schemathesis and \restler would map to WFC fault categories.

{
Due to the simplicity of implementing WFC reports, these are now fully supported by \cats, \evo, \schemathesis and \wuppie.
Currently, WFC provides definitions for \noracles different types of faults that existing fuzzers can automatically detect.
For example, seven of these categories (codes 123–125 and 313–316) originated from oracles already implemented in CATS, and were added to WFC through the same democratic process.
Since version 15.3.0, CATS maps its detected faults to 42 of these \noracles  categories.
}

%%%%%%%%%%%%%%%%%%%%%%%%%%%%%%%%%%%%%%%%%%%%%%%%%%%%%%%%%%%%%%%%%%%%%%%%%%%%
\subsection{Web HTML Report}
\label{sec:html}

When generating test cases with a fuzzer, it is important to enable user to visualize what has been covered.
This can help to visualize what is missing, and quickly check for any new detected faults.

%-------------------------
\begin{figure}[!t]
\includegraphics[width=1.0\textwidth]{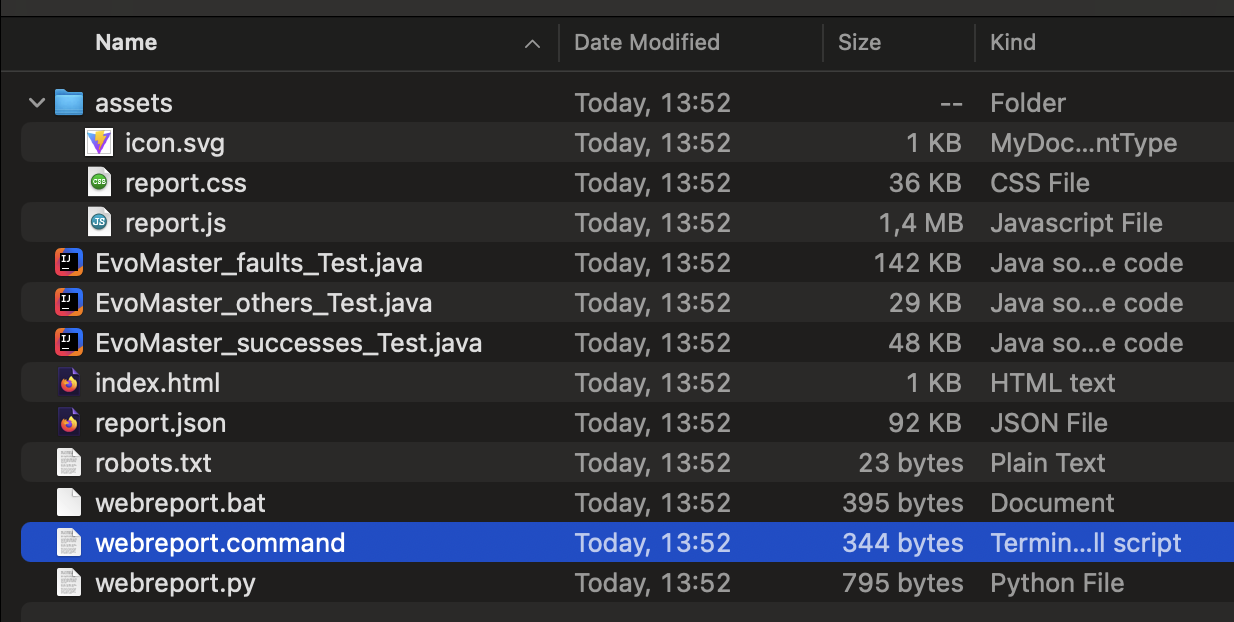}
\caption{
\label{fig:report:files}
Screenshot of files generated as output of \evo on the \emph{blogapi} API.
}
\end{figure}
%-------------------------

Given a set of test cases, and a \texttt{report.json} file with statistics about the fuzzing process, using the WFC format (Section~\ref{sec:report}), then the visualization of such data is independent from the fuzzer that generated these tests.
Such data can be visualized in a web application that takes those files as input, and build an interactive interface dynamically with JavaScript.
This web application, which would be a set of static files, can then be provided as output of the fuzzer together with the test suite files and the \texttt{report.json}.

Figure~\ref{fig:report:files} shows a screenshot of files generated as output by \evo when run for 1-minute on the \emph{blogapi} API.
The \texttt{.java} files are the test cases generated (e.g., using JUnit), together with the report file \texttt{report.json}.
All the other files are static resources for the web HTML visualization, like \texttt{index.html} and its asset files under the folder \texttt{assets} (e.g., images and JS/CSS files for the web application).
In modern browsers, an HTML page that is open on the local machine would not be able to access any other file, even in the same folder, due to security restrictions.
As such, the web application must be served via an HTTP server.
The Python script \texttt{webreport.py} starts an HTTP server with \texttt{index.html} as main page, and it automatically opens a new tab in the default browser on the local OS pointing to that page.
To simplify the starting of such server and opening of browser, e.g., with a mouse double-click, scripts for Windows (\texttt{webreport.bat}) and OSX (\texttt{webreport.command}) are provided.

%-------------------------
\begin{figure}[!t]
\includegraphics[width=1.0\textwidth]{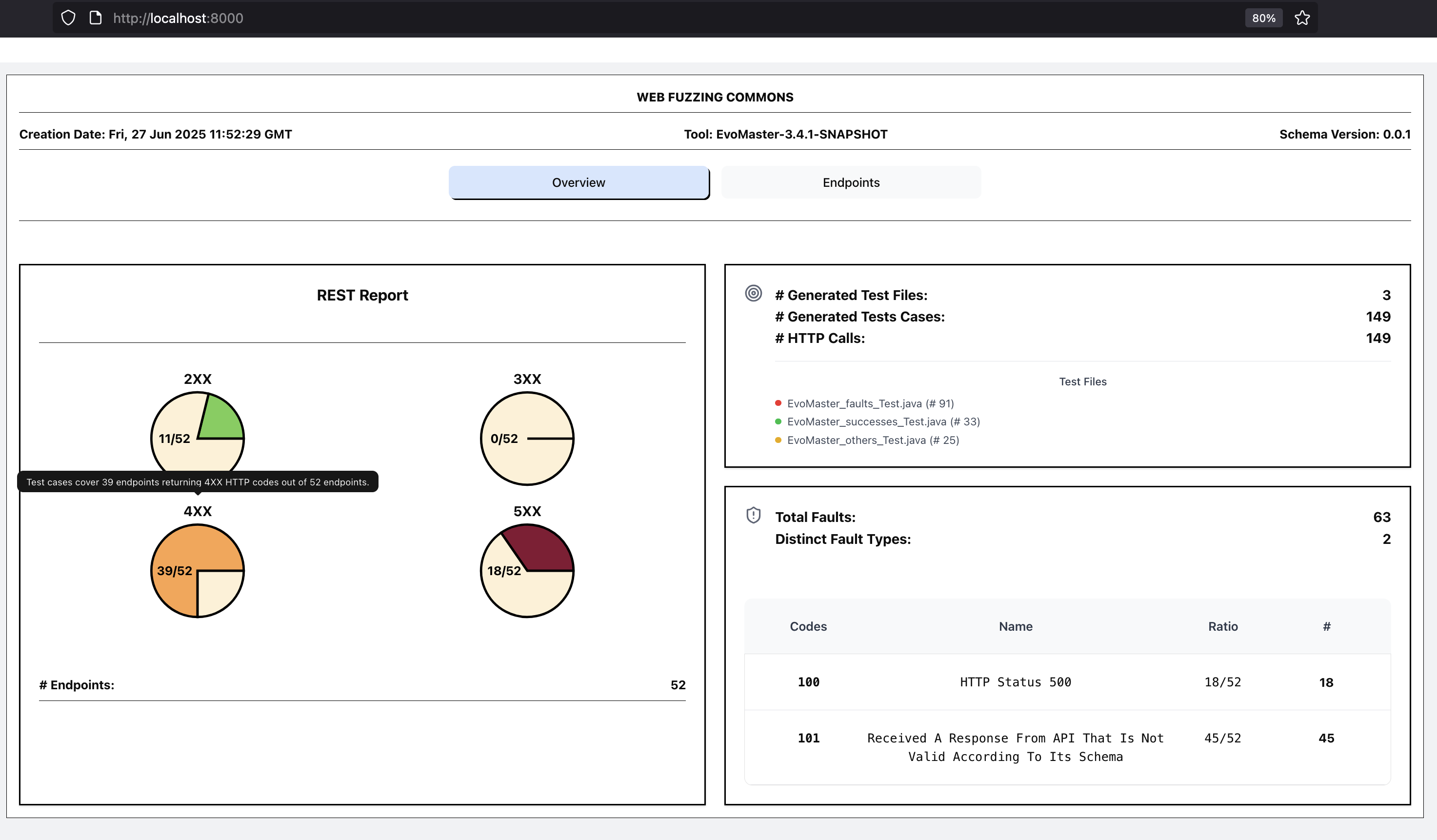}
\caption{
\label{fig:report:home}
Results of opening the \texttt{index.html} file in an HTTP server, e.g., by clicking on \texttt{webreport.command}.
}
\end{figure}
%-------------------------

Figure~\ref{fig:report:home} shows the home page for the web report, once opened it for example with \texttt{webreport.command}.
Different kinds of information are provided, based on the content of \texttt{report.json}.
This includes for example when the report was created, by which tool, and different properties of the tested API (e.g., number of
endpoints) and of the generated tests (e.g., covered HTTP status code per endpoint, and number and types of detected faults).
Hovering on the different elements does pop up informative tooltips.

%-------------------------
\begin{figure}[!t]
\includegraphics[width=1.0\textwidth]{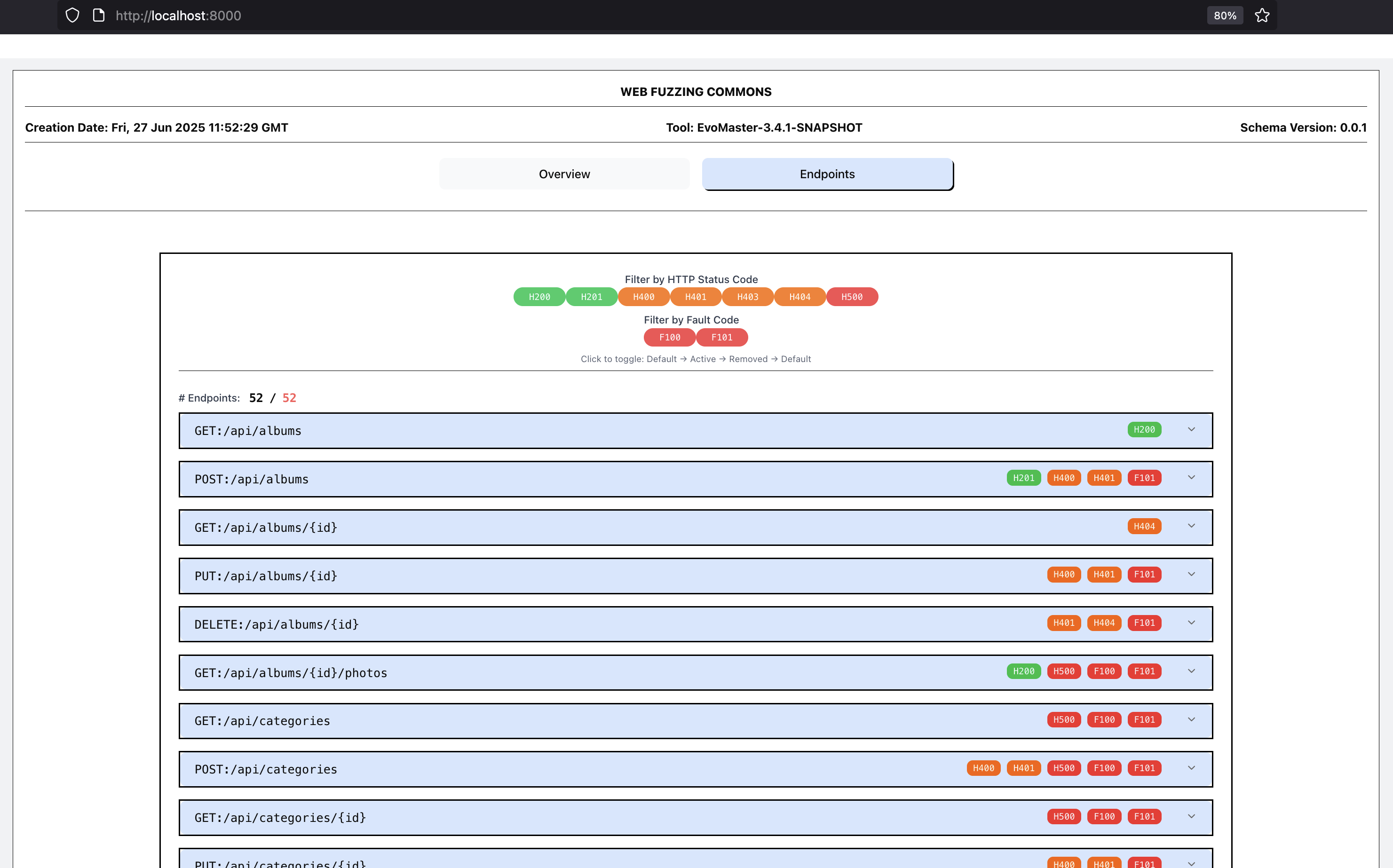}
\caption{
\label{fig:report:endpoints}
Results of opening the \texttt{Endpoints} tab, providing details per endpoint.
}
\end{figure}
%-------------------------

%-------------------------
\begin{figure}[!t]
\includegraphics[width=1.0\textwidth]{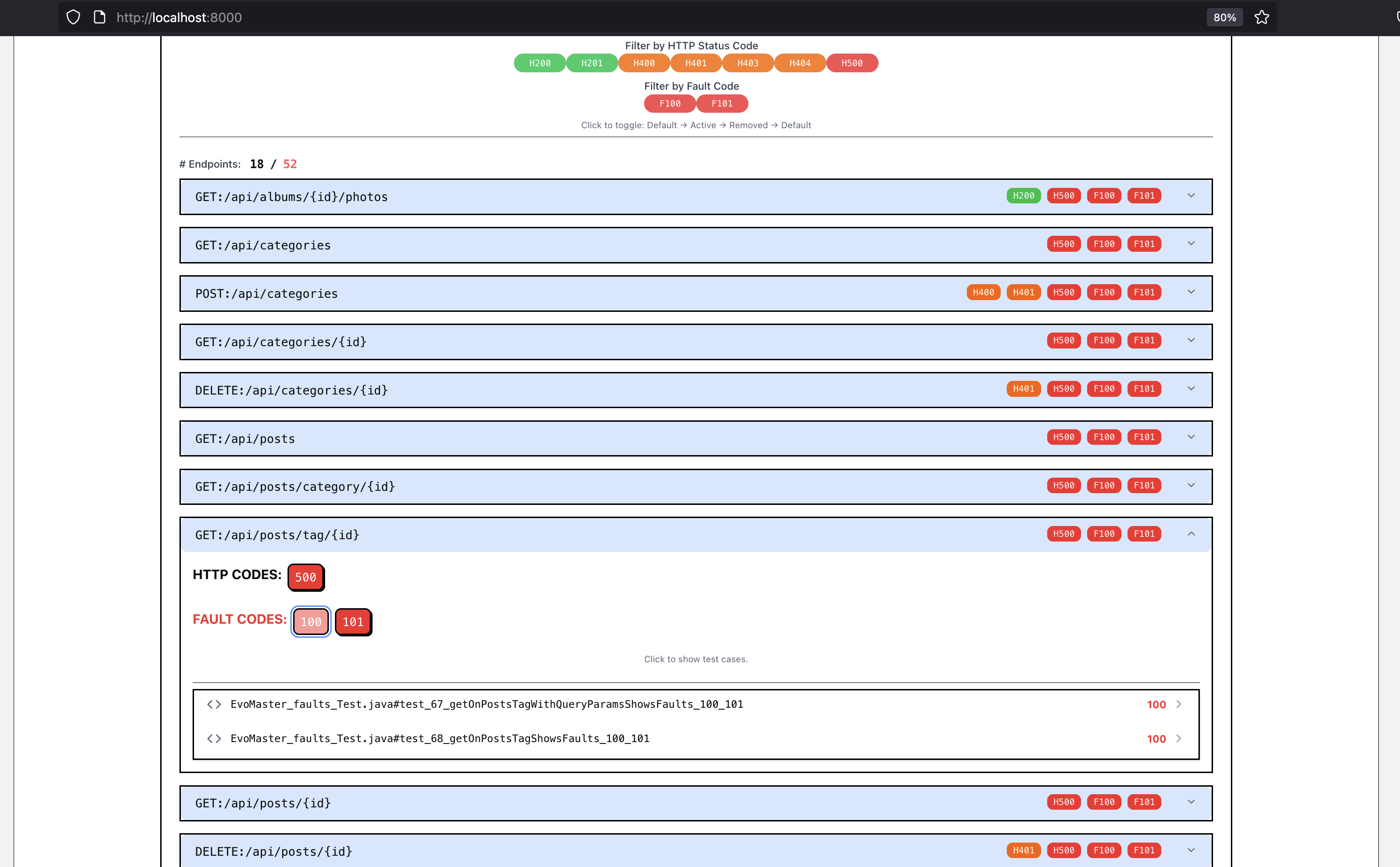}
\caption{
\label{fig:report:selection}
Results of opening the details for a specific endpoint.
}
\end{figure}
%-------------------------

By clicking on the \texttt{Endpoints} tab, results are shown for each endpoint (Figure~\ref{fig:report:endpoints}).
Endpoints can be filtered, e.g., based on HTTP status codes and faults that the generated test cases detect (or do not detect).
On each endpoint, and status/fault filter, it is possible to see the list of test cases that cover/not-cover them (Figure~\ref{fig:report:selection}).

%-------------------------
\begin{figure}[!t]
\includegraphics[width=1.0\textwidth]{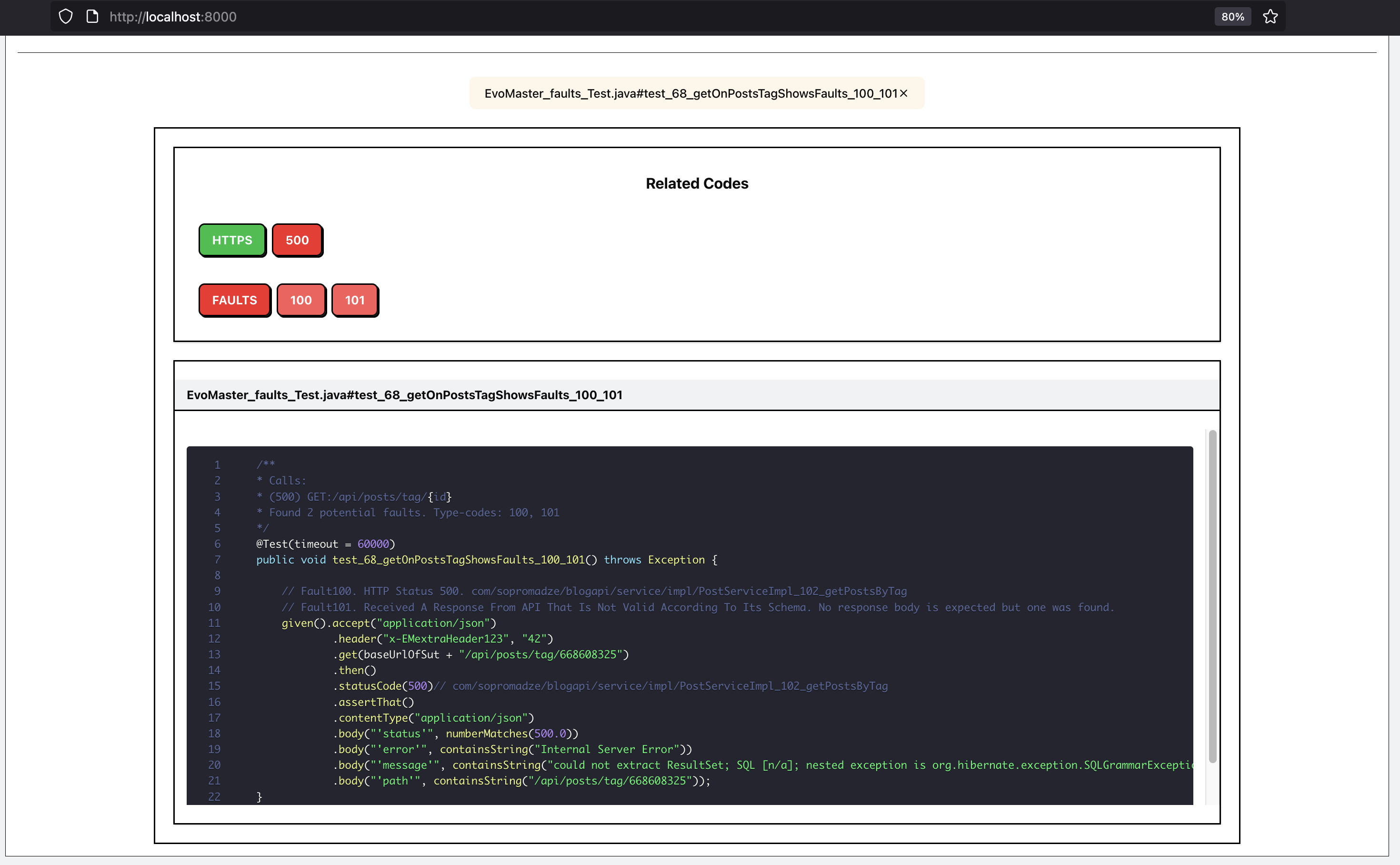}
\caption{
\label{fig:report:test}
Code of a test case, visualized in the WFC Report.
}
\end{figure}
%-------------------------

To ease visualization and review/analysis of the tests, each single test can be opened in the HTML application itself by clicking on it (e.g., see Figure~\ref{fig:report:test}).
Note, to be able to achieve this, the generated \texttt{report.json} needs to keep track of the location on the file system of where the test suite is (e.g., \texttt{./EvoMaster\_faults\_Test.java}) relatively to the \texttt{index.html} file, as well as the starting and end line of test case in that Java file.
The JavaScript code of the web application (i.e., \texttt{assets/report.js}) is then going to read that file and extract the needed text from it.
As long as a fuzzer generates a \texttt{report.json} file in WFC format, all these other steps are fully automated.

%%%%%%%%%%%%%%%%%%%%%%%%%%%%%%%%%%%%%%%%%%%%%%%%%%%%%%%%%%%%%%%%%%%%%%%%%%%%
\subsection{WFC Open-Source Tool Support}
\label{sec:wfc}

The Web Fuzzing Commons (WFC) repository (open-source on GitHub\footref{foot:wfc}) hosts the main contribution of this work.
In particular, it contains the
\texttt{auth.yaml} schema defining authentication information (Section~\ref{sec:auth}),
the \texttt{report.yaml} schema for report info together with the fault catalog \texttt{fault\_categories.json} (Section~\ref{sec:report}),
and the web application assets to display the report interactively (Section~\ref{sec:html}).

These schemas and web assets can be accessed directly from GitHub, and from its repository releases.\footnote{https://github.com/WebFuzzing/Commons/releases}
Also, each new release on GitHub is automatically uploaded on Zenodo~\cite{wfc010}.
To further simplify their access, all these static resources are also published as a library, e.g., for the JVM
on Maven Central,\footnote{https://central.sonatype.com/artifact/com.webfuzzing/commons}
{
 and for Python on PyPI.\footnote{\url{https://pypi.org/project/webfuzzing-commons/}}
}
Packaging these files as a library for other programming languages (e.g., {JavaScript}) is something worth to consider in the future.

If there is some authentication mechanism that we have not considered, we invite the community to report an issue on this repository.
The \texttt{auth.yaml} schema can be extended to support its declaration.
Furthermore, if novel oracles are designed to find new different kinds of faults in REST APIs, a new entry (with unique id) can be added to \texttt{fault\_categories.json}, together with a description of these new types of detected faults.
WFC is independent of {\cats, \evo, \schemathesis and \wuppie}.
New authentication mechanisms and fault categories can be added regardless of whether {these fuzzers} support them yet or not.

For a fuzzer, deciding to support WFC provides several benefits.
A common standard for authentication declarations simplifies the use of real-world APIs for experiments, as authentication info needs to be defined only once, and not for each different compared fuzzers.
If a fuzzer exports test reports in WFC format, and the detected faults are labeled with the same unique ids assigned in WFC, then comparisons of detected faults become simpler.
Furthermore, a further benefit is that interactive HTML reports can be automatically generated if using this WFC output format.

Due to the importance of the issues addressed by WFC,
{
in 2025 the authors of \evo
}
directly invited on GitHub the developers of the most used open-source fuzzers in industry (measured in number of GitHub stars) {at that time}, i.e.,
\schemathesis\footnote{\url{https://github.com/schemathesis/schemathesis/discussions/3123}}
(+2900 stars),
\restler\footnote{\url{https://github.com/microsoft/restler-fuzzer/issues/983}}
(+2900 stars)
and
CATS\footnote{\url{https://github.com/Endava/cats/issues/178}}
(+1300 stars).

At a later point in time, {it was decided} to also directly contact the authors of all the fuzzers that have at least 100 GitHub stars, and that had a new release within the last 12 months.
This process added only
\wuppie\footnote{\url{https://github.com/TNO-S3/WuppieFuzz/issues/262}}
to the list of fuzzer authors to contact.

{
At the time of writing in 2026,
the developers of \cats, \schemathesis,  and \wuppie have joined the WFC project, and already changes and improvements have been made to the specs to specify authentication information (an example is the handling of tokens\footnote{\url{https://github.com/WebFuzzing/Commons/pull/19}}).
Support for WFC in these fuzzers is currently fully implemented.

However, we received no reply from the maintainers of \restler.
With the passing of more time, it turned out that \restler is no longer maintained (last release $9.3.1$ is from January 2025).
This means that all of the most used, and still maintained, REST API fuzzers in industry are currently involved  and support WFC.
}

Changes in WFC are based on a democratic process of consensus among the representative of all the different fuzzers that have currently joined WFC.
There is no point in having specs that are only by name if then the different fuzzer maintainers do not agree in implementing such specs in their fuzzers.

Due to academic constraints related to conflicts of interest (e.g., any authors of fuzzers that joins WFC would have a conflict of interest with the academic reviewing of this work), we plan to directly contact the maintainers of all existing fuzzers to join WFC only once this article has passed peer-reviewing.

%%%%%%%%%%%%%%%%%%%%%%%%%%%%%%%%%%%%%%%%%%%%%%%%%%%%%%%%%%%%%%%%%%%%%%%%%%%%

\section{Web Fuzzing Dataset}
\label{sec:all_wfd}

%----------------------------------------------------------------------------
\subsection{API Selection Issues in Empirical Studies}
\label{sec:selection-issues}

To analyze the performance of a designed novel technique, and to compare it with the state-of-the-art, an empirical study on a set of artifacts is required.
In our context, those would be REST APIs.
Given $Z$ the set of all existing APIs, ideally for experiments one would select a \emph{random} subset  $z \subset Z$ with size $|z|=n$~\cite{FrA12b}, where $n$ is large enough to obtain statistical significance~\cite{Hitchhiker14}.

Unfortunately, this is not feasible, for a few reasons.
First, a large number (arguably the large majority) of APIs are in industry.
Researchers might have no access to those APIs, unless they established some sort of industry-academia collaboration.
And, even in this case, the number of collaborations would be necessarily limited in number, which would not lead to a good representative for $Z$.
Second, selecting open-source APIs is not trivial either.
Finding and identifying REST APIs among millions of projects hosted on open-source repositories such as GitHub has its set of challenges.
Furthermore, one would need to make sure to be able to compile and run any identified API, setting up all their needed environments (e.g., databases such as PostgreSQL, MySQL, MongoDB and Redis, as well as other services such as for example OAuth2).
This takes time and resources.
This is a reason why early work on fuzzing REST APIs only included a limited number of open-source APIs in the experiments.
For example, the earliest work in the literature, that we are aware of~\cite{golmohammadi2023testing},
is from 2009~\cite{chakrabarti2009test}, where only 1 in-house API was used in the experiments.
Old fuzzers used a limited number of open-source APIs when first introduced, like \evo in 2017~\cite{arcuri2017restful} using only 3 APIs, and RESTler in 2019~\cite{restlerICSE2019} using only 2 APIs.
The two largest studies with open-source projects, that we are aware of, came years later.
Those are from the authors of
\aratrl and \llamaresttest in 2022 (20 APIs employed in~\cite{Kim2022Rest})
and from us authors of \evo in 2023 (19 APIs employed in~\cite{zhang2023open}).

A possible alternative is to use \emph{online} APIs for the experiments.
At the time of writing, \texttt{apis.guru}\footref{foot:apisguru} lists more than 2 500 APIs online on internet with available schemas.
In the literature~\cite{golmohammadi2023testing}, there have been large studies using online APIs.
For example, the authors of RestTestGen used 107 online APIs for experimentation in~\cite{corradiniautomated2022} (as well as another 13 open-source APIs).
Unfortunately, using online APIs for experimentation has some major drawbacks.
First, online APIs might (most likely) have \emph{rate-limiters}, to avoid denial-of-service attacks.
This might significantly hinder experimentation with fuzzers.
Second, experiments on online APIs would be, most of the time, not \emph{repeatable}, as APIs could change at any time without warning.
Repeatability is essential for scientific research.
Furthermore, in experiments with online, remote APIs there would be no possibility to collect backend-related measures such as the achieved code coverage by the generated test cases.
Using  online APIs can provide extra information in an empirical study, but we argue they must still be accompanied  with repeatable experiments on open-source projects.

To select a set of artifacts for experiments in a new study $X$, if it is not feasible to sample artifacts from $Z$, a reasonable approach is to use what was employed in the state-of-the-art $Y$.
A good example is the work done for  \deeprest~\cite{corradini2024deeprest}, where the employed APIs are a superset of what was previously used when \aratrl~\cite{kim2023adaptive} was introduced.
Still, even in this case, there would be the question on whether the selection in $Y$ was \emph{biased} or \emph{unrepresentative}.
For example, the study in $Y$ could have focused only on APIs on which the introduced novel technique was performing best, and discarded all APIs in which it was giving worse results.
This could be done due to fearing that some \emph{negative results} could hinder the publication acceptance likelihood of $Y$.
Although this is clearly unethical, it is an issue that is not easy to spot.

Given $z_y$ the set of artifacts used in the experiments in the study $Y$, an approach when evaluating a new study $X$ would be to use a superset of $z_y$, i.e., $z_x=z_y \cup k$, for some new set of artifacts $k$.
This is for example what was done for \deeprest in~\cite{corradini2024deeprest}, where one new API was added to the set of experiments compared to~\cite{kim2023adaptive}.
If this approach is then repeated in future studies, using now $z_x$ instead of $z_y$, as long as the new $|k'|\geq 1$, this could solve the issues regarding bias and un-representativeness, as artifact selection would become community-driven.
Future studies that would not use $z_x$, or just use a subset of it, should then have a hard-time to justify such decision, unless there are good scientific reasons to do so.

Still, this approach has some side-effects that must be taken into account when reading and evaluating empirical studies.
Typically, research prototypes are not fully-fledged, commercial-grade products, especially when developed by only a single PhD student.
A new technique studied in $X$ could be better on the new APIs $k$ just because the tool presented in $Y$ is not \emph{robust}, e.g., crashing due to some unexpected non-supported feature, or simply due to a software fault.
Building a robust prototype that can reliably work on new APIs is no easy feat.
The technique/tool proposed in $X$ might crash as well on $k$, but, then, the authors of $X$ would have all the time to fix any issue in their tools regarding $k$ before attempting to publish $X$.
As such, when analyzing the results on $k$, this possible issue has to be taken into account and discussed.

Another issue, when it comes to API selection, is the technical scaffolding needed to run the APIs and collect different metrics (e.g., code coverage).
This is programming language dependent.
On the one hand, this is just technical work regarding setting up experiment infrastructures.
It is not a research problem.
On the other hand, if running experiments on APIs written in different programming languages provides no scientific benefits for black-box testing experiments, then this would be technical work and time that could had been better spent in other activities (e.g., designing novel techniques).
In the research literature~\cite{golmohammadi2023testing}, it seems that most of the empirical studies involving open-source projects employed APIs written in Java, using popular tools such as
JaCoCo\footnote{https://www.jacoco.org/}
to collect coverage metrics.
As long as it is possible to collect and employ a large number $n$ of APIs for experiments, concentrating on a single programming language is a reasonable compromise, at least for the time being.

%%%%%%%%%%%%%%%%%%%%%%%%%%%%%%%%%%%%%%%%%%%%%%%%%%%%%%%%%%%%%%%%%%%%%%%%%%%%
\subsection{Selection of APIs}
\label{sec:wfd}

To evaluate the performance of new techniques for fuzzing REST APIs, as previously discussed a set of APIs is needed for experimentation.
To enable the replicability of the studies, which is paramount for sound scientific research, ideally those APIs should be open-source.
APIs online could change at any time, so they do not lead to repeatable experiments.
Industrial, closed-source APIs can provide more realism (as we did with studies at Meituan~\cite{zhang2023rpc,zhang2024seeding,zhang2025fuzzing} and Volkswagen~\cite{poth2025technology,icst2025vw}), but still they do not enable replicated studies.
As such, using open-source APIs is a necessity, where online and industrial APIs can be added to provide more information.

In the first study published for \evo in 2017, only 2 open-source APIs were used~\cite{arcuri2017restful}, namely \emph{features-service} and \emph{scout-api}.
At that time, finding open-source APIs was difficult (e.g., lack of good search functionalities) and time consuming.
With the passing of the years, at each new study we added new APIs we could find, compile and run without major issues.
Also, when other research groups did studies on fuzzing REST APIs, like for example the study in~\cite{Kim2022Rest} where 20 open-source APIs were used, we checked those APIs and included several of them, if in Java or Kotlin, in our studies as well.
This was done to avoid the possible risk and concerns of cherry-picking APIs where \evo gives better results compared to the alternatives.

Throughout the years, we collected all these APIs in a repository called EMB (EvoMaster Benchmark), which is open-source on GitHub\footnote{https://github.com/WebFuzzing/EMB} since 2017.
In 2023, we wrote a ``tool paper'' in which we described EMB~\cite{icst2023emb}.
At that time, EMB contained 14 REST APIs for the JVM (i.e., written in either Java or Kotlin),
{
plus another 6 REST APIs in JavaScript, and 9 APIs for GraphQL and RPC.
}

One further reason to create a repository such as EMB was that, to do \emph{white-box} testing of REST APIs, there is the need to write \emph{driver} classes in Java/Kotlin~\cite{arcuri2019restful}.
Those (small) driver classes are needed to start and stop the APIs.
When they are started, their bytecode is automatically instrumented to do different kinds of white-box testing analyses~\cite{arcuri2020sql,arcuri2021enhancing,arcuri2024advanced}.
To be able to run experiments, we created those driver classes and included them in EMB.
EMB can be used to run experiments for \emph{black-box} testing as well, as we did in~\cite{zhang2023open}.
However, it was cumbersome, especially when dealing with non-embedded databases (e.g., PostgreSQL and MySQL instead of H2), and dealing with authentication configurations.

\begin{table*}[t]
	\centering
\caption{
{
Subset of APIs in WFD that have been used in some previous studies.
}
Note that those other studies might include as well non-JVM APIs, non-open-source (e.g., industrial) and online APIs.
These previous studies are EMB~\cite{icst2023emb},  the Comparison~\cite{Kim2022Rest} study of Kim et.~al, \aratrl~\cite{kim2023adaptive},  \emrest~\cite{xu2025effective}, \deeprest~\cite{corradini2024deeprest} and \llamaresttest~\cite{kim2025llamaresttest}.
\label{tab:tablerefs}
}
\vspace{-1.5\baselineskip}
\resizebox{.95\textwidth}{!}{
\begin{tabular}{l cccccc}\\
\toprule 
% SUT & EMB~\cite{icst2023emb} & Comparison~\cite{Kim2022Rest} & \aratrl~\cite{kim2023adaptive} & \emrest~\cite{xu2025effective} & \deeprest~\cite{corradini2024deeprest} & \llamaresttest~\cite{kim2025llamaresttest}  \\
SUT & EMB & Comparison & \aratrl & \emrest & \deeprest & \llamaresttest  \\
\midrule
% lightgrey
% \rowcolor{grey} \emph{bibliothek}              &             & & & & & \\
% \rowcolor{grey} \emph{familie-ba-sak}          &             & & & & & \\
% \rowcolor{grey} \emph{http-patch-spring}       &             & & & & & \\
% \rowcolor{grey} \emph{microcks}                &             & & & & & \\
% \rowcolor{grey} \emph{ohsome-api}              &             & & & & & \\
% \rowcolor{grey} \emph{pay-publicapi}           &             & & & & & \\
% \rowcolor{grey} \emph{quartz-manager}          &             & & & & & \\
% \rowcolor{grey} \emph{reservations-api}        &             & & & & & \\
% \rowcolor{grey} \emph{session-service}         &             & & & & & \\
% \rowcolor{grey} \emph{spring-ecommerce}        &             & & & & & \\
% \rowcolor{grey} \emph{swagger-petstore}        &             & & & & & \\
% \rowcolor{grey} \emph{tiltaksgjennomforing}    &             & & & & & \\
% \rowcolor{grey} \emph{webgoat}                 &             & & & & & \\
\rowcolor{lightgrey} \emph{erc20-rest-service}      &             & \checkmark & & & & \\
\rowcolor{lightgrey} \emph{spring-actuator-demo}    &             & \checkmark & & & & \\
\rowcolor{lightgrey} \emph{spring-batch-rest}       &             & \checkmark & & & & \\
\rowcolor{lightgrey} \emph{spring-rest-example}     &             & \checkmark & & & & \\
\rowcolor{grey} \emph{catwatch}                & \checkmark  &  & & & & \\
\rowcolor{grey} \emph{gestaohospital}          & \checkmark  &  & & & & \\
\rowcolor{lightgrey} \emph{cwa-verification}        & \checkmark  & \checkmark & & & & \\
\rowcolor{lightgrey} \emph{ocvn}                    & \checkmark  & \checkmark & & & & \\
\rowcolor{lightgrey} \emph{proxyprint}              & \checkmark  & \checkmark & & & & \\
\rowcolor{lightgrey} \emph{rest-news}               & \checkmark  & \checkmark & & & & \\
\rowcolor{lightgrey} \emph{scout-api}               & \checkmark  & \checkmark & & & & \\
\rowcolor{grey} \emph{rest-ncs}                & \checkmark  & \checkmark & \checkmark & \checkmark & \checkmark &            \\
\rowcolor{grey} \emph{rest-scs}                & \checkmark  & \checkmark & \checkmark & \checkmark & \checkmark &            \\
\rowcolor{lightgrey} \emph{features-service}        & \checkmark  & \checkmark & \checkmark & \checkmark & \checkmark & \checkmark \\
\rowcolor{lightgrey} \emph{languagetool}            & \checkmark  & \checkmark & \checkmark & \checkmark & \checkmark & \checkmark \\
\rowcolor{lightgrey} \emph{restcountries}           & \checkmark  & \checkmark & \checkmark & \checkmark & \checkmark & \checkmark \\
\rowcolor{lightgrey} \emph{genome-nexus}            & \checkmark  & \checkmark & \checkmark & \checkmark & \checkmark & \checkmark \\
\rowcolor{lightgrey} \emph{market}                  & \checkmark  & \checkmark & \checkmark & \checkmark & \checkmark & \checkmark \\
\rowcolor{grey} \emph{user-management}         &             & \checkmark & \checkmark & \checkmark & \checkmark & \checkmark \\
\rowcolor{grey} \emph{person-controller}       &             & \checkmark & \checkmark & \checkmark & \checkmark & \checkmark \\
\rowcolor{grey} \emph{tracking-system}         &             & \checkmark & \checkmark & \checkmark & \checkmark & \checkmark \\
\rowcolor{lightgrey} \emph{blogapi}                 &             &            &            &            & \checkmark &             \\
\rowcolor{grey} \emph{youtube-mock}            &             &            &            &            &            & \checkmark  \\
\midrule
Total 36 & 14 & 19 & 10 & 10 & 11 & 9 \\
\bottomrule 
\end{tabular} 

}
\end{table*}

Although the use of 14 JVM APIs provides a large case study, larger than in many recent studies (e.g., \cite{kim2023adaptive,corradini2024deeprest,foley2025apirl,kim2025llamaresttest}), we considered it still bit too small for sound empirical evidence, especially when considering experiment practices in other testing domains like unit testing~\cite{FrA12b}.
To address this issue, we have extended it with several more APIs.
Our selection criteria were threefold:
(1) we added any large (at least 20 000 LOCs), complex API that we could find searching on GitHub;
(2) we searched and added APIs for important technologies that we thought were underrepresented in EMB (e.g., Kafka and ElasticSearch);
(3) we looked at all the existing studies on REST APIs we are aware of, and added \emph{all} open-source Java APIs used in those studies.

{
As the work on EMB was bootstrapped by the authors of \evo, in theory in this selection there could be a potential bias in favor of \evo.
When the authors of \cats, \schemathesis and \wuppie joined in this work, one of the first actions was to discuss such potential issue.
New APIs were added to WFD, but none was deemed biased towards \evo and had to be removed.\footnote{\url{https://github.com/WebFuzzing/Evaluations/discussions/4}}
}

At the time of writing, our selection has now \nsuts REST APIs for the JVM, i.e., more than twice and half than what original used in EMB~\cite{icst2023emb}.
Table~\ref{tab:tablerefs}
shows a subset of these APIs, specifying some existing studies in which they have been used before.
Due to space constraints, there is some information that did not fit in the table.
In particular,
\emph{microcks} was used in VoAPI2~\cite{du2024vulnerability},
and \emph{swagger-petstore} in Morest~\cite{liu2022icse}.
The work for \llamaresttest~\cite{kim2025llamaresttest} used \emph{ohsome-api}, but its online version (and not its open-source version run locally).

%----------------------------------------------------------
\begin{table}[p]
	\centering
	\small
	\caption{ Statistics of the employed REST APIs in our empirical study, including number of source files, numbers of lines of code (LOCs), the number of HTTP endpoints, whether they require authentication credentials or not, {and which databases they employ (if any)}.
		\label{tab:suts}
	}
	\vspace{-1.5\baselineskip}
	\begin{adjustbox}{width=.95\textwidth,center}
	\begin{tabular}{l rrrrr}\\
\toprule
SUT & \#SourceFiles & \#LOCs & \#Endpoints & Auth& Databases\\
\midrule
\emph{adoptme} &  41 &  6158 &  14 &  &  Neo4j \\
\emph{arimaa} &  141 &  9460 &  38 & \checkmark &  MySQL;MongoDB;Neo4j \\
\emph{bibliothek} &  33 &  2176 &  8 &  &  MongoDB \\
\emph{blogapi} &  89 &  4787 &  52 & \checkmark &  MySQL \\
\emph{catwatch} &  106 &  9636 &  23 &  &  H2 \\
\emph{commafeed} &  201 &  21287 &  44 & \checkmark &  H2 \\
\emph{cwa-verification} &  47 &  3955 &  5 &  &  H2 \\
\emph{digitalbanking} &  27 &  2835 &  55 &  &  Redis;Cassandra \\
\emph{erc20-rest-service} &  7 &  1378 &  13 &  &   \\
\emph{familie-ba-sak} &  1089 &  143556 &  183 & \checkmark &  PostgreSQL \\
\emph{features-service} &  39 &  2275 &  18 &  &  H2 \\
\emph{flight-search-api} &  194 &  14440 &  40 & \checkmark &  MongoDB \\
\emph{gdpr-kv} &  62 &  6399 &  7 &  &  DynamoDB \\
\emph{genome-nexus} &  405 &  30004 &  23 &  &  MongoDB \\
\emph{gestaohospital} &  33 &  3506 &  20 &  &  MongoDB \\
\emph{gravitee} &  7349 &  886852 &  1007 & \checkmark &  MongoDB \\
\emph{http-patch-spring} &  30 &  1450 &  6 &  &   \\
\emph{jasper} &  337 &  46791 &  61 & \checkmark &  PostgreSQL \\
\emph{joinus} &  101 &  7906 &  37 & \checkmark &  MongoDB;Neo4j \\
\emph{kafka-rest} &  475 &  81418 &  50 &  &  Kafka \\
\emph{komga} &  543 &  62303 &  174 & \checkmark &  SQLite \\
\emph{languagetool} &  1385 &  174781 &  2 &  &   \\
\emph{lovemining} &  25 &  1793 &  18 & \checkmark &  MongoDB;Neo4j \\
\emph{management-cassandra} &  225 &  35096 &  63 &  &  Cassandra \\
\emph{market} &  124 &  9861 &  13 & \checkmark &  H2 \\
\emph{microcks} &  471 &  66186 &  88 & \checkmark &  MongoDB \\
\emph{movies-xml} &  22 &  2502 &  7 &  &  H2 \\
\emph{notebook-manager} &  10 &  826 &  5 &  &  MySQL \\
\emph{ocvn} &  526 &  45521 &  258 & \checkmark &  H2;MongoDB \\
\emph{ohsome-api} &  87 &  14166 &  134 &  &  OSHDB \\
\emph{pay-publicapi} &  377 &  34576 &  16 & \checkmark &  Redis \\
\emph{person-controller} &  16 &  1112 &  12 &  &  MongoDB \\
\emph{petclinic-rest} &  109 &  8675 &  43 &  &  H2 \\
\emph{pn-paper-tracker} &  254 &  24197 &  8 &  &  DynamoDB \\
\emph{proxyprint} &  73 &  8338 &  115 & \checkmark &  H2 \\
\emph{quartz-manager} &  129 &  5068 &  11 & \checkmark &   \\
\emph{redis-sample} &  18 &  1006 &  4 &  &  Redis \\
\emph{reservations-api} &  39 &  1853 &  7 & \checkmark &  MongoDB \\
\emph{rest-ncs} &  9 &  605 &  6 &  &   \\
\emph{rest-news} &  11 &  857 &  7 &  &  H2 \\
\emph{rest-scs} &  13 &  862 &  11 &  &   \\
\emph{restcountries} &  24 &  1977 &  22 &  &   \\
\emph{scout-api} &  93 &  9736 &  49 & \checkmark &  H2 \\
\emph{session-service} &  15 &  1471 &  8 &  &  MongoDB \\
\emph{spring-actuator-demo} &  5 &  117 &  2 & \checkmark &   \\
\emph{spring-batch-rest} &  65 &  3668 &  5 &  &   \\
\emph{spring-ecommerce} &  58 &  2223 &  27 & \checkmark &  MongoDB;Redis;Elasticsearch \\
\emph{spring-rest-example} &  32 &  1426 &  9 &  &  MySQL \\
\emph{swagger-petstore} &  23 &  1631 &  19 &  &   \\
\emph{testing-system} &  72 &  3692 &  11 & \checkmark &  PostgreSQL;Redis \\
\emph{tiltaksgjennomforing} &  472 &  27316 &  79 & \checkmark &  PostgreSQL \\
\emph{traccar} &  1478 &  141962 &  117 & \checkmark &  H2 \\
\emph{tracking-system} &  87 &  5947 &  67 & \checkmark &  H2 \\
\emph{user-management} &  69 &  4274 &  21 &  &  MySQL \\
\emph{webgoat} &  355 &  27638 &  204 & \checkmark &  H2 \\
\emph{youtube-mock} &  29 &  3229 &  1 &  &   \\
\midrule
Total 56 & 18149 & 2022760 & 3347 & 25 & 45 \\
\bottomrule
\end{tabular}
	\end{adjustbox}
\end{table}
%----------------------------------------------------------

Regarding the properties of these selected \nsuts REST APIs,
Table~\ref{tab:suts}
shows some statistics on these  APIs.
Full details (including source code and links to their original repositories)
are available in our open-source
repository.\footnote{\url{https://github.com/WebFuzzing/Dataset}}
There is a large variety of size and complexity in these APIs.
Some are just small example APIs, others are real-world APIs.
For example, few governments around the world have their public sector IT systems published as open-source,
like for example the cases of Germany (\emph{cwa-verification}), Norway (\emph{familie-ba-sak} and \emph{tiltaksgjennomforing}), the United Kingdom (\emph{pay-publicapi}) and Vietnam (\emph{ocvn}).
Also, there are commercial tools such as \emph{languagetool} that have open-source REST APIs.
These APIs use a variety of databases (e.g., PostgreSQL, MySQL, MongoDB and Redis), and some use OAuth2 services.

In this paper, to enable and simplify experiments for black-box testing, regardless of the fuzzer involved, we have now created Docker Compose files for every single API in our dataset.
These Docker Compose files not only start the API (with the right version of their needed JDK, e.g., LTS versions such as 8, 11, 17, 21 {and 25}), but also all their needed databases, with initializing data where needed (e.g, for users and administrators).
Each API is configured to run with JaCoCo, to enable collecting code-coverage metrics after the fuzzing sessions are completed.
Each API is also run behind a mitmproxy instance, where we log all HTTP requests and status responses.
This is needed to evaluate which endpoints are covered, and which 500-status-code faults are found per endpoint.
Furthermore, for each API that requires authentication, we provide authentication info in WFC format (recall Section~\ref{sec:auth}).
The data in these authentication configurations is based on what the databases are initialized with.
Furthermore, to enable running multiple experiments in parallel on the same machine without TCP port conflicts,
all exposed TCP ports in Docker Compose are configurable via parameters.

As this experiment scaffolding is independent of \evo, using a name such as EMB would be misleading.
We have now renamed it into
Web Fuzzing Dataset (WFD).\footnote{\url{https://github.com/WebFuzzing/Dataset}}
A black-box fuzzer that wants to use WFD for experiments just needs to Docker Compose \emph{up} an API, run the fuzzer, collect JaCoCo metrics, and then Docker Compose \emph{down} to stop the API.

WFD contains
more than {2 millions}
lines of code (excluding third-party libraries)
and
more than {3 300}
endpoints,
as shown in Table~\ref{tab:suts}.
{
WFD now represents the largest selection of open-source REST APIs with scaffolding to run and analyze fuzzing experiments.
}

To simplify the running of experiments on all the APIs and collect JaCoCo and mitmproxy metrics, we also provide Python scripts to do all these steps automatically, {as it will be presented in more details in Section~\ref{sub:wfe}}.

%%%%%%%%%%%%%%%%%%%%%%%%%%%%%%%%%%%%%%%%%%%%%%%%%%%%%%%%%%%%%%%%%%%%%%%%%%%%
\section{{Fuzzer Selection}}
\label{sec:selection-criteria}

%-----------------------------------------------------
\subsection{First Selection Criteria}
\label{sec:usability}

To compare techniques presented in different studies, especially to evaluate whether a novel technique provides some benefits over the current state-of-the-art, tool comparisons are necessary.
Ideally, one would like to compare ``techniques'' and not ``tools'', as the performance of these latter might be strongly impacted by engineering concerns.
However, re-implementing from scratch a tool/technique based on the high-level description of a research article is not a scalable option.
Tool comparisons are hence a necessity, not just when fuzzing REST APIs.

To be able to compare with a tool, such tool must be available.
Best if it is open-source, which is the case for most REST API fuzzers in the literature (recall Table~\ref{tab:auth}).
If not open-source, or if in need to satisfy double-blind review constraints, long-term storage solutions such as Zenodo are a viable option.

To demonstrate the use of WFC and WFD, we have carried out a series of experiments (which will be explained in more details in Section~\ref{sec:experiments}), where we compare different fuzzers.
There are many fuzzers available (recall Table~\ref{tab:auth}), and comparing all of them in each new study is not viable.
A typical approach in the literature is to employ 4-5 fuzzers, possibly representing the current ``state-of-the-art''.

There are at least four possible complementary approaches to select fuzzers for comparisons:

\begin{enumerate}

\item  The fuzzer provides the best results in {\emph{peer-reviewed independent studies}}.

\item  The fuzzer is popular among practitioners in industry.

\item  The fuzzer has been presented in recent research work, showing better results compared to the state-of-the-art.

\item  The fuzzer provided results of particular interest on some specific SUTs.

\end{enumerate}

When carrying an empirical study to evaluate a novel technique, there is always a potential bias and threat to validity, as the authors have direct incentives to show their techniques work better.
This applies to all studies, our included.
Studies where authors have no connection to any of the employed tools would not suffer of such issue.
In the case of fuzzing REST APIs, we are aware of only two existing studies of this kind.
These include the work of Kim \etal~\cite{Kim2022Rest}, published in 2022, where 10 fuzzers where evaluated on 20 open-source APIS.
The other is the work of Sartaj \etal,
published first on arXiv in 2024~\cite{sartaj2024restapitestingdevops},
and then on a peer-reviewed journal in 2025~\cite{sartaj2025rest},
where 5 fuzzers were compared on the regression testing of 17 APIs through  14 different releases of an industrial IoT Application.

Regarding tool competitions, e.g., hosted at international conferences and workshops, those are important for pushing forward the state-of-the-art.
Examples in software engineering research exist at the
SBFT workshop,\footnote{https://sbft25.github.io/}
with competitions on for example JUnit test generation, self-driving car testing and C/C++ library fuzzing.

In 2026, the first tool competition at SBFT for REST API fuzzing
{
was run by the authors of RestGym.
Results were not officialized at the time we made a selection of fuzzers for our first set of experiments in this paper.
}

The better and more usable a fuzzer is, the more likely it will be used by practitioners in industry.
If the fuzzer gives bad results, unlikely a practitioner will keep using it, or recommend it to other colleagues.
As such, it is important in any study to consider some of the most popular tools, regardless of how much ``novelty'' or other academic metric score would they have.
At times, ``popularity'' can be non-trivial to determine rigorously.
In case of open-source projects, metrics such as the number of tool downloads  and GitHub ``stars'' can be used as reasonable indicative.
However, those metrics can be manipulated.
Furthermore, popularity might also be strongly dependent on ``marketing'' efforts, and not on any intrinsic quality of the tool.
As such, popularity can be used to select some fuzzers for comparison, but by no means it should be the only criterion to make such a selection.

Novel presented techniques could provide results that are better than the state-of-the-art.
As such, those should be used for comparisons, and see if a replicated study does indeed confirm such results.
However, based on the venues in which those novel fuzzers are presented, there is always the risk of missing the fact that those have been published.
Furthermore, a research study might take months, and new fuzzers could be presented during such period.
Including them in the comparisons at the last moment could be cumbersome.
Although aiming at including the latest work is important, it is just on a best-effort.

At times, even if overall a fuzzer might not work best on average on a large selection of SUTs, it might provide interesting results on some specific SUTs that are warranted of further investigation.
In such cases, it is advisable to include those fuzzers as well in the comparisons.

%----------------------------------------------------------------------
\subsection{Fuzzer Usability Requirements}
\label{subsec:fuzzer_usability}

Regardless of which criteria are used to select fuzzers for comparisons, those selected fuzzers must be \emph{usable}.
Being open-source, and with a research article published in a respected venue, does not imply that a tool is usable, or that it is easy to configure in a comparison study.
For example,
regardless of how complex or easy it is to download and install a fuzzer, this should be a \emph{constant} time/effort cost $O(1)$ which is \emph{independent} of the number $n$ of APIs in the experiments.
Otherwise, the more APIs you have in the experiments, the more effort would be required to setup comparisons.

On the one hand,
 \aratrl, as an example, is a single Python file that, after a simple \texttt{pip3 install -r requirements.txt} command, it can be directly run on the command-line.
\evo provides different ways to be installed, like via a Windows \texttt{msi} installer.
\schemathesis can be installed via \emph{brew}.
\emrest requires to install \emph{miniconda}.
\llamaresttest requires to manually download two large model files (3GB each).
All those are ``one-time'' costs, as needed to be done only once.

On the other hand, there are examples of fuzzers where the cost is not
constant.\footnote{Due to the challenges of correctly interpreting constructive feedback, we shall not name those tools in this article.}
For example, fuzzers that require to manually execute operations before being able to be executed (e.g., to call an LLM to generate custom dictionary files based on the OpenAPI schema for each different tested API), do not have a constant installation cost.
We argue that, for doing sound comparisons in new scientific studies, any extra manual steps should not really be required, as it is not the case for example for all the tools used in our study.
For example, if an LLM needs to be called, the tool could automatically do it at runtime.
 The \emph{usability} of a tool should be taken into account when doing comparisons with the state-of-the-art.

%----------------------------------------------------------------
\begin{figure}[!t]
\begin{result}
List of \emph{usability requirements}.
\begin{enumerate}
\item \label{enum:open} The tool should be open-source, or available in a long-term storage solution like Zenodo.
\item \label{enum:constant} Installing the tool should be a one-time cost, independent of the number of APIs used in the experiments.
\item \label{enum:terminal} The tool should have at least 3 command-line input parameters to specify:
    (a) where the OpenAPI schema is located on the file system;
    (b) the URL of where the API is up and running;
    (c) for how long to run the fuzzing session.
\item \label{enum:llm} If the tool makes use of an LLM, there {must be} an option to employ a locally run one.
\end{enumerate}
\end{result}
\caption{
\label{fig:usability}
List of \emph{usability requirements} for selection of tools in empirical comparisons of REST APIs.
}
\end{figure}
%----------------------------------------------------------------

Figure~\ref{fig:usability} provides our formalization of a list of minimal \emph{usability requirements} which we evaluate as essential for choosing tools for comparisons.
These guidelines are based on 20 years of experience in carrying out experiments with academic prototypes (e.g., for unit test generation~\cite{FrA12b}), dealing specifically with the testing of REST APIs for more than 10 years~\cite{arcuri2017restful,arcuri2018experience}.

As previously stated, to be usable, a tool should be available (Requirement~\ref{enum:open}), e.g., either as open-source or on long term storage systems such as Zenodo.
Also, to enable large empirical experiments, the installation of the tool should be a one-time cost, regardless of the number of used APIs in the study (Requirement~\ref{enum:constant}).

In an empirical study involving $N$ APIs, in which experiments are repeated $R$ times to deal with randomness in these tools, a fuzzer would need to be run $N*R$ times.
In our case, considering $N=\nsutsOld$ and $R=10$, it would mean 360 times.
These experiments should be configured via scripts (e.g., Bash or Python), and not through a GUI.
As such, the fuzzer would need to be able to handle input parameters to configure how it is going to be run (Requirement~\ref{enum:terminal}).
Also, an important aspect of generating separated scripts for each experiment is that it makes easier to debug them (if needed), as they can be run and repeated individually.
Furthermore, many advance clusters for experimentation available to universities and research centers worldwide schedule jobs based on submitted Bash scripts.

The first needed input parameter is the location of where the OpenAPI schema is located on the file system (Requirement~\ref{enum:terminal}(a)).
The fuzzer could also provide other methods to locate schemas (e.g., in \evo we support automated download of schemas via provided URLs), but those are just an extra for usability.

Where an API is running, and it is accessible to, is specified in the OpenAPI schema itself.
However, typically this would include an actual hostname available on internet, such as for example \texttt{https://petstore.swagger.io}.
This would not work when an API is running locally for testing purposes, as the API would run on the \texttt{localhost} or \texttt{127.0.0.1} IP address.
Modifying the schema manually is possible, as long as it is a one-time cost.
However, when running experiments, and each run of a fuzzer on an API needs to be repeated several times to take into account the randomness of the fuzzing, these experiments must be run in parallel to be able to be completed in a reasonable amount of time.
This would not work if all instances of the tested API try to bind on the same TCP port.
Creating a different OpenAPI schema file specifying different ports for each experiment is feasible, but quite cumbersome.
For example, considering \nsutsOld APIs and experiments repeated 10 times, it would end up with 360 experiments per fuzzer.
Having to create 360 different OpenAPI schemas pointing to different TCP ports would be extremely tedious to do manually.
It could be automated (by writing scripts that parse JSON/YAML OpenAPI files, and output modified versions with different port values), but we argue that this feature should be rather handled directly by the fuzzer (as most fuzzers do), and not by the experiment scaffolding used to compare tools.
The ability to specify hostname (e.g., \emph{localhost}) and port of where the API is up and running is essential for enabling fuzzer comparisons (Requirement~\ref{enum:terminal}(b)).

Another important parameter is for how long a fuzzer is going to be left running (Requirement~\ref{enum:terminal}(c)).
The more time is left for running a fuzzer, the better results one can expect.
How long to run a fuzzer depends on the practical constraints of the test engineers.
There is no one-size-fits-all solution.
In the literature, typically experiments are compared by running fuzzers for 1 hour~\cite{golmohammadi2023testing}.
Regardless of how long experiments should be run for, a tool should provide mechanisms to stop within a specified amount of time.
Note, however, that such stopping mechanism do not need to be precise, as achieving it might be cumbersome.
For example, in \evo a user can specify to run the search for 1 hour, but that does not include the time needed to output test cases or minimizing them, just the search.
This means that it could run for 1 or 2 minutes extra.
Also, some tools might use a deterministic approach to generate tests without specifying a timeout.
In those cases, such tools could still be used for comparisons, but they should be preemptively stopped if they take too long.
{
Hybrid approaches are possible as well.
For example, CATS first executes its deterministic catalog of fuzzers and, if these complete before the given time budget, it can continue with seeded random mutations until the budget is consumed (option \texttt{--fillTimeBudget}).
}
In our case, we configured 1 hour in all tools that provide such timeout option, and run each script inside a \texttt{timeout} Bash command of 90 minutes.

The use of LLMs is becoming popular in software engineering research, especially in software testing~\cite{wang2024software}.
However, from a scientific point of view, there can be major issues when LLMs are employed in empirical studies~\cite{sallou2024breaking}.
We strongly agree with the points raised by Sallou \etal~\cite{sallou2024breaking}.
Regarding fuzzing REST APIs, we argue that, if an LLM is needed, it should be able to run them locally (Requirement~\ref{enum:llm}).
Otherwise, results might become impossible to replicate.
{
This is also in line with emerging guidelines on using LLMs in software engineering research~\cite{baltes2027guidelines}.
}
Furthermore, costs of commercial LLMs might end up in the order of hundreds or thousands of dollars in large empirical studies, which is not a viable approach when comparing with fuzzers that can be run on a laptop for no extra cost.
Experiments using commercial LLMs can still be useful, but there should always be the option to rather use a locally run one.

{
LLMs can be very useful, and hybrid approaches that can exploit all the available CPUs and GPUs are likely going to provide better results in the long run.
Still, how to proper do fair comparisons, under the same constraints of used resources, is not trivial~\cite{sartaj2026search}.
As commercial LLMs run on remote servers can be rented, the same can be technically said for cloud resources (e.g., AWS and Azure) to run traditional fuzzers in parallel (using the same amount of money).
This is a complex topic, and guidelines supported by sound empirical evidence will need to be established in the future to address it.
}

%----------------------------------------------------------------
\begin{figure}[!t]
\begin{result}
List of \emph{desirable features}.
\begin{enumerate}
\item \label{enum:auth} Provide configurations to setup authentication information.
\item \label{enum:output} Provide a configuration to specify where the output of the fuzzer should be saved into (e.g., a specific folder).
\item \label{enum:tests} The tool should be able to generate executable test cases. Ideally, in Python, to simplify comparisons.
\end{enumerate}
\end{result}
\caption{
\label{fig:desirable-features}
List of \emph{desirable features} for selection of tools in empirical comparisons of REST APIs.
}
\end{figure}
%----------------------------------------------------------------

{
Besides minimal usability \emph{requirements} to select fuzzers for comparisons, there are a few \emph{desirable features} that are worth considering when selecting and excluding fuzzers.
Figure~\ref{fig:desirable-features} lists them.
}

As previously discussed in Section~\ref{sec:auth}, authentication is very important.
Technically, though, the ability of enabling authentication is \emph{optional}, in the sense that a fuzzer can still be used even if it has no support for authentication.
It will simply get worse results compared to fuzzers that do support authentication.
However, preparing custom authentication scripts and configuration files for each different fuzzer is not a scalable solution for large empirical studies.
For example, in WFD there 25 APIs that use authentication mechanism (recall Table~\ref{tab:suts}).
This was one of the main motivations to propose the WFC format for authentication information (Section~\ref{sec:auth}), with prepared configuration files for all the 25 APIs in WFD that need it.
As command-line input parameter, it should be possible to specify the location of the WFC authentication configuration file ({Feature}~\ref{enum:auth}).

Another useful, but technically optional, feature is the ability of specifying where, on the file-system, to save the results of the fuzzer once its run is completed ({Feature}~\ref{enum:output}).
This can significantly simplify the scaffolding needed to run large sets of experiments.
This feature is technically optional, because, if a tool simply writes on the working folder it is run from, in experiment scripts it is just a matter of changing the working folder before running the experiments.
Tedious, but it is not a showstopper (e.g., this was the case for \aratrl).

Regarding the ability of generating \emph{executable test cases} ({Feature}~\ref{enum:tests}) (e.g., in JUnit or PyTest formats),
{
 this is important to study how such tests can be used for \emph{regression testing}.
 There are several research challenges involved, like how to select a minimized output test suite, and how to deal with potential flakiness when creating assertions on the returned responses.
 However, some fuzzers might only focus on finding faults in the current versions of the APIs (e.g., by creating repeatable cURL commands, as done by \cats and \schemathesis).
 Other academic prototypes might not generate any directly usable output at all (e.g., just summaries), especially if the employed comparison metrics only focus on what achieved during the fuzzing process.
 Still, outputting executable test suites provide extra benefits to the users of a fuzzer, without having to change its internal fuzzing engine.
 As this feature can then enable the study of further research questions regarding regression testing, it is a desirable feature when selecting fuzzers for empirical comparisons.
}
We will go back on these points in more details in Section~\ref{sec:generation}.

%----------------------------------------------------------------------------
\subsection{Example of Fuzzer Selection}
\label{sub:tool-selection-example}

{
To show how our guidelines for tool selection, discussed in Sections~\ref{sec:usability} and~\ref{subsec:fuzzer_usability}, can be used, we provide a walk-through example.
As tools evolve, and new fuzzers and empirical studies are published, such selection is time dependent.
The example discussed next is based on what available in 2025.

The selected tools discussed here are then going to be empirically compared in our first set of experiments in Section~\ref{sub:exp-first}.
}
First, we selected \evo.
The reason is that it seems the one that gave the best results in \emph{independent} studies~\cite{Kim2022Rest,sartaj2024restapitestingdevops}.
For example, Kim \etal~\cite{Kim2022Rest} wrote:
\emph{``In one hour, the best-performing tool (EvoMasterWB) achieved \ldots The other black-box tools achieved lower coverage; the best-performing tool among them is EvoMasterBB''}.
Sartaj \etal~\cite{sartaj2024restapitestingdevops} wrote: \emph{``Notably, EvoMaster-BB demonstrated superior performance over all REST API testing tools''}.
Note that in these quotes ``BB'' refers to black-box, whereas ``WB'' refers to white-box.
In this paper, we only use its black-box version, and make no experiments with its white-box version.

{Second, } considering their number of stars on GitHub, \restler and \schemathesis seem the most popular fuzzers in industry, considering over 2 900 stars at the time of writing {in 2025}.
As such, we included these fuzzers in our experiments in this paper.

Regarding the most recent work on REST API fuzzing, Table~\ref{tab:auth} shows that, at the time of writing in the end of 2025, there have been 7 newly proposed fuzzers in 2025:
\apirl~\cite{foley2025apirl},
ASTRA~\cite{sondhi2025utilizing},
AutoRestTest~\cite{kim2025autoresttest},
\emrest~\cite{xu2025effective},
\llamaresttest~\cite{kim2025llamaresttest},
LogiAgent~\cite{zhang2025logiagent}
and
WuppieFuzz~\cite{rooijakkers2025wuppiefuzz}.
Based on the guidelines outlined in Section~\ref{sec:usability},
ASTRA~\cite{sondhi2025utilizing} and LogiAgent~\cite{zhang2025logiagent} were excluded because they are not available.
AutoRestTest~\cite{kim2025autoresttest} was excluded because it requires a commercial subscription to an LLM called ChatGPT (this at the time {this selection was made}; most recent versions of AutoRestTest seem to support using locally deployed LLMs, based on its most recent documentation at the time of writing).
We also excluded \apirl, as it does not enable specifying where the API is up and running.
At that time, WuppieFuzz provided no comparisons with any other tools, and it was evaluated only on a single API, so it was excluded.
This left us with \emrest~\cite{xu2025effective} and \llamaresttest~\cite{kim2025llamaresttest}, which fulfill all the requirements outlined in Section~\ref{sec:usability}.

In one of the recent work of \evo on white-box testing of REST APIs that use MongoDB~\cite{ghianni2025search}, we carried out fuzzer comparisons including the fuzzer \aratrl~\cite{kim2023adaptive}.
This fuzzer provided the best results on 2 APIs in such a study, surprisingly even compared to white-box fuzzing.
As such, we include \aratrl~\cite{kim2023adaptive} as well.

Based on this selection, we end up with 6 fuzzers to compare:
\evo, \restler, \schemathesis, \emrest, \llamaresttest and \aratrl.
We argue these are a good representative of the current state-of-the-art in fuzzing REST APIs.

%-------------------
\begin{table}[t]
	\centering
	\small
	\caption{New fuzzers proposed in 2025, showing for each of them the tools used in the empirical comparisons.}
	\label{tab:2025}
	\setlength{\tabcolsep}{4pt}
	\begin{tabular}{c l ccccccc c}
		\toprule
		\multirow{2}{*}{}
		& \multirow{2}{*}{\shortstack[l]{Existing\\Fuzzers\\(EF)}}
		& \multicolumn{7}{c}{New Fuzzers (NF) Proposed in 2025}
		& \multirow{2}{*}{\shortstack[c]{Selected\\ by\#NF}} \\
		\cmidrule(lr){3-9}
		& & \apirl & ASTRA & ART & ER & LRT & LA & WF & \\
		\midrule
		1 & \aratrl
		& \checkmark & \checkmark & \checkmark & \checkmark & \checkmark & \checkmark &
		& \databar{85.7}{\textbf{6/7}} \\
		2 & \evo
		& \checkmark & \checkmark & \checkmark & \checkmark & \checkmark & \checkmark &
		& \databar{85.7}{\textbf{6/7}} \\
		3 & DeepREST
		& \checkmark &            &            &            &            &            &
		& \databar{14.3}{\textbf{1/7}} \\
		4 & MINER
		& \checkmark &            &            & \checkmark &            &            &
		& \databar{28.6}{\textbf{2/7}} \\
		5 & MOREST
		&            & \checkmark & \checkmark & \checkmark & \checkmark & \checkmark &
		& \databar{71.4}{\textbf{5/7}} \\
		6 & RestCT
		&            &            &            & \checkmark &            &            &
		& \databar{14.3}{\textbf{1/7}} \\
		7 & RESTler
		&            &            & \checkmark &            & \checkmark & \checkmark &
		& \databar{42.9}{\textbf{3/7}} \\
		8 & RestTestGen
		& \checkmark &            &            &            &            &            &
		& \databar{14.3}{\textbf{1/7}} \\
		9 & Schemathesis
		&            &            &            & \checkmark &            &            &
		& \databar{14.3}{\textbf{1/7}} \\
		\midrule

		& Used \#EF
		& \databar{55.6}{\textbf{5/9}}
		& \databar{33.3}{\textbf{3/9}}
		& \databar{44.4}{\textbf{4/9}}
		& \databar{66.7}{\textbf{6/9}}
		& \databar{44.4}{\textbf{4/9}}
		& \databar{44.4}{\textbf{4/9}}
		& \databar{0.0}{\textbf{0/9}}
		&  \\
		\bottomrule
	\end{tabular}

	%\vspace{2pt}
	\raggedright
	\footnotesize\textit{Table note:}
	Columns denote papers published in 2025:
	\apirl~\cite{foley2025apirl},
	ASTRA~\cite{sondhi2025utilizing},
	ART(AutoRestTest)~\cite{kim2025autoresttest},
	ER (\emrest)~\cite{xu2025effective},
	LRT (\llamaresttest)~\cite{kim2025llamaresttest},
	LA (LogiAgent)~\cite{zhang2025logiagent}, and
	WF (WuppieFuzz)~\cite{rooijakkers2025wuppiefuzz}.
	Rows denote tools published before 2025.
\end{table}
%-------------------

If we look at what used in empirical comparisons by the community in 2025 (i.e., see Table~\ref{tab:2025}), we see several similarities, but also some clear differences.
Excluding the study presenting WuppieFuzz that included no comparisons~\cite{rooijakkers2025wuppiefuzz}, \emph{all} other presented  fuzzers compared with \aratrl and \evo.
However, a significant difference is that only half of the studies included \restler, and only one included \schemathesis.

\schemathesis is developed in industry by non-academics.
This is as well the case for popular tools such as CATS, and previously Dredd (this latter no longer maintained).
We believe that ``use in industry'' is an important criterion to consider for selection, especially when academic results want to have an impact on industrial practice.
Excluding or ignoring the most used fuzzers in industry would be, in our opinion, detrimental.
Our guidelines are meant to avoid such issues.
Note that there might other, better metrics to try to quantify usage and impact in industry.
The use of GitHub ``stars'' is just an easy, quick to verify metric.
Furthermore, popularity and usage might vary.
What is most common today might not be tomorrow.
So, these metrics need to be checked at each new empirical study.

{
In academic peer-reviewing processes, it is not uncommon that peers criticize not using a specific tool in the comparisons.
For example, a previous version of this article was rejected because not including RestTestGen, without providing any scientific argument for its inclusion.
It is not in our intention to single out a tool like RestTestGen,
but, let us consider it as an example of why it would not be selected when following the criteria expressed in Section~\ref{sec:usability}.
These discussions are provided in Appendix~\ref{sec:resttestgen}, as they can be useful to the research community when dealing with similar situations.

Any empirical study in which fuzzers are compared would have some discussion on how such fuzzers were selected.
It is not our goal to claim whether the criteria presented here are better or worse than what has been used in the literature.
Our goal is to provide a more formal, citable definition with in-depth discussions on their implications, especially in new empirical studies if they choose to use WFC and WFD.
This is aimed at avoiding \emph{unclear} and \emph{arbitrary} decisions, both for the researchers that carry out such studies, and for the researchers that will peer-review them.
For example, criticizing an empirical study because it lacks comparisons with a specific tool $X$, without providing any objective, measurable metric for inclusion/exclusion, would be scientifically unsound.

}

%%%%%%%%%%%%%%%%%%%%%%%%%%%%%%%%%%%%%%%%%%%%%%%%%%%%%%%%%%%%%%%%%%%%%%%%%%%%%
\section{{Web Fuzzing Evaluations}}
\label{sub:wfe}

{
To enable replicable experiments, not only EMB stored a set of APIs, but also all needed configuration scaffolding to run experiments in which different fuzzers can be compared~\cite{icst2023emb}.
}
We use a Python script to generate a set of experiments using the different tools and APIs, each one instantiated with a different Bash script~\cite{arcuri2023building}.
For example, given the Python script to setup experiments on 6 tools on 36 APIs with 10 repetitions, such Python script will create $6 \times 36 \times 10 = 2160$ Bash scripts as output, each one using different TCP ports.
Those scripts can then be run in parallel (e.g., 15 at a time) using a scheduler for Bash scripts {(which we provide in our experiment scaffolding)}.

{
In this paper, we have updated our experiment scaffolding to run both the APIs and fuzzers via Docker, in a similar way as done in RestGym (recall Section~\ref{sec:relatedwork}).
Although there are a few technical challenges when running large sets of experiments in which thousands of Docker images are created and destroyed,
an advantage of using Docker is that constraints can be put on the resources (e.g., memory and number of employed CPUs) used by each fuzzer.
For example, in our case we currently use a limit of 4GB of RAM, and up to 2 CPUs per fuzzer.
This prevents issues of resource contention between different fuzzers run in parallel.
Furthermore, this enables running existing fuzzers in the literature without having to manually download and configure them.

Another feature is that, now, we provide all the R scripts needed to automatically generate tables and figures doing all different kinds of analyses on the data resulting from these experiments.
Tables are not only outputted in LaTeX (so can be directly used in scientific articles, like this one), but also in Markdown.
This enables to publish such results directly on sites such as GitHub, as Markdown pages are directly interpreted and visualized without having to download any file.

All these experiment scaffolding and analyses has been moved out of EMB/WFD, and put in a new open-source repository called
Web Fuzzing Evaluations\footnote{\url{https://github.com/WebFuzzing/Evaluations}}
(WFE).
Each time new experiments are carried out, this repository also contains the results of such experiments (sorted in chronological order), with all the needed visualization aids (e.g., tables in Markdown format, automatically generated from the experiment results via the provided R scripts).

The reason to move WFE out of WFD is to enable both projects to update independently.
New versions of WFE with new added fuzzers and new experiments could be released in a new version even if the APIs in WFD have not changed.
Similarly, new APIs can be added to WFD without having to wait for and a publish a new full set of experiments with WFE.

WFE can be used as it is to study how current state-of-the-art fuzzers compare (e.g., \cats, \evo, \schemathesis and \wuppie).
As all Docker images are published on Docker Hub (including all the APIs and fuzzers), experiments can be run without need of any manual installation, besides having Docker and Python installed.
}

%---------------------------------------
\begin{figure}
	\centering
% 	%\resizebox{.8\textwidth}{!}{
	\begin{lstlisting}[language=bash,basicstyle=\footnotesize]
FROM amazoncorretto:25-jdk

ARG VERSION=15.6.0
ARG TAR_FILE=/cats.tar.gz

RUN yum install -y tar gzip && yum clean all

ADD https://github.com/Endava/cats/releases/download/cats-${VERSION}/cats_uberjar_${VERSION}.tar.gz  ${TAR_FILE}

RUN tar -xzf ${TAR_FILE} -C / && rm ${TAR_FILE}

ENTRYPOINT ["/bin/sh", "-c", \
    "java  \
    -jar /cats.jar \
    --blackbox \
    --reportFormat WFC_REPORT \
    --contract ${SCHEMA_FILE_PATH} \
    --output ${OUTPUT_FOLDER} \
    --server http://${HOST}:${PORT} \
    ${TIME_SECONDS:+--stopAfterTimeInSec} ${TIME_SECONDS} \
    ${TIME_SECONDS:+--fillTimeBudget} \
    ${AUTH_FILE_PATH:+--wfcAuth} ${AUTH_FILE_PATH} \
    ${EXTRA_CONFIGS} \
    " \
]
	\end{lstlisting}
	%}
	\caption{\label{fig:docker-cats}
		Docker file to run \cats in WFE on WFD.
	}
\end{figure}
%---------------------------------------

{
Developers of new fuzzers that want to use WFE/WFD for their studies, need to provide a Docker image for their fuzzer, and add an entry in the experiment Python script.
Figure~\ref{fig:docker-cats} shows the current Docker file for \cats.
All fuzzers are run in the same way, where configurations are provided via environment variables.
These include for example where the OpenAPI schema is located for the tested API (\texttt{SCHEMA\_FILE\_PATH}),
and for how many second the fuzzing process should run (\texttt{TIME\_SECONDS}).
In this latter case, Docker containers are also preemptively stopped if exceeding a global timeout specified in the generated Bash scripts.

Fuzzers are configured to output WFC Report files (e.g., with the custom option \texttt{--reportFormat} in \cats), and conditionally with WFC Auth files if the APIs need them (\texttt{AUTH\_FILE\_PATH}).
Note that, in Docker, the operator \texttt{:+} is used for conditional interpolations (i.e., the right-side is not applied if the left-side is not set).

Another important aspect is that, in all of our Docker configurations, tools are downloaded and installed automatically (e.g., \texttt{pip install} for \schemathesis).
Updating a Docker image to the latest version of a fuzzer just requires changing a version number in the Docker image (e.g., $15.6.0$ in Figure~\ref{fig:docker-cats}).
}

If a tool fulfills the requirements expressed in Figure~\ref{fig:usability}, then adding a new tool to this experiment infrastructure becomes simple, regardless of the APIs present in WFD, as it would just require to write a {Docker file} such as the one in Figure~\ref{fig:docker-cats}.

%%%%%%%%%%%%%%%%%%%%%%%%%%%%%%%%%%%%%%%%%%%%%%%%%%%%%%%%%%%%%%%%%%%%%%%%%%%%
\section{Empirical Study}
\label{sec:experiments}

{
To showcase the use of WFC, WFD, WFE and our experiment guidelines, we have carried two sets of experiments.

In the first set of experiments carried out in 2025, we ran experiments on WFD using a selection of fuzzers following our guidelines explained in Section~\ref{sec:all_wfd}.
In particular, as discussed in Section~\ref{sub:tool-selection-example}, those six compared fuzzers are
\aratrl, \emrest, \evo, \llamaresttest, \restler and \schemathesis.
At that point in time, only \evo supported the new authentication format provided by WFC.
So no authentication information was used in these experiments.
Also, at that time WFD contained only \nsutsOld APIs.

Once the authors of \cats, \schemathesis and \wuppie joined this work, and added full WFC support to their fuzzers, a second set of experiments including \evo was carried out in 2026.
These four fuzzers were run on the latest version of WFD, which includes \nsuts APIs.
Fuzzers were provided authentication credentials in WFC format for all the APIs that need it.
As all of these fuzzers can output fault reports in WFC format, those were used as well in the result analyses.

The following sections describe these two set of experiments in more details.
}

%-----------------------------------------------------
\subsection{{First Set of Experiments}}
\label{sub:exp-first}

\subsubsection{{First Experiment Setup}}

We ran each one of the 6 {selected} fuzzers on each of the \nsutsOld APIs in WFD for 1 hour (with a forced safety timeout of 90 minutes, although, looking at the logs, it seems such timeout was never reached in any of the experiments).
Each experiment was repeated 10 times to take into account the randomness of these tools.
In total, this required $6 \times \nsutsOld \times 10 = 2160$ experiments, for roughly 90 days of computation if run in sequence.
Between logs, tests and coverage results, more than 300GB of data was generated in these experiments.

Experiments were run in parallel (15 at a time) on a high level server machine with 200GB of RAM, using Windows 11.
{
At that point in time, APIs were run in Docker, but support for running the fuzzers in Docker was not ready yet.
Each fuzzer was manually installed on that Windows machine, with all their needed dependencies.
}

Unfortunately, it seems like that \emrest does not work on Windows, due to {some incompatible} binary dependencies.
Experiments for \emrest were hence run on MacBook Pro M4 with 128GB of RAM.
{
As this second machine is much faster than the first, this technically gives an advantage to \emrest.
However, as we will show in our experiments, still \emrest gave worse results compared to the other fuzzers.
As such, we did not consider it as a major threat to the validity of our study.
}

No authentication information was provided and used in these experiments.
Authentication information in the WFC format, presented in this paper, {at that time was} supported only by \evo.
Using WFC for \evo would give an unfair advantage over other tools that have their own custom methods for authentication, like \restler and \schemathesis.
The use of a standardized way to provide authentication information, like the candidate proposal WFC format, is one of the main messages we want to promote in this work.
Note that \emrest, \llamaresttest and \aratrl have no support for authentication (recall Table~\ref{tab:auth}).
However, our goal in this study is not to show which tool is best, but rather to show what kind of results can be obtained when using WFD as case study.
Experiments using WFC authentication format will be presented in Section~\ref{sec:auh:experiments}, only for \evo,
{
and then for second set of experiments in Section~\ref{sub:experiments-second}.
}

To compare the fuzzers, we used 3 different metrics:
(1) percentage of endpoints for which at least 1 test case has HTTP response in the 2xx range (using \texttt{mitmproxy});
(2) number of endpoints in which at least one response with HTTP status code 500 was obtained (using \texttt{mitmproxy});
(3) achieved line code coverage (measured with \texttt{JaCoCo}).

{
Note that, for fault detection, we do not try to distinguish between different 500 crashes for the same endpoint, based on some heuristics on the returned responses.
As such responses can be flaky and contain sources of non-determinism, any heuristics is bound to fail on some APIs, especially when a large set is used for experiments.
We wanted to avoid unsound situations like in RestGym used in the SBFT'26 REST API fuzzing competition, where on some of  the used APIs a fuzzer could find as many ``unique'' faults as it wanted by just repeating executing the same HTTP call with no modification, or with modifications that had no impact on the API execution~\cite{arcuri2026sbft,hatfield2026schemathesis}.
In those cases, the resulting fault results are just noise in the experiment analyses.
}

%---------------------------------------------------------------------------
\begin{table}[t]
	\centering
	\small
\caption{
\label{tab:2xx}
Average results, out of 10 runs, for 2xx endpoint coverage percentage over the 6 compared fuzzers, on all the \nsutsOld APIs.
On each API, tool are compared by rank, where rank 1 is the best.
Rank values are presented in `()' parentheses after the average values.
In case of ties, ranks are averaged.
%The best fuzzers on each API are highlighted in bold.
Green indicates the best result, yellow the second-best result, and red text the worst result. }

\vspace{-1.5\baselineskip}
\resizebox{0.99\textwidth}{!}{
   \begin{tabular}{ l r r r r r r  }\\ 
\toprule 
SUT & \aratrl & \emrest & \evomasterbbnoauth & \llamaresttest & \restler & \schemathesis \\ 
\midrule 
\emph{bibliothek} & \ranksecond{0.0 (4.5)} & \ranksecond{0.0 (4.5)} & \rankbest{12.5 (1.5)} & \ranksecond{0.0 (4.5)} & \ranksecond{0.0 (4.5)} & \rankbest{12.5 (1.5)} \\ 
\emph{blogapi} & \ranksecond{15.1 (3.0)} & \rankbest{20.9 (1.0)} & \ranksecond{15.1 (3.0)} & \rankworst{6.9 (6.0)} & \ranksecond{15.1 (3.0)} & 9.9 (5.0) \\ 
\emph{catwatch} & \rankworst{0.0 (5.0)} & \rankworst{0.0 (5.0)} & \rankbest{42.6 (1.0)} & \rankworst{0.0 (5.0)} & \ranksecond{39.1 (2.5)} & \ranksecond{39.1 (2.5)} \\ 
\emph{cwa-verification} & \rankworst{0.0 (4.5)} & \rankworst{0.0 (4.5)} & \rankbest{60.0 (1.0)} & \rankworst{0.0 (4.5)} & \rankworst{0.0 (4.5)} & \ranksecond{20.0 (2.0)} \\ 
\emph{erc20-rest-service} & \rankworst{0.0 (4.5)} & \rankworst{0.0 (4.5)} & \rankbest{7.7 (1.0)} & \rankworst{0.0 (4.5)} & \rankworst{0.0 (4.5)} & \ranksecond{6.9 (2.0)} \\ 
\emph{familie-ba-sak} & \rankworst{0.0 (4.5)} & \rankworst{0.0 (4.5)} & \rankbest{2.2 (1.0)} & \rankworst{0.0 (4.5)} & \rankworst{0.0 (4.5)} & \ranksecond{1.9 (2.0)} \\ 
\emph{features-service} & \rankworst{0.0 (4.5)} & \rankworst{0.0 (4.5)} & \rankbest{88.9 (1.0)} & \rankworst{0.0 (4.5)} & \rankworst{0.0 (4.5)} & \ranksecond{29.2 (2.0)} \\ 
\emph{genome-nexus} & \rankworst{0.0 (5.0)} & \rankworst{0.0 (5.0)} & \rankbest{73.9 (1.0)} & \rankworst{0.0 (5.0)} & 47.8 (3.0) & \ranksecond{64.3 (2.0)} \\ 
\emph{gestaohospital} & \ranksecond{57.5 (2.0)} & \rankworst{0.0 (5.5)} & \rankbest{62.5 (1.0)} & 35.5 (3.0) & \rankworst{0.0 (5.5)} & 4.4 (4.0) \\ 
\emph{http-patch-spring} & 22.9 (3.0) & \rankworst{0.0 (6.0)} & \rankbest{100.0 (1.0)} & 16.7 (4.0) & \ranksecond{66.7 (2.0)} & 14.8 (5.0) \\ 
\emph{languagetool} & \rankworst{0.0 (4.5)} & \rankworst{0.0 (4.5)} & \rankbest{95.0 (1.0)} & \rankworst{0.0 (4.5)} & \ranksecond{50.0 (2.0)} & \rankworst{0.0 (4.5)} \\ 
\emph{market} & \ranksecond{13.8 (4.0)} & \rankworst{0.0 (6.0)} & \rankbest{15.4 (2.0)} & \rankbest{15.4 (2.0)} & \rankbest{15.4 (2.0)} & 5.4 (5.0) \\ 
\emph{microcks} & \rankworst{0.0 (4.5)} & \rankworst{0.0 (4.5)} & \rankbest{23.9 (1.0)} & \rankworst{0.0 (4.5)} & \rankworst{0.0 (4.5)} & \ranksecond{19.1 (2.0)} \\ 
\emph{ocvn} & \rankworst{0.0 (4.5)} & \rankworst{0.0 (4.5)} & \rankbest{65.6 (1.0)} & \rankworst{0.0 (4.5)} & \rankworst{0.0 (4.5)} & \ranksecond{6.7 (2.0)} \\ 
\emph{ohsome-api} & \ranksecond{0.0 (4.0)} & \ranksecond{0.0 (4.0)} & \rankbest{23.4 (1.0)} & \ranksecond{0.0 (4.0)} & \ranksecond{0.0 (4.0)} & \ranksecond{0.0 (4.0)} \\ 
\emph{pay-publicapi} & 0.0 (3.5) & 0.0 (3.5) & 0.0 (3.5) & 0.0 (3.5) & 0.0 (3.5) & 0.0 (3.5) \\ 
\emph{person-controller} & \rankworst{0.0 (5.0)} & \rankworst{0.0 (5.0)} & \rankbest{50.0 (1.0)} & \rankworst{0.0 (5.0)} & 25.0 (3.0) & \ranksecond{25.9 (2.0)} \\ 
\emph{proxyprint} & \rankworst{0.0 (4.5)} & \rankworst{0.0 (4.5)} & \rankbest{40.7 (1.0)} & \rankworst{0.0 (4.5)} & \rankworst{0.0 (4.5)} & \ranksecond{20.4 (2.0)} \\ 
\emph{quartz-manager} & 0.0 (3.5) & 0.0 (3.5) & 0.0 (3.5) & 0.0 (3.5) & 0.0 (3.5) & 0.0 (3.5) \\ 
\emph{reservations-api} & 0.0 (3.5) & 0.0 (3.5) & 0.0 (3.5) & 0.0 (3.5) & 0.0 (3.5) & 0.0 (3.5) \\ 
\emph{rest-ncs} & \rankworst{0.0 (5.0)} & \rankworst{0.0 (5.0)} & \rankbest{100.0 (1.0)} & \rankworst{0.0 (5.0)} & \ranksecond{83.3 (2.0)} & 45.0 (3.0) \\ 
\emph{rest-news} & \rankworst{0.0 (5.0)} & \rankworst{0.0 (5.0)} & \rankbest{85.7 (1.0)} & \rankworst{0.0 (5.0)} & \ranksecond{28.6 (2.5)} & \ranksecond{28.6 (2.5)} \\ 
\emph{rest-scs} & 87.5 (3.5) & \rankworst{0.0 (6.0)} & \rankbest{100.0 (1.0)} & 87.5 (3.5) & \ranksecond{90.0 (2.0)} & 77.8 (5.0) \\ 
\emph{restcountries} & \rankworst{0.0 (5.0)} & \rankbest{100.0 (1.0)} & \ranksecond{97.7 (2.0)} & \rankworst{0.0 (5.0)} & 8.0 (3.0) & \rankworst{0.0 (5.0)} \\ 
\emph{scout-api} & \rankworst{0.0 (4.5)} & \rankworst{0.0 (4.5)} & \rankbest{28.6 (1.0)} & \rankworst{0.0 (4.5)} & \ranksecond{16.3 (2.0)} & \rankworst{0.0 (4.5)} \\ 
\emph{session-service} & \ranksecond{88.9 (2.0)} & \rankworst{0.0 (6.0)} & 50.0 (3.5) & \rankbest{93.8 (1.0)} & 50.0 (3.5) & 22.5 (5.0) \\ 
\emph{spring-actuator-demo} & 80.0 (5.0) & \rankworst{0.0 (6.0)} & \rankbest{100.0 (1.5)} & \ranksecond{90.0 (3.0)} & \rankbest{100.0 (1.5)} & 87.5 (4.0) \\ 
\emph{spring-batch-rest} & \rankbest{71.1 (1.0)} & \rankworst{0.0 (5.5)} & \ranksecond{60.0 (2.5)} & 53.3 (4.0) & \rankworst{0.0 (5.5)} & \ranksecond{60.0 (2.5)} \\ 
\emph{spring-ecommerce} & \rankworst{0.0 (5.0)} & \rankworst{0.0 (5.0)} & \rankbest{22.2 (1.5)} & \rankworst{0.0 (5.0)} & \rankbest{22.2 (1.5)} & \ranksecond{16.7 (3.0)} \\ 
\emph{spring-rest-example} & \ranksecond{0.0 (4.5)} & \ranksecond{0.0 (4.5)} & \rankbest{44.4 (1.5)} & \ranksecond{0.0 (4.5)} & \rankbest{44.4 (1.5)} & \ranksecond{0.0 (4.5)} \\ 
\emph{swagger-petstore} & \rankworst{0.0 (5.0)} & 78.9 (3.0) & \ranksecond{81.6 (2.0)} & \rankworst{0.0 (5.0)} & \rankbest{89.5 (1.0)} & \rankworst{0.0 (5.0)} \\ 
\emph{tiltaksgjennomforing} & \ranksecond{0.0 (4.0)} & \ranksecond{0.0 (4.0)} & \rankbest{8.9 (1.0)} & \ranksecond{0.0 (4.0)} & \ranksecond{0.0 (4.0)} & \ranksecond{0.0 (4.0)} \\ 
\emph{tracking-system} & \rankworst{0.0 (4.5)} & \rankworst{0.0 (4.5)} & \rankbest{73.4 (1.0)} & \rankworst{0.0 (4.5)} & \ranksecond{19.6 (2.0)} & \rankworst{0.0 (4.5)} \\ 
\emph{user-management} & 60.7 (3.0) & \rankworst{0.0 (6.0)} & \rankbest{76.2 (1.0)} & \ranksecond{67.1 (2.0)} & 38.1 (4.0) & 21.2 (5.0) \\ 
\emph{webgoat} & 0.0 (3.5) & 0.0 (3.5) & 0.0 (3.5) & 0.0 (3.5) & 0.0 (3.5) & 0.0 (3.5) \\ 
\emph{youtube-mock} & 0.0 (3.5) & 0.0 (3.5) & 0.0 (3.5) & 0.0 (3.5) & 0.0 (3.5) & 0.0 (3.5) \\ 
\midrule 
Average  & 13.8 (4.0) & \rankworst{5.6 (6.0)} & \rankbest{47.4 (1.0)} & 12.9 (5.0) & \ranksecond{23.6 (2.0)} & 17.8 (3.0) \\ 
Median  & \rankworst{0.0 (5.0)} & \rankworst{0.0 (5.0)} & \rankbest{47.2 (1.0)} & \rankworst{0.0 (5.0)} & \ranksecond{11.5 (2.0)} & 8.4 (3.0) \\ 
\midrule 
Friedman Test & \multicolumn{6}{r}{$\chi^2$ = 76.190, $p$-value = $< 0.001$} \\ 
\bottomrule 
\end{tabular} 

}
\end{table}
%---------------------------------------------------------------------------

%---------------------------------------------------------------------------
\begin{table}[t]
	\centering
\caption{
\label{tab:500}
Average results, out of 10 runs, for the number of endpoints in which at least on test case leading to an HTTP 500 status was generated.
	Results are presented for all the 6 compared fuzzers, on all the \nsutsOld APIs.
	On each API, tool are compared by rank, where rank 1 is the best.
	Rank values are presented in `()' parentheses after the average values.
	In case of ties, ranks are averaged.
	%The best fuzzers on each API are highlighted in bold.
	Green indicates the best result, yellow the second-best result, and red text the worst result. }

\vspace{-1.5\baselineskip}
\resizebox{0.99\textwidth}{!}{
    \begin{tabular}{ l r r r r r r  }\\ 
\toprule 
SUT & \aratrl & \emrest & \evomasterbbnoauth & \llamaresttest & \restler & \schemathesis \\ 
\midrule 
\emph{bibliothek} & 0.0 (3.5) & 0.0 (3.5) & 0.0 (3.5) & 0.0 (3.5) & 0.0 (3.5) & 0.0 (3.5) \\ 
\emph{blogapi} & \rankbest{22.0 (1.0)} & \ranksecond{14.3 (2.0)} & 9.2 (4.0) & 9.1 (5.0) & \rankworst{8.0 (6.0)} & 11.6 (3.0) \\ 
\emph{catwatch} & \rankworst{0.0 (5.0)} & \rankworst{0.0 (5.0)} & \rankbest{10.9 (1.0)} & \rankworst{0.0 (5.0)} & 9.0 (3.0) & \ranksecond{10.5 (2.0)} \\ 
\emph{cwa-verification} & 0.0 (3.5) & 0.0 (3.5) & 0.0 (3.5) & 0.0 (3.5) & 0.0 (3.5) & 0.0 (3.5) \\ 
\emph{erc20-rest-service} & \rankworst{0.0 (4.5)} & \rankworst{0.0 (4.5)} & \rankbest{12.0 (1.0)} & \rankworst{0.0 (4.5)} & \rankworst{0.0 (4.5)} & \ranksecond{10.8 (2.0)} \\ 
\emph{familie-ba-sak} & \rankworst{0.0 (4.5)} & \rankworst{0.0 (4.5)} & \rankbest{15.0 (1.0)} & \rankworst{0.0 (4.5)} & \rankworst{0.0 (4.5)} & \ranksecond{13.3 (2.0)} \\ 
\emph{features-service} & \rankworst{0.0 (4.5)} & \rankworst{0.0 (4.5)} & \rankbest{13.0 (1.0)} & \rankworst{0.0 (4.5)} & \rankworst{0.0 (4.5)} & \ranksecond{10.5 (2.0)} \\ 
\emph{genome-nexus} & 0.0 (3.5) & 0.0 (3.5) & 0.0 (3.5) & 0.0 (3.5) & 0.0 (3.5) & 0.0 (3.5) \\ 
\emph{gestaohospital} & \ranksecond{9.2 (2.0)} & \rankworst{0.0 (5.0)} & \rankworst{0.0 (5.0)} & \rankbest{13.3 (1.0)} & \rankworst{0.0 (5.0)} & 1.8 (3.0) \\ 
\emph{http-patch-spring} & \rankbest{4.5 (1.0)} & \rankworst{0.0 (4.5)} & \rankworst{0.0 (4.5)} & \ranksecond{2.2 (2.0)} & \rankworst{0.0 (4.5)} & \rankworst{0.0 (4.5)} \\ 
\emph{languagetool} & \ranksecond{0.0 (4.0)} & \ranksecond{0.0 (4.0)} & \rankbest{1.0 (1.0)} & \ranksecond{0.0 (4.0)} & \ranksecond{0.0 (4.0)} & \ranksecond{0.0 (4.0)} \\ 
\emph{market} & \ranksecond{9.9 (2.0)} & \rankworst{0.0 (6.0)} & 3.0 (4.5) & \rankbest{12.0 (1.0)} & 3.0 (4.5) & 4.9 (3.0) \\ 
\emph{microcks} & \rankworst{0.0 (4.5)} & \rankworst{0.0 (4.5)} & \rankbest{4.4 (1.0)} & \rankworst{0.0 (4.5)} & \rankworst{0.0 (4.5)} & \ranksecond{3.2 (2.0)} \\ 
\emph{ocvn} & \rankworst{0.0 (4.5)} & \rankworst{0.0 (4.5)} & \rankbest{110.0 (1.0)} & \rankworst{0.0 (4.5)} & \rankworst{0.0 (4.5)} & \ranksecond{4.2 (2.0)} \\ 
\emph{ohsome-api} & 0.0 (3.5) & 0.0 (3.5) & 0.0 (3.5) & 0.0 (3.5) & 0.0 (3.5) & 0.0 (3.5) \\ 
\emph{pay-publicapi} & \rankworst{0.0 (5.5)} & \rankbest{1.0 (2.0)} & \ranksecond{0.9 (4.0)} & \rankworst{0.0 (5.5)} & \rankbest{1.0 (2.0)} & \rankbest{1.0 (2.0)} \\ 
\emph{person-controller} & \rankworst{0.0 (5.0)} & \rankworst{0.0 (5.0)} & \rankbest{10.0 (1.0)} & \rankworst{0.0 (5.0)} & \ranksecond{7.0 (2.0)} & 6.2 (3.0) \\ 
\emph{proxyprint} & \rankworst{0.0 (4.5)} & \rankworst{0.0 (4.5)} & \rankbest{16.0 (1.0)} & \rankworst{0.0 (4.5)} & \rankworst{0.0 (4.5)} & \ranksecond{12.3 (2.0)} \\ 
\emph{quartz-manager} & 0.0 (3.5) & 0.0 (3.5) & 0.0 (3.5) & 0.0 (3.5) & 0.0 (3.5) & 0.0 (3.5) \\ 
\emph{reservations-api} & \rankbest{3.6 (1.0)} & \rankworst{0.0 (5.0)} & 1.0 (3.0) & \ranksecond{1.5 (2.0)} & \rankworst{0.0 (5.0)} & \rankworst{0.0 (5.0)} \\ 
\emph{rest-ncs} & 0.0 (3.5) & 0.0 (3.5) & 0.0 (3.5) & 0.0 (3.5) & 0.0 (3.5) & 0.0 (3.5) \\ 
\emph{rest-news} & \ranksecond{0.0 (4.0)} & \ranksecond{0.0 (4.0)} & \rankbest{1.0 (1.0)} & \ranksecond{0.0 (4.0)} & \ranksecond{0.0 (4.0)} & \ranksecond{0.0 (4.0)} \\ 
\emph{rest-scs} & 0.0 (3.5) & 0.0 (3.5) & 0.0 (3.5) & 0.0 (3.5) & 0.0 (3.5) & 0.0 (3.5) \\ 
\emph{restcountries} & \ranksecond{0.0 (4.5)} & \rankbest{1.0 (1.5)} & \rankbest{1.0 (1.5)} & \ranksecond{0.0 (4.5)} & \ranksecond{0.0 (4.5)} & \ranksecond{0.0 (4.5)} \\ 
\emph{scout-api} & \rankworst{0.0 (4.5)} & \rankworst{0.0 (4.5)} & \rankbest{3.0 (1.0)} & \rankworst{0.0 (4.5)} & \ranksecond{2.0 (2.0)} & \rankworst{0.0 (4.5)} \\ 
\emph{session-service} & \ranksecond{6.2 (2.0)} & \rankworst{0.0 (6.0)} & 3.0 (3.0) & \rankbest{7.0 (1.0)} & 1.0 (4.0) & 0.9 (5.0) \\ 
\emph{spring-actuator-demo} & \ranksecond{0.0 (4.0)} & \ranksecond{0.0 (4.0)} & \ranksecond{0.0 (4.0)} & \rankbest{0.5 (1.0)} & \ranksecond{0.0 (4.0)} & \ranksecond{0.0 (4.0)} \\ 
\emph{spring-batch-rest} & \rankbest{3.0 (1.0)} & \rankworst{0.0 (5.5)} & 1.0 (4.0) & \ranksecond{2.2 (2.0)} & \rankworst{0.0 (5.5)} & 1.1 (3.0) \\ 
\emph{spring-ecommerce} & \rankworst{0.0 (5.0)} & \rankworst{0.0 (5.0)} & \rankbest{4.0 (1.5)} & \rankworst{0.0 (5.0)} & \rankbest{4.0 (1.5)} & \ranksecond{3.0 (3.0)} \\ 
\emph{spring-rest-example} & \rankworst{0.0 (5.0)} & \rankworst{0.0 (5.0)} & \rankbest{1.0 (1.5)} & \rankworst{0.0 (5.0)} & \rankbest{1.0 (1.5)} & \ranksecond{0.9 (3.0)} \\ 
\emph{swagger-petstore} & \rankworst{0.0 (4.5)} & \ranksecond{11.6 (2.0)} & \rankbest{12.3 (1.0)} & \rankworst{0.0 (4.5)} & \rankworst{0.0 (4.5)} & \rankworst{0.0 (4.5)} \\ 
\emph{tiltaksgjennomforing} & \ranksecond{0.0 (4.0)} & \ranksecond{0.0 (4.0)} & \rankbest{2.0 (1.0)} & \ranksecond{0.0 (4.0)} & \ranksecond{0.0 (4.0)} & \ranksecond{0.0 (4.0)} \\ 
\emph{tracking-system} & 0.0 (3.5) & 0.0 (3.5) & 0.0 (3.5) & 0.0 (3.5) & 0.0 (3.5) & 0.0 (3.5) \\ 
\emph{user-management} & \ranksecond{5.6 (2.0)} & \rankworst{0.0 (5.5)} & 1.0 (3.0) & \rankbest{16.1 (1.0)} & \rankworst{0.0 (5.5)} & 0.9 (4.0) \\ 
\emph{webgoat} & 0.0 (3.5) & 0.0 (3.5) & 0.0 (3.5) & 0.0 (3.5) & 0.0 (3.5) & 0.0 (3.5) \\ 
\emph{youtube-mock} & 0.0 (3.5) & 0.0 (3.5) & 0.0 (3.5) & 0.0 (3.5) & 0.0 (3.5) & 0.0 (3.5) \\ 
\midrule 
Average  & 1.8 (3.0) & \rankworst{0.8 (6.0)} & \rankbest{6.5 (1.0)} & 1.8 (4.0) & 1.0 (5.0) & \ranksecond{2.7 (2.0)} \\ 
Median  & \ranksecond{0.0 (4.0)} & \ranksecond{0.0 (4.0)} & \rankbest{1.0 (1.0)} & \ranksecond{0.0 (4.0)} & \ranksecond{0.0 (4.0)} & \ranksecond{0.0 (4.0)} \\ 
\midrule 
Friedman Test & \multicolumn{6}{r}{$\chi^2$ = 26.986, $p$-value = $< 0.001$} \\ 
\bottomrule 
\end{tabular} 

}
\end{table}
%---------------------------------------------------------------------------

%---------------------------------------------------------------------------
\begin{table}[t]
	\centering
\caption{
\label{tab:jacoco}
Average results, out of 10 runs, for code coverage percentage, measured with JaCoCo, over the 6 compared fuzzers, on all the \nsutsOld APIs.
	As some code coverage is achieved by simply booting up the APIs, the \emph{Base} configuration shows the results when no fuzzing is executed.
	On each API, tool are compared by rank, where rank 1 is the best.
	Rank values are presented in `()' parentheses after the average values.
	In case of ties, ranks are averaged.
	%The best fuzzers on each API are highlighted in bold.
	Green indicates the best result, yellow the second-best result, and red text the worst result. }

\vspace{-1.5\baselineskip}
\resizebox{0.99\textwidth}{!}{
    \begin{tabular}{ l r r r r r r r  }\\ 
\toprule 
SUT & \base & \aratrl & \emrest & \evomasterbbnoauth & \llamaresttest & \restler & \schemathesis \\ 
\midrule 
\emph{bibliothek} & \ranksecond{32.9 (5.0)} & \ranksecond{32.9 (5.0)} & \ranksecond{32.9 (5.0)} & \rankbest{39.3 (1.5)} & \ranksecond{32.9 (5.0)} & \ranksecond{32.9 (5.0)} & \rankbest{39.3 (1.5)} \\ 
\emph{blogapi} & \rankworst{7.1 (7.0)} & \rankbest{19.4 (1.5)} & \rankbest{19.4 (1.5)} & \ranksecond{18.1 (3.0)} & 15.8 (6.0) & 16.0 (4.0) & 15.9 (5.0) \\ 
\emph{catwatch} & \rankworst{9.4 (5.5)} & \rankworst{9.4 (5.5)} & \rankworst{9.4 (5.5)} & \rankbest{45.7 (1.0)} & \rankworst{9.4 (5.5)} & \ranksecond{34.7 (2.0)} & 31.2 (3.0) \\ 
\emph{cwa-verification} & \rankworst{21.9 (5.0)} & \rankworst{21.9 (5.0)} & \rankworst{21.9 (5.0)} & \rankbest{49.3 (1.0)} & \rankworst{21.9 (5.0)} & \rankworst{21.9 (5.0)} & \ranksecond{40.6 (2.0)} \\ 
\emph{erc20-rest-service} & \rankworst{7.9 (5.0)} & \rankworst{7.9 (5.0)} & \rankworst{7.9 (5.0)} & \rankbest{26.7 (1.0)} & \rankworst{7.9 (5.0)} & \rankworst{7.9 (5.0)} & \ranksecond{23.4 (2.0)} \\ 
\emph{familie-ba-sak} & 15.8 (4.5) & 15.8 (4.5) & 15.8 (4.5) & \rankbest{16.6 (1.0)} & \rankworst{15.1 (7.0)} & 15.8 (4.5) & \ranksecond{16.5 (2.0)} \\ 
\emph{features-service} & \rankworst{19.0 (5.0)} & \rankworst{19.0 (5.0)} & \rankworst{19.0 (5.0)} & \rankbest{65.0 (1.0)} & \rankworst{19.0 (5.0)} & \rankworst{19.0 (5.0)} & \ranksecond{36.3 (2.0)} \\ 
\emph{genome-nexus} & 12.8 (5.0) & 12.8 (5.0) & \rankworst{0.0 (7.0)} & \rankbest{33.4 (1.0)} & 12.8 (5.0) & \ranksecond{24.4 (2.0)} & 21.1 (3.0) \\ 
\emph{gestaohospital} & \rankworst{19.9 (6.5)} & \ranksecond{49.5 (2.0)} & \rankworst{19.9 (6.5)} & \rankbest{58.7 (1.0)} & 43.1 (3.0) & 21.3 (5.0) & 34.8 (4.0) \\ 
\emph{http-patch-spring} & \rankworst{4.9 (6.5)} & 14.7 (3.5) & \rankworst{4.9 (6.5)} & \rankbest{78.5 (1.0)} & 14.7 (3.5) & \ranksecond{45.6 (2.0)} & 14.6 (5.0) \\ 
\emph{languagetool} & \rankworst{1.3 (5.5)} & \rankworst{1.3 (5.5)} & \rankworst{1.3 (5.5)} & \rankbest{10.8 (1.0)} & \rankworst{1.3 (5.5)} & \ranksecond{1.9 (2.0)} & 1.7 (3.0) \\ 
\emph{market} & \rankworst{7.0 (6.5)} & 13.7 (3.0) & \rankworst{7.0 (6.5)} & \ranksecond{14.3 (2.0)} & \rankbest{14.4 (1.0)} & 12.5 (4.0) & 11.5 (5.0) \\ 
\emph{microcks} & 3.1 (4.5) & \rankworst{3.1 (7.0)} & 3.1 (4.5) & \rankbest{7.2 (1.0)} & 3.1 (4.5) & 3.1 (4.5) & \ranksecond{5.8 (2.0)} \\ 
\emph{ocvn} & 5.5 (6.0) & 6.8 (4.0) & \rankworst{0.0 (7.0)} & \rankbest{18.7 (1.0)} & 6.7 (5.0) & 6.9 (3.0) & \ranksecond{8.6 (2.0)} \\ 
\emph{ohsome-api} & \rankworst{6.1 (5.0)} & \rankworst{6.1 (5.0)} & \rankworst{6.1 (5.0)} & \rankbest{37.7 (1.0)} & \rankworst{6.1 (5.0)} & \rankworst{6.1 (5.0)} & \ranksecond{13.6 (2.0)} \\ 
\emph{pay-publicapi} & \rankworst{9.0 (7.0)} & 13.8 (5.0) & \rankbest{15.7 (1.0)} & 14.6 (4.0) & 10.6 (6.0) & 15.3 (3.0) & \ranksecond{15.4 (2.0)} \\ 
\emph{person-controller} & \rankworst{12.8 (5.5)} & \rankworst{12.8 (5.5)} & \rankworst{12.8 (5.5)} & \rankbest{72.5 (1.0)} & \rankworst{12.8 (5.5)} & \ranksecond{21.3 (2.0)} & 21.3 (3.0) \\ 
\emph{proxyprint} & \rankworst{4.2 (5.0)} & \rankworst{4.2 (5.0)} & \rankworst{4.2 (5.0)} & \rankbest{31.5 (1.0)} & \rankworst{4.2 (5.0)} & \rankworst{4.2 (5.0)} & \ranksecond{26.0 (2.0)} \\ 
\emph{quartz-manager} & \rankworst{34.4 (5.0)} & \rankworst{34.4 (5.0)} & \rankworst{34.4 (5.0)} & \rankbest{37.0 (1.0)} & \rankworst{34.4 (5.0)} & \rankworst{34.4 (5.0)} & \ranksecond{36.7 (2.0)} \\ 
\emph{reservations-api} & \rankworst{21.1 (6.5)} & 25.6 (3.0) & \rankworst{21.1 (6.5)} & \rankbest{27.6 (1.0)} & 24.5 (5.0) & \ranksecond{27.2 (2.0)} & 25.0 (4.0) \\ 
\emph{rest-ncs} & \rankworst{5.1 (5.5)} & \rankworst{5.1 (5.5)} & \rankworst{5.1 (5.5)} & \rankbest{64.4 (1.0)} & \rankworst{5.1 (5.5)} & \ranksecond{40.7 (2.0)} & 22.8 (3.0) \\ 
\emph{rest-news} & \rankworst{13.9 (5.5)} & \rankworst{13.9 (5.5)} & \rankworst{13.9 (5.5)} & \rankbest{84.2 (1.0)} & \rankworst{13.9 (5.5)} & 42.4 (3.0) & \ranksecond{44.4 (2.0)} \\ 
\emph{rest-scs} & \rankworst{4.7 (6.5)} & \ranksecond{57.5 (2.0)} & \rankworst{4.7 (6.5)} & \rankbest{65.1 (1.0)} & 55.9 (3.0) & 52.9 (4.0) & 47.6 (5.0) \\ 
\emph{restcountries} & \rankworst{3.5 (5.5)} & \rankworst{3.5 (5.5)} & \rankbest{77.9 (1.0)} & \ranksecond{77.5 (2.0)} & \rankworst{3.5 (5.5)} & 46.3 (3.0) & \rankworst{3.5 (5.5)} \\ 
\emph{scout-api} & \rankworst{12.0 (5.0)} & \rankworst{12.0 (5.0)} & \rankworst{12.0 (5.0)} & \rankbest{18.9 (1.0)} & \rankworst{12.0 (5.0)} & \ranksecond{16.5 (2.0)} & \rankworst{12.0 (5.0)} \\ 
\emph{session-service} & \rankworst{21.4 (6.5)} & \ranksecond{80.5 (2.0)} & \rankworst{21.4 (6.5)} & 57.9 (3.0) & \rankbest{85.5 (1.0)} & 50.9 (4.0) & 49.1 (5.0) \\ 
\emph{spring-actuator-demo} & \rankworst{67.7 (6.5)} & 84.2 (4.0) & \rankworst{67.7 (6.5)} & \rankbest{87.1 (1.0)} & \ranksecond{85.2 (2.0)} & 80.6 (5.0) & 84.7 (3.0) \\ 
\emph{spring-batch-rest} & \rankworst{15.0 (6.0)} & \ranksecond{31.5 (2.0)} & \rankworst{15.0 (6.0)} & \rankbest{34.3 (1.0)} & 29.4 (4.0) & \rankworst{15.0 (6.0)} & 30.2 (3.0) \\ 
\emph{spring-ecommerce} & \rankworst{14.2 (5.5)} & \rankworst{14.2 (5.5)} & \rankworst{14.2 (5.5)} & \rankbest{33.0 (1.5)} & \rankworst{14.2 (5.5)} & \rankbest{33.0 (1.5)} & \ranksecond{21.5 (3.0)} \\ 
\emph{spring-rest-example} & \rankworst{11.1 (5.5)} & \rankworst{11.1 (5.5)} & \rankworst{11.1 (5.5)} & \rankbest{50.2 (1.0)} & \rankworst{11.1 (5.5)} & \ranksecond{43.7 (2.0)} & 25.0 (3.0) \\ 
\emph{swagger-petstore} & \rankworst{29.5 (5.5)} & \rankworst{29.5 (5.5)} & \rankbest{71.3 (1.0)} & \ranksecond{70.2 (2.0)} & \rankworst{29.5 (5.5)} & 62.9 (3.0) & \rankworst{29.5 (5.5)} \\ 
\emph{tiltaksgjennomforing} & \ranksecond{7.2 (4.5)} & \ranksecond{7.2 (4.5)} & \ranksecond{7.2 (4.5)} & \rankbest{9.0 (1.0)} & \ranksecond{7.2 (4.5)} & \ranksecond{7.2 (4.5)} & \ranksecond{7.2 (4.5)} \\ 
\emph{tracking-system} & \rankworst{19.5 (5.0)} & \rankworst{19.5 (5.0)} & \rankworst{19.5 (5.0)} & \rankbest{39.7 (1.0)} & \rankworst{19.5 (5.0)} & \ranksecond{26.4 (2.0)} & \rankworst{19.5 (5.0)} \\ 
\emph{user-management} & \rankworst{6.5 (6.5)} & 49.9 (3.0) & \rankworst{6.5 (6.5)} & \rankbest{58.8 (1.0)} & \ranksecond{56.3 (2.0)} & 43.1 (4.0) & 37.9 (5.0) \\ 
\emph{webgoat} & 19.6 (5.0) & \rankworst{19.5 (7.0)} & 19.6 (6.0) & \rankbest{19.7 (1.0)} & \ranksecond{19.6 (3.0)} & \ranksecond{19.6 (3.0)} & \ranksecond{19.6 (3.0)} \\ 
\emph{youtube-mock} & \rankworst{36.9 (5.0)} & \rankworst{36.9 (5.0)} & \rankworst{36.9 (5.0)} & \rankbest{48.6 (1.0)} & \rankworst{36.9 (5.0)} & \ranksecond{40.8 (2.0)} & \rankworst{36.9 (5.0)} \\ 
\midrule 
Average  & \rankworst{15.1 (7.0)} & 22.3 (4.0) & 18.4 (6.0) & \rankbest{42.3 (1.0)} & 22.1 (5.0) & \ranksecond{27.6 (2.0)} & 25.9 (3.0) \\ 
Median  & \rankworst{12.4 (7.0)} & 14.4 (5.0) & 14.0 (6.0) & \rankbest{38.5 (1.0)} & 14.6 (4.0) & \ranksecond{23.2 (2.0)} & 23.1 (3.0) \\ 
\midrule 
Friedman Test & \multicolumn{7}{r}{$\chi^2$ = 115.770, $p$-value = $< 0.001$} \\ 
\bottomrule 
\end{tabular} 

}
\end{table}
%---------------------------------------------------------------------------

%---------------------------------------------------------------------------
\subsubsection{Result Analysis}
\label{sub:first-analysis}

Table~\ref{tab:2xx} shows results for the first metric on 2xx coverage, Table~\ref{tab:500} for fault detection of 500 responses, and finally Table~\ref{tab:jacoco} shows the code coverage results.
To show if there is any statistical difference among the 6 compared fuzzers, a Friedman Test is executed.

%---------------------------------------------------------------------------
\begin{table}[!t]
	\centering
	\small
\caption{
\label{tab:schemathesis}
Pairwise comparison on JaCoCo code coverage between no fuzzing ($B$), \evo ($X$) and \schemathesis ($Y$),
with standardized effect sizes and p-values.
When p-values are equal or lower than the threshold $0.05$, the effect sizes are highlighted in bold.
}
\vspace{-1.5\baselineskip}
%\resizebox{0.98\textwidth}{!}{
    \begin{tabular}{ l r|  rr|  rr| rrr}\\ 
\toprule 
 SUT   & $B$ & $X$ & $X-B$ & $Y$  & $Y-B$ &  $X-Y$  & $\hat{A}_{XY}$ & p-value  \\ 
\midrule 
\emph{bibliothek} & 32.9 & 39.3 & +6.3 & 39.3 & +6.3 & 0.0 & 0.50 & 1.000 \\ 
\emph{blogapi} & 7.1 & 18.1 & +10.9 & 15.9 & +8.8 & +2.1 & {\bf 1.00} & $< 0.001$ \\ 
\emph{catwatch} & 9.4 & 45.7 & +36.3 & 31.2 & +21.8 & +14.5 & {\bf 1.00} & $< 0.001$ \\ 
\emph{cwa-verification} & 21.9 & 49.3 & +27.3 & 40.6 & +18.7 & +8.6 & {\bf 1.00} & $< 0.001$ \\ 
\emph{erc20-rest-service} & 7.9 & 26.7 & +18.8 & 23.4 & +15.5 & +3.3 & {\bf 1.00} & $< 0.001$ \\ 
\emph{familie-ba-sak} & 15.8 & 16.6 & +0.8 & 16.5 & +0.7 & +0.1 & 0.42 & 0.567 \\ 
\emph{features-service} & 19.0 & 65.0 & +45.9 & 36.3 & +17.2 & +28.7 & {\bf 1.00} & $< 0.001$ \\ 
\emph{genome-nexus} & 12.8 & 33.4 & +20.6 & 21.1 & +8.3 & +12.3 & {\bf 1.00} & $< 0.001$ \\ 
\emph{gestaohospital} & 19.9 & 58.7 & +38.8 & 34.8 & +14.9 & +23.9 & {\bf 1.00} & $< 0.001$ \\ 
\emph{http-patch-spring} & 4.9 & 78.5 & +73.6 & 14.6 & +9.7 & +63.9 & {\bf 1.00} & $< 0.001$ \\ 
\emph{languagetool} & 1.3 & 10.8 & +9.4 & 1.7 & +0.4 & +9.0 & {\bf 1.00} & $< 0.001$ \\ 
\emph{market} & 7.0 & 14.3 & +7.3 & 11.5 & +4.5 & +2.8 & {\bf 1.00} & $< 0.001$ \\ 
\emph{microcks} & 3.1 & 7.2 & +4.1 & 5.8 & +2.7 & +1.3 & {\bf 1.00} & $< 0.001$ \\ 
\emph{ocvn} & 5.5 & 18.7 & +13.1 & 8.6 & +3.0 & +10.1 & {\bf 1.00} & $< 0.001$ \\ 
\emph{ohsome-api} & 6.1 & 37.7 & +31.6 & 13.6 & +7.5 & +24.1 & {\bf 1.00} & $< 0.001$ \\ 
\emph{pay-publicapi} & 9.0 & 14.6 & +5.7 & 15.4 & +6.5 & -0.8 & {\bf 0.00} & $< 0.001$ \\ 
\emph{person-controller} & 12.8 & 72.5 & +59.7 & 21.3 & +8.5 & +51.2 & {\bf 1.00} & $< 0.001$ \\ 
\emph{proxyprint} & 4.2 & 31.5 & +27.2 & 26.0 & +21.8 & +5.5 & {\bf 1.00} & $< 0.001$ \\ 
\emph{quartz-manager} & 34.4 & 37.0 & +2.6 & 36.7 & +2.3 & +0.3 & 0.56 & 0.314 \\ 
\emph{reservations-api} & 21.1 & 27.6 & +6.5 & 25.0 & +3.9 & +2.5 & {\bf 1.00} & $< 0.001$ \\ 
\emph{rest-ncs} & 5.1 & 64.4 & +59.3 & 22.8 & +17.7 & +41.7 & {\bf 1.00} & $< 0.001$ \\ 
\emph{rest-news} & 13.9 & 84.2 & +70.3 & 44.4 & +30.6 & +39.7 & {\bf 1.00} & $< 0.001$ \\ 
\emph{rest-scs} & 4.7 & 65.1 & +60.4 & 47.6 & +42.8 & +17.5 & {\bf 1.00} & $< 0.001$ \\ 
\emph{restcountries} & 3.5 & 77.5 & +74.0 & 3.5 & 0.0 & +74.0 & {\bf 1.00} & $< 0.001$ \\ 
\emph{scout-api} & 12.0 & 18.9 & +6.9 & 12.0 & 0.0 & +6.9 & {\bf 1.00} & $< 0.001$ \\ 
\emph{session-service} & 21.4 & 57.9 & +36.5 & 49.1 & +27.7 & +8.7 & {\bf 1.00} & $< 0.001$ \\ 
\emph{spring-actuator-demo} & 67.7 & 87.1 & +19.4 & 84.7 & +16.9 & +2.4 & 0.56 & 0.314 \\ 
\emph{spring-batch-rest} & 15.0 & 34.3 & +19.3 & 30.2 & +15.2 & +4.2 & {\bf 1.00} & $< 0.001$ \\ 
\emph{spring-ecommerce} & 14.2 & 33.0 & +18.8 & 21.5 & +7.3 & +11.5 & {\bf 1.00} & $< 0.001$ \\ 
\emph{spring-rest-example} & 11.1 & 50.2 & +39.1 & 25.0 & +13.9 & +25.2 & {\bf 1.00} & $< 0.001$ \\ 
\emph{swagger-petstore} & 29.5 & 70.2 & +40.7 & 29.5 & 0.0 & +40.7 & {\bf 1.00} & $< 0.001$ \\ 
\emph{tiltaksgjennomforing} & 7.2 & 9.0 & +1.8 & 7.2 & 0.0 & +1.8 & {\bf 1.00} & $< 0.001$ \\ 
\emph{tracking-system} & 19.5 & 39.7 & +20.2 & 19.5 & 0.0 & +20.2 & {\bf 1.00} & $< 0.001$ \\ 
\emph{user-management} & 6.5 & 58.8 & +52.3 & 37.9 & +31.4 & +20.8 & {\bf 0.91} & 0.002 \\ 
\emph{webgoat} & 19.6 & 19.7 & +0.1 & 19.6 & +0.0 & +0.1 & {\bf 1.00} & $< 0.001$ \\ 
\emph{youtube-mock} & 36.9 & 48.6 & +11.7 & 36.9 & 0.0 & +11.7 & {\bf 1.00} & $< 0.001$ \\ 
\midrule 
Mean  & 15.1 & 42.3 & +27.1 & 25.9 & +10.7 & +16.4 & 0.92 &  \\ 
Median  & 12.4 & 38.5 & +26.1 & 23.1 & +10.7 & +15.4 & 1.00 &  \\ 
\bottomrule 
\end{tabular} 

%}
\end{table}
%---------------------------------------------------------------------------

In these experiments, it would appear that \evo gives the best results, followed by \schemathesis and \restler having similar results.
When showing results on a set of different tools, then it is also import to show pair-wise comparisons with more in details statistical analysis.
For example,
Table~\ref{tab:schemathesis} shows a pair-comparison for JaCoCo results between \evo and \schemathesis.
On each API, we shows Vargha-Delaney $\hat{A}_{12}$ statistics, with p-values for Mann-Whitney-Wilcoxon U-tests,
following the guidelines in~\cite{Hitchhiker14}.
For reason of space, we do not provide such kind of table for each paired tool and target metric.

Based on the results of these experiments, caution needs to be taken.
Not only \evo gives better results, but also by a very large margin.
Considering that the work on
\aratrl\cite{kim2023adaptive},
\emrest~\cite{xu2025effective}
and
\llamaresttest~\cite{kim2025llamaresttest}
claimed better results compared to \evo, we need to analyze the discrepancies in these results.

The first possible issue is that tools might be misconfigured.
However, it does not seem the case here.
Each tool gave the best results on at least 1 API.
If they were misconfigured, that would unlikely happen.
There is no specific, custom configuration per API, and all tools get exactly the same information as input.
Still, to reduce the risks of misconfigurations, we contacted all the authors of the different fuzzers on their GitHub repositories,
particularly for
\aratrl,\footnote{\url{https://github.com/codingsoo/ARAT-RL/issues/7}}
\emrest,\footnote{\url{https://github.com/GIST-NJU/EmRest/issues/1}}
\llamaresttest,\footnote{\url{https://github.com/codingsoo/LlamaRestTest/issues/2}}
\restler,\footnote{\url{https://github.com/microsoft/restler-fuzzer/issues/982}}
and
\schemathesis.\footnote{\url{https://github.com/schemathesis/schemathesis/discussions/3122}}
No issue was reported regarding how we ran those tools, apart from \emrest stating it supports static authentication (which is not discussed in its documentation).

At the time {of these experiments in 2025}, \aratrl has not been updated in 2 years.
On the other hand, \emrest~\cite{xu2025effective} and \llamaresttest~\cite{kim2025llamaresttest} are among the most recent published works.
Interestingly,  \llamaresttest~\cite{kim2025llamaresttest} compared with \restler (but not \schemathesis),
and \emrest~\cite{xu2025effective} compared with \schemathesis (but not \restler), stating to provide better results.
However, in our experiments they provide worse results compared to \restler and \schemathesis.

By looking at the logs of these tools, and by considering the properties of WFD, these results are rather obvious in hindsight.
In most of the runs, those tools either crashed or were not able to make any calls.
It is not in the scope of this paper to analyse the specific reasons for each of the $5 \times \nsutsOld = 180$ different cases.
However, a random selection of some of those logs pointed out two common issues:

\begin{itemize}
\item  Few of these tools assume the OpenAPI schemas to be 100\% valid.
       As APIs can have faults, schemas can have to.
       This is not a hypothetical issue, as it happens quite often.
       Robust fuzzers can still try to generate valuable tests even if there are some minor issues in the schema.
       Straight out crashing without making any HTTP call is not a viable approach for the use of fuzzing in industry~\cite{icst2025vw}.
\item  What endpoints can be called in an API is not necessarily trivial.
       In OpenAPI schemas, there is an array field called \texttt{path} with such information.
       However, those endpoints might have a common prefix, defined elsewhere, that needs to be added on each call.
       In version 2 of the OpenAPI schema, there is an optional field called \texttt{basePath}.
        It might be left empty, or with default value \texttt{/}.
        However, in cases like \emph{languagetool}, it has the value \texttt{/v2}.
        On the other hand, in version 3 of OpenAPI, such info is under \texttt{servers.url}.
        This could be a path entry (e.g., like \texttt{/api/v3} for \emph{swagger-petstore}) or a full
        URL (e.g., like \texttt{http://localhost:8080/rest} for \emph{restcountries}).
        If a tool is not able to automatically infer the right path prefix from the schema, then it will not be able
        to make any valid call.
\end{itemize}

These issues are technical, and have little to do with research.
Robust fuzzers, under development for years, and used by thousands of people in industry, have higher chances to work fine on new APIs.
On the other hand, academic prototypes might straight out crash as soon as used on a different SUT from their original studies.
Creating a basic corpus like EMB, and now WFD, is aimed to avoid this kind of issues in future studies.
Otherwise, if one wants to show better results compared to the most recent state-of-the-art in academic prototypes, a simple approach is to simply choose a different set of APIs for experiments.
{
Note that fuzzers developed in industry are not immune to these issues either.
As we will discuss in Section~\ref{sec:second-setup}, CATS was also affected by the inference of the path prefix in preliminary runs on WFD.
}

{
In this first set of experiments, only the authors of \evo were involved in running them.
This can create a potential bias.
}
The reason is straightforward: although we have not purposely defined any technique or heuristics tailored to any API in WFD, we made sure that, if there was any crash, we fixed them before finalizing the experiments.
Bugs in \evo that show up in WFD might happen as well for the APIs of thousands of practitioners in industry that download and use  \evo.
As such, we have a strong incentive to fix all these found issues.

{
However, this is nothing peculiar to these experiments, and actually it represents the current common practice among empirical studies, at least in the context of fuzzing REST APIs.
Authors of new fuzzers, when empirically evaluating their techniques, do not typically involve the authors of the other compared fuzzers in their studies.
And they can fix any issue they find in their tools on their selection of the APIs before trying to publish their final experiment results.
Although, at the time of writing, \evo has been available open-source for nearly 10 years~\cite{arcuri2017restful}, and used in tens of studies by other researchers, not a single time we were contacted about whether it was correctly set up, or if there was any  crash in the experiments that were run in those comparisons when using new APIs that \evo has not been run on before.
}
{
On the other hand,
}
as EMB/WFD are publicly available, with EMB open-source since 2017, any fuzzer developer can use these APIs and verify their tools work reliably on them.
{
Many of the fuzzer crashes in these comparisons are on APIs that have been available in EMB for many years (e.g., \emph{scout-api} was added in EMB in 2017).
Furthermore, in the sake of scientific integrity, we directly contacted the authors of all compared fuzzers, by opening new issues/discussions on the open-source repositories of these fuzzers, linked here in this paper, which everyone can review.
}

%---------------------------------------------------------------------------
\begin{table}[!t]
	\centering
\caption{
\label{tab:2xxbias}
Same kind of results as presented in Table~\ref{tab:2xx} for 2xx coverage.
	However, instead of using \nsutsOld APIs, results are presented only for a cherry-picked selection of 6 APIs.}

\vspace{-1.5\baselineskip}
\resizebox{0.99\textwidth}{!}{
    \begin{tabular}{ l r r r r r r  }\\ 
\toprule 
SUT & \aratrl & \emrest & \evomasterbbnoauth & \llamaresttest & \restler & \schemathesis \\ 
\midrule 
\emph{blogapi} & \ranksecond{15.1 (3.0)} & \rankbest{20.9 (1.0)} & \ranksecond{15.1 (3.0)} & \rankworst{6.9 (6.0)} & \ranksecond{15.1 (3.0)} & 9.9 (5.0) \\ 
\emph{market} & \ranksecond{13.8 (4.0)} & \rankworst{0.0 (6.0)} & \rankbest{15.4 (2.0)} & \rankbest{15.4 (2.0)} & \rankbest{15.4 (2.0)} & 5.4 (5.0) \\ 
\emph{pay-publicapi} & 0.0 (3.5) & 0.0 (3.5) & 0.0 (3.5) & 0.0 (3.5) & 0.0 (3.5) & 0.0 (3.5) \\ 
\emph{spring-ecommerce} & \rankworst{0.0 (5.0)} & \rankworst{0.0 (5.0)} & \rankbest{22.2 (1.5)} & \rankworst{0.0 (5.0)} & \rankbest{22.2 (1.5)} & \ranksecond{16.7 (3.0)} \\ 
\emph{spring-rest-example} & \ranksecond{0.0 (4.5)} & \ranksecond{0.0 (4.5)} & \rankbest{44.4 (1.5)} & \ranksecond{0.0 (4.5)} & \rankbest{44.4 (1.5)} & \ranksecond{0.0 (4.5)} \\ 
\emph{swagger-petstore} & \rankworst{0.0 (5.0)} & 78.9 (3.0) & \ranksecond{81.6 (2.0)} & \rankworst{0.0 (5.0)} & \rankbest{89.5 (1.0)} & \rankworst{0.0 (5.0)} \\ 
\midrule 
Average  & 4.8 (5.0) & 16.6 (3.0) & \ranksecond{29.8 (2.0)} & \rankworst{3.7 (6.0)} & \rankbest{31.1 (1.0)} & 5.3 (4.0) \\ 
Median  & \rankworst{0.0 (5.0)} & \rankworst{0.0 (5.0)} & \rankbest{18.8 (1.5)} & \rankworst{0.0 (5.0)} & \rankbest{18.8 (1.5)} & \ranksecond{2.7 (3.0)} \\ 
\midrule 
Friedman Test & \multicolumn{6}{r}{$\chi^2$ = 13.503, $p$-value = 0.019} \\ 
\bottomrule 
\end{tabular} 

}
\end{table}
%---------------------------------------------------------------------------

Even if the APIs were used in previous studies, what happens if only a subset is selected, cherry-picking APIs of interest, and discarding the rest based on some criterion?
Table~\ref{tab:2xxbias} shows a subset of the same results as Table~\ref{tab:2xx} on HTTP 2xx coverage.
However, this time we cherry-picked only 6 APIs.
In terms of number of APIs, this is no different than what done for APIRL~\cite{foley2025apirl} or Morest~\cite{liu2022icse}.
Here, in this cherry-picked subset, \restler gives the best results on average.
We argue that, once a corpus such as WFD is employed, cherry-picking APIs is unsound.
If there is any implicit bias in the selection of APIs for WFD, then, instead of removing APIs, new ones should be added to edge the potential bias, if any.

Note that the issue of selecting artifacts for experimentation is, technically speaking, a very well known issue in many scientific domains, not just in software engineering.
A recent example on fuzz testing in which these issues are discussed is~\cite{wolff2025fuzzing}.
Here we provide more empirical evidence, specific to the fuzzing of REST APIs, to stress how serious the problem is, and how it has been handled in the literature.

%%%%%%%%%%%%%%%%%%%%%%%%%%%%%%%%%%%%%%%%%%%%%%%%%%%%%%%%%%%%%%%%%%%%%%%%%%%%
\subsubsection{WFC Authentication}
\label{sec:auh:experiments}

%----------------------------
\begin{table*}[!t]
	\centering
	\small
	\caption{ Average code coverage comparisons, per API needing authentication, between the base booting ($B$) of the API, the fuzzer \evo ($X$) using WFC authentication, and without it ($Y$). We also reports the delta differences, e.g., of 1-hour  over base ($X-B$ and $Y-B$), and 1-hour \evo with authentication over no authentication ($X-Y$).
		\label{tab:authnoauth}
	}
	\vspace{-1.5\baselineskip}
% 	\resizebox{0.98\textwidth}{!}{
		\begin{tabular}{ l r|  rr|  rr| rrr}\\ 
\toprule 
 SUT   & $B$ & $X$ & $X-B$ & $Y$  & $Y-B$ &  $X-Y$  & $\hat{A}_{XY}$ & p-value  \\ 
\midrule 
\emph{blogapi} & 7.1 & 30.5 & +23.4 & 18.1 & +10.9 & +12.5 & {\bf 1.00} & $< 0.001$ \\ 
\emph{familie-ba-sak} & 15.8 & 16.6 & +0.8 & 16.6 & +0.8 & +0.0 & 0.53 & 0.868 \\ 
\emph{market} & 7.0 & 25.8 & +18.7 & 14.3 & +7.3 & +11.5 & {\bf 1.00} & $< 0.001$ \\ 
\emph{microcks} & 3.1 & 12.2 & +9.1 & 7.2 & +4.1 & +5.0 & {\bf 0.97} & $< 0.001$ \\ 
\emph{ocvn} & 5.5 & 19.8 & +14.3 & 18.7 & +13.1 & +1.2 & {\bf 1.00} & $< 0.001$ \\ 
\emph{pay-publicapi} & 9.0 & 14.7 & +5.8 & 14.6 & +5.7 & +0.1 & {\bf 0.86} & 0.007 \\ 
\emph{proxyprint} & 4.2 & 52.3 & +48.0 & 31.5 & +27.2 & +20.8 & {\bf 1.00} & $< 0.001$ \\ 
\emph{quartz-manager} & 34.4 & 50.1 & +15.7 & 37.0 & +2.6 & +13.1 & {\bf 1.00} & $< 0.001$ \\ 
\emph{reservations-api} & 21.1 & 39.8 & +18.6 & 27.6 & +6.5 & +12.2 & 0.44 & 0.687 \\ 
\emph{scout-api} & 12.0 & 40.2 & +28.2 & 18.9 & +6.9 & +21.3 & {\bf 1.00} & $< 0.001$ \\ 
\emph{spring-actuator-demo} & 67.7 & 87.1 & +19.4 & 87.1 & +19.4 & 0.0 & 0.50 & 1.000 \\ 
\emph{spring-ecommerce} & 14.2 & 41.0 & +26.9 & 33.0 & +18.8 & +8.1 & {\bf 1.00} & $< 0.001$ \\ 
\emph{tiltaksgjennomforing} & 7.2 & 9.0 & +1.8 & 9.0 & +1.8 & 0.0 & 0.50 & 1.000 \\ 
\emph{tracking-system} & 19.5 & 39.9 & +20.4 & 39.7 & +20.2 & +0.3 & {\bf 0.81} & 0.022 \\ 
\emph{webgoat} & 19.6 & 52.8 & +33.2 & 19.7 & +0.1 & +33.1 & {\bf 1.00} & $< 0.001$ \\ 
\midrule 
Mean  & 16.5 & 35.5 & +18.9 & 26.2 & +9.7 & +9.3 & 0.84 &  \\ 
Median  & 12.0 & 39.8 & +27.8 & 18.9 & +6.9 & +20.9 & 1.00 &  \\ 
\bottomrule 
\end{tabular} 

% 	}
	\vspace{-0.5\baselineskip}
\end{table*}
%----------------------------

In this section, we show the results of using authentication information, with WFC format.
We ran \evo with auth info on the 15 APIs that require authentication
{ out of \nsutsOld used in the first set of experiments}.
Experiments were run with the same settings, i.e., 1-hour per experiment, on the same hardware, with 10 repetitions.
Results for code coverage with and without authentication are shown in Table~\ref{tab:authnoauth}.

Note that, as these experiments are based only on \evo,  these should be considered only as demonstration of WFD/WFC's utility rather than as a general comparison of fuzzer capability
{
 (which will be presented in Section~\ref{sub:experiments-second}).
}

On average, using authentication leads to an improvement to code coverage of $9.3$\%.
Using authentication itself does not guarantee higher code coverage, as still need to generate the right input data to pass the first layer of input validation.
An interesting example is \texttt{pay-publicapi}.
With, authentication only a $0.1$\% improvement is obtained.
However, in previous work, by using advanced \emph{white-box} techniques, improvements were by over 20\% on this API~\cite{seran2025handling}.

%%%%%%%%%%%%%%%%%%%%%%%%%%%%%%%%%%%%%%%%%%%%%%%%%%%%%%%%%%%%%%%%%%%%%%%%%%%%
\subsubsection{Test Generation}
\label{sec:generation}

Table~\ref{tab:schemathesis} shows a comparison between 2 fuzzers based on code coverage measured with JaCoCo.
For the sake of discussion, which tools they are is not relevant.
Let us just call them $X$ and $Y$.

This type of comparisons, in which code coverage metrics are computed during the fuzzing session, is common in the literature.
 All studies in the literature of black-box fuzzing~\cite{golmohammadi2023testing} have done this, as far as we know, including our own previous work on fuzzer comparisons~\cite{zhang2023open}.
However, in this paper we claim this kind of analysis is \emph{biased}, and potential \emph{misleading}, as we will argue in more details next.

Assume that, when running experiments on a local machine, an HTTP call takes 5 milliseconds (this of course will vary from API to API and hardware involved).
During 1 hour, a fuzzer would be able to make 720 000 calls toward the API.
In our experiments in this paper, we have seen cases of up to 572 000 calls within 1 hour (recall though that the APIs were running inside Docker, behind a mitmproxy instance, which might create some delay).
No test engineer in their right state of mind  would be interested in obtaining, as output of the fuzzer, a test suite containing more than half-million HTTP calls.
Handling such gargantuan test files would be infeasible.
Somehow, a fuzzer needs to output a small test suite, based on some criteria.
What strategy to use would be up to each fuzzer.
For example, if a fuzzer detects a fault based on some automated oracle (e.g., whether it is a 500 HTTP status code, or some security faults such as mass assignments~\cite{corradini2023automated}), tests revealing such faults could be saved for the final output test suite.
However, \emph{this does not work for code coverage metrics}.

%----------------------------
\begin{table*}[!t]
	\centering
	\small
	\caption{ Average code coverage comparisons, per API, between the base booting ($B$) of the API, the fuzzer \evo ($X$) during the 1-hour fuzzing session, and the execution of its generated tests ($Y$). We also reports the delta differences, e.g., of 1-hour \evo  over base ($X-B$), generated tests over base ($Y-B$), and 1-hour \evo over generated tests ($X-Y$).
		\label{tab:tests}
	}
	\vspace{-1.5\baselineskip}
% 	\resizebox{0.98\textwidth}{!}{
		\begin{tabular}{ l r|  rr|  rr| rrr}\\ 
\toprule 
 SUT   & $B$ & $X$ & $X-B$ & $Y$  & $Y-B$ &  $X-Y$  & $\hat{A}_{XY}$ & p-value  \\ 
\midrule 
\emph{bibliothek} & 32.9 & 37.5 & +4.5 & 35.5 & +2.5 & +2.0 & 0.66 & 0.237 \\ 
\emph{blogapi} & 7.1 & 30.5 & +23.4 & 28.4 & +21.3 & +2.1 & 0.42 & 0.623 \\ 
\emph{catwatch} & 9.4 & 43.8 & +34.4 & 28.6 & +19.2 & +15.2 & {\bf 0.91} & 0.002 \\ 
\emph{cwa-verification} & 21.9 & 49.2 & +27.3 & 43.7 & +21.8 & +5.5 & {\bf 1.00} & $< 0.001$ \\ 
\emph{erc20-rest-service} & 7.9 & 24.3 & +16.5 & 20.1 & +12.3 & +4.2 & {\bf 0.89} & 0.003 \\ 
\emph{familie-ba-sak} & 15.8 & 16.6 & +0.8 & 16.4 & +0.5 & +0.2 & {\bf 0.80} & 0.019 \\ 
\emph{features-service} & 19.0 & 65.0 & +45.9 & 57.0 & +38.0 & +7.9 & {\bf 1.00} & $< 0.001$ \\ 
\emph{genome-nexus} & 12.8 & 31.7 & +18.8 & 24.8 & +11.9 & +6.9 & {\bf 1.00} & $< 0.001$ \\ 
\emph{gestaohospital} & 19.9 & 62.6 & +42.7 & 48.0 & +28.1 & +14.6 & {\bf 0.99} & $< 0.001$ \\ 
\emph{http-patch-spring} & 4.9 & 78.5 & +73.6 & 64.1 & +59.2 & +14.3 & {\bf 1.00} & $< 0.001$ \\ 
\emph{languagetool} & 1.3 & 10.8 & +9.5 & 8.1 & +6.7 & +2.7 & 0.72 & 0.113 \\ 
\emph{market} & 7.0 & 25.8 & +18.7 & 22.1 & +15.0 & +3.7 & 0.40 & 0.495 \\ 
\emph{microcks} & 3.1 & 12.2 & +9.1 & 9.6 & +6.5 & +2.5 & {\bf 0.90} & 0.005 \\ 
\emph{ocvn} & 5.5 & 19.8 & +14.3 & 17.3 & +11.7 & +2.6 & 0.65 & 0.115 \\ 
\emph{ohsome-api} & 6.1 & 33.1 & +27.0 & 6.1 & 0.0 & +27.0 & {\bf 0.93} & $< 0.001$ \\ 
\emph{pay-publicapi} & 9.0 & 14.7 & +5.8 & 13.0 & +4.0 & +1.7 & 0.63 & 0.294 \\ 
\emph{person-controller} & 12.8 & 65.0 & +52.3 & 54.3 & +41.5 & +10.8 & {\bf 0.89} & 0.005 \\ 
\emph{proxyprint} & 4.2 & 52.3 & +48.0 & 6.3 & +2.1 & +45.9 & {\bf 1.00} & $< 0.001$ \\ 
\emph{quartz-manager} & 34.4 & 50.1 & +15.7 & 52.2 & +17.8 & -2.1 & {\bf 0.00} & $< 0.001$ \\ 
\emph{reservations-api} & 21.1 & 39.8 & +18.6 & 52.4 & +31.3 & -12.7 & 0.30 & 0.079 \\ 
\emph{rest-ncs} & 5.1 & 64.4 & +59.3 & 46.9 & +41.8 & +17.5 & {\bf 1.00} & $< 0.001$ \\ 
\emph{rest-news} & 13.9 & 84.4 & +70.5 & 55.7 & +41.8 & +28.7 & {\bf 1.00} & $< 0.001$ \\ 
\emph{rest-scs} & 4.7 & 65.1 & +60.4 & 57.6 & +52.9 & +7.5 & {\bf 1.00} & $< 0.001$ \\ 
\emph{restcountries} & 3.5 & 69.3 & +65.8 & 61.4 & +57.9 & +7.9 & {\bf 0.90} & 0.003 \\ 
\emph{scout-api} & 12.0 & 40.2 & +28.2 & 29.3 & +17.3 & +10.9 & {\bf 1.00} & $< 0.001$ \\ 
\emph{session-service} & 21.4 & 57.9 & +36.5 & 57.5 & +36.1 & +0.4 & 0.55 & 0.368 \\ 
\emph{spring-actuator-demo} & 67.7 & 87.1 & +19.4 & 84.9 & +17.2 & +2.2 & 0.56 & 0.374 \\ 
\emph{spring-batch-rest} & 15.0 & 34.3 & +19.3 & 30.3 & +15.3 & +4.0 & {\bf 0.90} & 0.002 \\ 
\emph{spring-ecommerce} & 14.2 & 41.0 & +26.9 & 31.5 & +17.3 & +9.6 & {\bf 1.00} & $< 0.001$ \\ 
\emph{spring-rest-example} & 11.1 & 50.2 & +39.0 & 41.0 & +29.9 & +9.2 & {\bf 0.94} & 0.002 \\ 
\emph{swagger-petstore} & 29.5 & 70.2 & +40.7 & 66.3 & +36.8 & +3.9 & {\bf 1.00} & $< 0.001$ \\ 
\emph{tiltaksgjennomforing} & 7.2 & 9.0 & +1.8 & 9.0 & +1.8 & 0.0 & 0.50 & 1.000 \\ 
\emph{tracking-system} & 19.5 & 39.9 & +20.4 & 34.1 & +14.6 & +5.8 & {\bf 1.00} & $< 0.001$ \\ 
\emph{user-management} & 6.5 & 58.2 & +51.7 & 40.7 & +34.2 & +17.5 & {\bf 0.90} & 0.004 \\ 
\emph{webgoat} & 19.6 & 52.8 & +33.2 & 39.0 & +19.4 & +13.8 & {\bf 1.00} & $< 0.001$ \\ 
\emph{youtube-mock} & 36.9 & 51.5 & +14.6 & 43.7 & +6.7 & +7.9 & {\bf 0.91} & 0.003 \\ 
\midrule 
Mean  & 15.1 & 45.5 & +30.4 & 37.1 & +22.0 & +8.4 & 0.81 &  \\ 
Median  & 12.4 & 46.5 & +34.1 & 37.2 & +24.8 & +9.3 & 0.90 &  \\ 
\bottomrule 
\end{tabular} 

% 	}
	\vspace{-0.5\baselineskip}
\end{table*}
%----------------------------

Table~\ref{tab:tests} shows the results of \evo during the 1 hour fuzzing sessions (exactly the same data as in Table~\ref{tab:schemathesis}).
However, this time \evo is compared with the execution of the Python test cases it \emph{generated}, on each API.
To avoid bias, the approach of computing code metrics is the same as described in Section~\ref{sec:experiments}.
In other words, for each SUT, we started the API with Docker Compose, run all the generated test cases (e.g., using commands such as \texttt{py -m pytest <folder>}), collect coverage metrics with JaCoCo, and then stop Docker Compose.
As we can see in Table~\ref{tab:tests}, results of the generated test cases are much worse.
Average coverage drops from $45.5$\% to $37.1$\%.

There can be few explanations to explain these results.
First, there could be faults in the tool itself, in which the generated test files are erroneous.
But, from a scientific point, the ``elephant-in-the-room'' is that, in black-box testing, you have no access to code metrics.
Running an API with JaCoCo is done just for scientific reasons, as higher code coverage can lead to more effective regression test suites.
If a new sampled test case leads to cover more code, the fuzzer would have no direct way to determine it.
However, if in practical scenarios one would measure code coverage, and use such results to select which test suites to generate as output, then it would no longer be \emph{black-box} testing, but rather \emph{grey-box} testing.
And, in these cases, more advanced and performant techniques like \emph{white-box} testing could be used~\cite{arcuri2020blackbox}.

In other words, tables such as Table~\ref{tab:schemathesis}
{
  in a black-box context need to be carefully interpreted.
Measuring code coverage can give an extra insight on the capability of a black-box fuzzer.
And a fuzzer that can achieve higher code coverage during the fuzzing sessions might be preferred.
However, there is no guarantee that any achieved code coverage during the fuzzing session will be still achieved by executing the output test cases generated by the fuzzer.
}
Measuring code coverage is important, but {ideally} it should be calculated on the generated tests, and not during the fuzzing sessions.
What strategies to design to increase the code coverage of the \emph{small}, final test suites, \emph{without computing code coverage}, is an important research problem that has been ignored {so far} in the literature of fuzzing REST APIs.

%---------------------------------------------------------------------------
\begin{table}[!t]
	\centering
\caption{ Status results of the Python tests generated by \evo, and run with PyTest.
It is reported when they failed due to runtime errors (\emph{Errors}), or due to assertion failures (\emph{Failures}).
We also report the number of \emph{Skipped} tests, if any.
The values \emph{Tests} represent the total number of tests (which included all the other three categories plus the tests that were successful).
The metric \emph{Problem \%} is based on the ratio of \emph{Errors} plus \emph{Failures}, divided by the total \emph{Tests}.
\label{tab:pytest}
}
\vspace{-1.5\baselineskip}
% \resizebox{0.98\textwidth}{!}{
\begin{tabular}{ l  rrrr r}\\ 
\toprule 
 SUT & Errors & Failures & Skipped & Tests & Problems \% \\ 
\midrule 
\emph{bibliothek} & 0.0 & 0.0 & 0.0 & 4.0 & 0.0 \\ 
\emph{blogapi} & 0.0 & 14.1 & 0.0 & 157.5 & 9.0 \\ 
\emph{catwatch} & 0.0 & 1.9 & 0.0 & 51.4 & 3.7 \\ 
\emph{cwa-verification} & 0.0 & 2.7 & 0.0 & 6.3 & 42.9 \\ 
\emph{erc20-rest-service} & 0.0 & 0.0 & 0.0 & 9.1 & 0.0 \\ 
\emph{familie-ba-sak} & 0.0 & 7.7 & 0.0 & 1106.1 & 0.7 \\ 
\emph{features-service} & 0.0 & 1.8 & 0.0 & 27.0 & 6.7 \\ 
\emph{genome-nexus} & 0.0 & 7.5 & 0.0 & 35.8 & 20.9 \\ 
\emph{gestaohospital} & 0.0 & 15.3 & 0.0 & 35.6 & 43.0 \\ 
\emph{http-patch-spring} & 0.0 & 4.5 & 0.0 & 20.5 & 22.0 \\ 
\emph{languagetool} & 0.0 & 0.0 & 0.0 & 3.8 & 0.0 \\ 
\emph{market} & 0.0 & 7.8 & 0.0 & 41.7 & 18.7 \\ 
\emph{microcks} & 0.0 & 34.7 & 0.0 & 226.2 & 15.3 \\ 
\emph{ocvn} & 0.0 & 337.6 & 0.0 & 1550.8 & 21.8 \\ 
\emph{ohsome-api} & 0.7 & 0.0 & 0.0 & 0.7 & 100.0 \\ 
\emph{pay-publicapi} & 0.0 & 0.0 & 0.0 & 37.2 & 0.0 \\ 
\emph{person-controller} & 0.0 & 3.9 & 0.0 & 11.2 & 34.8 \\ 
\emph{proxyprint} & 0.0 & 371.9 & 0.0 & 444.9 & 83.6 \\ 
\emph{quartz-manager} & 0.0 & 8.0 & 0.0 & 27.0 & 29.6 \\ 
\emph{reservations-api} & 0.0 & 9.1 & 0.0 & 21.2 & 42.9 \\ 
\emph{rest-ncs} & 0.0 & 0.0 & 0.0 & 13.8 & 0.0 \\ 
\emph{rest-news} & 0.0 & 3.5 & 0.0 & 14.4 & 24.3 \\ 
\emph{rest-scs} & 0.0 & 0.0 & 0.0 & 13.4 & 0.0 \\ 
\emph{restcountries} & 0.0 & 0.0 & 0.0 & 49.5 & 0.0 \\ 
\emph{scout-api} & 0.0 & 46.5 & 0.0 & 248.9 & 18.7 \\ 
\emph{session-service} & 0.0 & 0.0 & 0.0 & 66.2 & 0.0 \\ 
\emph{spring-actuator-demo} & 0.0 & 0.7 & 0.0 & 7.5 & 9.3 \\ 
\emph{spring-batch-rest} & 0.0 & 0.0 & 0.0 & 9.8 & 0.0 \\ 
\emph{spring-ecommerce} & 0.0 & 1.8 & 0.0 & 56.4 & 3.2 \\ 
\emph{spring-rest-example} & 0.0 & 0.0 & 0.0 & 15.4 & 0.0 \\ 
\emph{swagger-petstore} & 0.0 & 24.1 & 0.0 & 55.1 & 43.7 \\ 
\emph{tiltaksgjennomforing} & 0.0 & 0.0 & 0.0 & 459.0 & 0.0 \\ 
\emph{tracking-system} & 0.0 & 154.2 & 0.0 & 210.4 & 73.3 \\ 
\emph{user-management} & 0.0 & 14.9 & 0.0 & 38.2 & 39.0 \\ 
\emph{webgoat} & 0.1 & 133.5 & 0.0 & 371.8 & 35.9 \\ 
\emph{youtube-mock} & 0.0 & 0.1 & 0.0 & 14.0 & 0.7 \\ 
\midrule 
Mean  & 0.0 & 33.5 & 0.0 & 151.7 & 20.7 \\ 
Median  & 0.0 & 3.1 & 0.0 & 35.7 & 12.3 \\ 
\bottomrule 
\end{tabular} 

% }
\end{table}
%---------------------------------------------------------------------------

Unfortunately, how code coverage is used in comparisons is not the only problem.
Table~\ref{tab:pytest} shows the status results collected by executing the tests with PyTest (using the option \texttt{--junitxml} to collect this data).
Considering an average of $151.7$ tests generated per API, and \nsutsOld APIs with 10 repetitions per experiment, a total of
54 600 test cases were generated.
In these tests, none were skipped (there would be no much point for a fuzzer to generate tests and then mark them as disabled), and very few crashed due to runtime errors.
However, 20\% of these tests failed due to failing assertions.
In the generated tests, fuzzers such as \evo add assertions on the responses received from each HTTP call, to capture the current behavior of the API.
These include assertions on the returned HTTP status codes, and assertions on the content of body payloads of the responses.
This is done to enable the generated tests to be used for regression testing.

It is not in the goals of this paper to provide an in-depth analysis of why, for each of the more than 10 000 tests that failed, they failed.
However, a first look at some of those failures point to a few different reasons.
The easiest to spot are assertions on time-related properties (e.g., time-stamps).
Those are flaky.
However, few test cases fail due to receiving a different status code.
This is not unexpected.
A test case that   returns a successful 204 status code on a \texttt{DELETE} operation, for a resource that was part of the initializing data for the database, might return a different 404, not found status code, when executed after another test case that deletes that same resource first.
Such tests that depend on test execution order will fail.
In this particular case, to be usable for practitioners in industry, tests that delete resources should make sure such resources are created in the same test (e.g., with a previous \texttt{POST} operation).
This kind of issue will not be visible if the generated tests are ignored in the experiments.

After the realization of this major issue, we made a follow up study on flakiness in REST API fuzzing.%, for both black-box and white-box testing.
We generated 10 test suites for black-box testing and 10 for white-box testing for 36 REST APIs from WFD, and executed each suite 100 times in two different environments to examine stable and unstable failing tests as well as the impact of execution environments.
To further investigate the sources of flakiness, we conducted a manual analysis of nearly 3,000 tests and the execution reports produced when each test was run 100 times across two environments.
%However, due to the scope of such study and the large quantity of data collected and analyzed, such work is discussed separately in a different article~\cite{zhang2026flakinessrest}.
However, given the large quantity of data collected and analyzed, this investigation does not fit within a section of this study.
We therefore discussed it separately in another article~\cite{zhang2026flakinessrest}.
Based on our analysis in~\cite{zhang2026flakinessrest}, a taxonomy of nine distinct sources of flakiness was defined, and remedial techniques that can be applied for both black-box and white-box fuzzing were implemented and evaluated in \evo.
Results showed that our technique can reduce the number of generated flaky tests, however,
to complete avoid the flakiness, further techniques that access the source code or internal states of APIs are needed.

An any rate,
an option to avoid having failing assertions would be to not generate any assertion in the first place.
But assertions are needed for practitioners.
In experiments, and in tool comparisons, there is a need to evaluate the presence and quality of the assertions in the generated tests.
In the literature of software testing research, there is already a solution for this problem: \emph{mutation testing}~\cite{mutation2011}.
However, to the best of our knowledge, no empirical study in the literature of fuzzing REST APIs has carried out any kind of mutation testing analysis.
This is a problem that needs to be addressed in the literature.
However, mutation testing tools aimed at unit testing (e.g., like PIT~\cite{pit2016}) might not be ideal for system testing, especially for REST APIs.
There might be avenues for novel research in mutation testing for this specific domain.

Table~\ref{tab:tests} shows the coverage results of the tests generated by \evo.
Regarding the tests generated by the other tools in this study, there are none.
Some tools might generate some report with some information and logs of the calls, but no executable, automated test case with assertions on the returned status codes and payloads.
In the case of \schemathesis, it  reports \texttt{curl} commands in its console logs to be able to replicate its requests.
Note however that, besides \evo, other fuzzers in the literature are able to generate executable test cases,
like for example RestTestGen~\cite{viglianisi2020resttestgen}.
{
Among the fuzzers used in the second set of experiments (Section~\ref{sub:experiments-second}), CATS does not generate test cases in a programming language either.
However, it stores each executed test case (i.e., request, response and oracle verdict) in a JSON file, which can be re-executed (\texttt{cats replay}) or exported as a \texttt{cURL} command.
To avoid hardcoding short-lived credentials, authentication values in these files are replaced with references to environment variables.
}

%%%%%%%%%%%%%%%%%%%%%%%%%%%%%%%%%%%%%%%%%%%%%%%%%%%%%%%%%%%%%%%%%%%%%%%%%%%%
\subsection{{Second Set of Experiments}}
\label{sub:experiments-second}

%---------------------------------------------------------
\subsubsection{{Second Experiment Setup}}
\label{sec:second-setup}

{
The second set of experiments was carried out in late 2026, using four fuzzers:
\cats (version $15.6.0$),
\evo (version $6.2.0$),
\schemathesis (version $4.33.4$),
and
\wuppie (version $1.8.1$).
These are the current fuzzers that have full support for WFC.
For these experiments, we used WFE version $0.1.0$ and all the \nsuts APIs in WFD version $4.4.0$, using WFC authentication, where needed.

Due to unavailability of the server used in previous experiments, these experiments were run on a
MacBook Pro M4 with 128GB of RAM (the same one used to previously run \emrest).
Each experiment was run for 10 minutes, and repeated 10 times, for a total of
$4 \times 56 \times 10 = 2240$ experiments, taking a total duration of more than 15 days, if run in sequence.
Experiments were run in parallel, 10 at a time.

Fuzzers were compared on the same metrics used in the first set of experiments, i.e.,
2xx endpoint coverage, number of endpoints returning 500 at least once, and achieved code coverage.
In this second set of experiments, we also introduced a 4th metric: \emph{best on}.
Considering \nsuts APIs and three metrics, there are $56 \times 3 = 168$ distinct cases in which a fuzzer could provide results better than any other.
This new metric counts how often this happens.
The rational behind this metric is as follows:
fuzzing REST APIs is a complex problem, where many challenges need to be handled;
a  fuzzer might not be best on average, but based on some of its novel techniques it might provide the best results on some classes of APIs with some specific properties.
Being able to highlight when a fuzzer provides the best results on some APIs can help in identifying important techniques and ideas that could be reused and adapted.
Even if a fuzzer is worse on average, as long as it is best on some cases, it could be worth to be included in future studies.
On the other hand, a no longer maintained fuzzer that is fully subsumed by all others could be excluded in new comparisons.

As these four fuzzers generate as outputs reports in WFC format, we also provide comparisons on \emph{all} the faults they find, besides 500 responses.
However, as the presence of these faults is self-reported, and might contain false positives, these comparisons are analyzed separately.

As several new APIs were introduced in WFD, to avoid misleading results due to crashes and faults in the fuzzers, several dry-runs of the experiments (e.g., only 3 repetitions with a 2-minute budget) were carried beforehand.
All fuzzer authors had access to the results of these preliminary experiments, and fixed any misconfiguration and clear fault encountered, before the final experiments were finalized.
For example, \schemathesis had a limitation that its \texttt{--url} option to specify where the API is located requires as well the base path of the API's endpoints.
 This would make it not working in our experiment infrastructure (recall the discussion on base paths in Section~\ref{sub:first-analysis}), which can also partially explain the results of \schemathesis in the first set of experiments.
 Providing to the fuzzers as input a base path for each API, besides their OpenAPI schemas,
  would unnecessarily complicate our  experiment scaffolding, considering that information is already present in the OpenAPI schemas.
 Therefore, \schemathesis was enhanced with a new option \texttt{--origin}, which works as \texttt{--url}, but it does then automatically infer the base path from the input OpenAPI schema.

 \cats had a similar issue with the handling of base paths, which was then fixed after those dry-run experiments.
 For a user in industry testing a single API, explicitly providing the base path as input is not particularly problematic.
 As such, neither \cats nor \schemathesis could handle it automatically before doing these experiments.
 However, as it is still beneficial for the final users (i.e., one parameter less to manually configure), the decision of adding such feature was straightforward, especially as it makes empirical comparisons easier.
}

%------------------------------------------------------------------------------------
\subsubsection{{Result Analysis}}

\begin{table}[!t]
\centering
\caption{ \label{tab:second-overall}
{
Overall results of the four compared fuzzers based on the four different metrics.
Results are the average of 10 runs of the average of the \nsuts.
Values in parentheses are the average ranks, where lower values are better (i.e., 1 is best, wheres 4 is the worst).
The final overall rank is the average of the four ranks on the different four metrics.
}
}
%\vspace{-1.5\baselineskip}
% \resizebox{0.98\textwidth}{!}{
\begin{tabular}{ l rrrrr   }\\ 
\toprule 
Tools & \%2xx & \#500 & \%Lines & Best-On & Overall \\ 
\midrule 
\emph{cats} & \cellcolor{gray!30} 72.4 (1.9) & \cellcolor{gray!30} 34.3 (1.7) & \cellcolor{gray!30} 50.0 (1.8) & \cellcolor{gray!30} 59 (1) & \cellcolor{gray!30} 1.6 \\ 
\emph{evomaster} & 58.5 (2.6) & 14.0 (3.1) & 43.6 (2.7) & 13 (3) & 2.9 \\ 
\emph{schemathesis} & 69.5 (2.0) & 28.2 (1.7) & 49.4 (1.9) & 38 (2) & 1.9 \\ 
\emph{wuppiefuzz} & 31.0 (3.5) & 4.3 (3.5) & 30.0 (3.7) & 2 (4) & 3.7 \\ 
\bottomrule 
\end{tabular} 

\end{table}

{

Table~\ref{tab:second-overall} provides a summary of the comparisons between the four fuzzers.
Details on three metrics in isolation, per API, are reported in appendix,
in Table~\ref{tab:second-2xx},
Table~\ref{tab:second-500}
and Table~\ref{tab:second-coverage}.

In these experiments,
\cats
provides the best results,
followed by
\schemathesis,
\evo
and then
\wuppie.
}

\begin{table}[!t]
\centering
\caption{ \label{tab:second-faults}
{
Detected faults in the experiments, based on the generated WFC reports.
Each entry is the average (out of 10 runs) on each API, summed together (i.e., the sum of \nsuts averages).
Values in parentheses `()' are the number of distinct APIs in which at least one fault of that type was found.
In the total over the \noracles fault categories (which is the sum of those summed averages), values in
brackets `[]' are the number of distinct fault categories for which at least one fault was found.
}
}
%\vspace{-1.5\baselineskip}
 \resizebox{0.98\textwidth}{!}{
\begin{tabular}{ l r r r r r  }\\ 
\toprule 
Category & Code  &  \cats &  \evomaster &  \schemathesis &  \wuppiefuzz \\ 
\midrule 
\emph{HTTP\_STATUS\_500} & 100 & 5999.7 (50) & 537.3 (39) & 1849.6 (53) & 199.7 (27) \\ 
\rowcolor{gray!30} \emph{HTTP\_STATUS\_NO\_NON\_STANDARD\_CODES} & 101 & 0.0 (0) & 0.0 (0) & 0.0 (0) & 0.0 (0) \\ 
\emph{HTTP\_STATUS\_NO\_201\_IF\_DELETE} & 102 & 0.0 (0) & 0.0 (0) & 0.0 (0) & 0.0 (0) \\ 
\rowcolor{gray!30} \emph{HTTP\_STATUS\_NO\_201\_IF\_GET} & 103 & 1.0 (1) & 1.0 (1) & 0.0 (0) & 0.0 (0) \\ 
\emph{HTTP\_STATUS\_NO\_201\_IF\_PATCH} & 104 & 0.0 (0) & 0.0 (0) & 0.0 (0) & 0.0 (0) \\ 
\rowcolor{gray!30} \emph{HTTP\_STATUS\_NO\_204\_IF\_CONTENT} & 105 & 0.0 (0) & 0.0 (0) & 0.0 (0) & 0.0 (0) \\ 
\emph{HTTP\_STATUS\_NO\_413\_IF\_NO\_PAYLOAD} & 106 & 0.0 (0) & 0.0 (0) & 0.0 (0) & 0.0 (0) \\ 
\rowcolor{gray!30} \emph{HTTP\_STATUS\_NO\_415\_IF\_NO\_PAYLOAD} & 107 & 77.5 (12) & 7.0 (2) & 0.0 (0) & 0.0 (0) \\ 
\emph{HTTP\_STATUS\_NO\_304\_IF\_NO\_GET\_OR\_HEAD} & 108 & 0.0 (0) & 0.0 (0) & 0.0 (0) & 0.0 (0) \\ 
\rowcolor{gray!30} \emph{HTTP\_STATUS\_NO\_401\_IF\_NO\_WWW\_AUTHENTICATE} & 109 & 828.2 (15) & 511.1 (15) & 0.0 (0) & 0.0 (0) \\ 
\emph{HTTP\_STATUS\_NO\_405\_IF\_NO\_ALLOW} & 110 & 2820.7 (9) & 0.0 (0) & 280.0 (11) & 0.0 (0) \\ 
\rowcolor{gray!30} \emph{HTTP\_STATUS\_NO\_205\_IF\_CONTENT} & 111 & 0.0 (0) & 0.0 (0) & 0.0 (0) & 0.0 (0) \\ 
\emph{HTTP\_STATUS\_NO\_426\_IF\_NO\_UPGRADE} & 112 & 0.0 (0) & 0.0 (0) & 0.0 (0) & 0.0 (0) \\ 
\rowcolor{gray!30} \emph{HTTP\_NONWORKING\_DELETE} & 113 & 50.0 (22) & 20.9 (9) & 2.5 (3) & 0.0 (0) \\ 
\emph{HTTP\_SIDE\_EFFECTS\_FAILED\_MODIFICATION} & 114 & 0.0 (0) & 0.0 (0) & 0.0 (0) & 0.0 (0) \\ 
\rowcolor{gray!30} \emph{HTTP\_REPEATED\_CREATE\_PUT} & 115 & 0.0 (0) & 0.5 (1) & 0.0 (0) & 0.0 (0) \\ 
\emph{HTTP\_MISLEADING\_CREATE\_PUT} & 116 & 0.3 (1) & 0.5 (1) & 0.0 (0) & 0.0 (0) \\ 
\rowcolor{gray!30} \emph{HTTP\_PARTIAL\_UPDATE\_PUT} & 117 & 0.3 (1) & 21.0 (11) & 0.0 (0) & 0.0 (0) \\ 
\emph{HTTP\_NON\_IDEMPOTENT\_PUT} & 118 & 3.2 (2) & 6.7 (1) & 0.0 (0) & 0.0 (0) \\ 
\rowcolor{gray!30} \emph{HTTP\_INVALID\_MERGE\_PATCH} & 119 & 0.0 (0) & 0.4 (1) & 0.0 (0) & 0.0 (0) \\ 
\emph{HTTP\_INVALID\_LOCATION} & 120 & 0.0 (0) & 4.9 (2) & 0.0 (0) & 0.0 (0) \\ 
\rowcolor{gray!30} \emph{HTTP\_OTHER\_STATUS\_5XX} & 121 & 854.3 (49) & 0.0 (0) & 25.5 (5) & 2.3 (4) \\ 
\emph{HTTP\_CREATED\_RESOURCE\_NOT\_FOUND} & 122 & 1.0 (1) & 0.0 (0) & 1.8 (4) & 0.0 (0) \\ 
\rowcolor{gray!30} \emph{HTTP\_BROKEN\_READ} & 123 & 1.6 (1) & 0.0 (0) & 0.0 (0) & 0.0 (0) \\ 
\emph{HTTP\_UPDATE\_READ\_INCONSISTENCY} & 124 & 0.3 (1) & 0.0 (0) & 0.0 (0) & 0.0 (0) \\ 
\rowcolor{gray!30} \emph{HTTP\_STALE\_ETAG\_ACCEPTED} & 125 & 0.0 (0) & 0.0 (0) & 0.0 (0) & 0.0 (0) \\ 
\emph{SCHEMA\_INVALID\_RESPONSE} & 200 & 1537.9 (53) & 1009.5 (31) & 2792.1 (55) & 1942.7 (53) \\ 
\rowcolor{gray!30} \emph{SCHEMA\_INVALID\_ALLOW} & 201 & 36.8 (2) & 177.0 (20) & 56.0 (8) & 0.0 (0) \\ 
\emph{SCHEMA\_STATUS\_NO\_401\_IF\_NO\_AUTH} & 202 & 0.0 (0) & 239.3 (10) & 0.0 (0) & 0.0 (0) \\ 
\rowcolor{gray!30} \emph{SCHEMA\_STATUS\_NO\_403\_IF\_NO\_401} & 203 & 0.0 (0) & 86.4 (6) & 0.0 (0) & 0.0 (0) \\ 
\emph{SCHEMA\_STATUS\_HAS\_406\_IF\_ACCEPT} & 204 & 45.0 (7) & 14.0 (3) & 8.5 (3) & 0.0 (0) \\ 
\rowcolor{gray!30} \emph{SCHEMA\_STATUS\_NO\_501\_IF\_IMPLEMENTED} & 205 & 2.0 (1) & 1.0 (1) & 1.0 (1) & 0.0 (0) \\ 
\emph{SCHEMA\_VALIDATION\_BYPASS} & 206 & 998.0 (52) & 0.0 (0) & 291.8 (43) & 0.0 (0) \\ 
\rowcolor{gray!30} \emph{SCHEMA\_VALID\_INPUT\_REJECTED} & 207 & 681.3 (39) & 0.0 (0) & 1084.4 (44) & 0.0 (0) \\ 
\emph{SECURITY\_SQL\_INJECTION} & 300 & 0.0 (0) & 0.2 (2) & 0.0 (0) & 0.0 (0) \\ 
\rowcolor{gray!30} \emph{SECURITY\_XSS\_INJECTION} & 301 & 101.4 (26) & 68.4 (18) & 0.0 (0) & 4.1 (5) \\ 
\emph{SECURITY\_SSRF} & 302 & 6.1 (4) & 0.0 (0) & 0.0 (0) & 0.0 (0) \\ 
\rowcolor{gray!30} \emph{SECURITY\_MASS\_ASSIGNMENT} & 303 & 0.2 (1) & 0.0 (0) & 0.0 (0) & 0.0 (0) \\ 
\emph{SECURITY\_EXISTENCE\_LEAKAGE} & 304 & 0.0 (0) & 5.0 (3) & 0.0 (0) & 0.0 (0) \\ 
\rowcolor{gray!30} \emph{SECURITY\_NOT\_RECOGNIZED\_AUTHENTICATED} & 305 & 0.0 (0) & 5.1 (2) & 0.0 (0) & 0.0 (0) \\ 
\emph{SECURITY\_INCONSISTENT\_WRITE\_AUTHORIZATION} & 306 & 40.5 (12) & 2.4 (1) & 0.0 (0) & 0.0 (0) \\ 
\rowcolor{gray!30} \emph{SECURITY\_IGNORE\_ANONYMOUS} & 307 & 0.0 (0) & 4.0 (2) & 0.0 (0) & 0.0 (0) \\ 
\emph{SECURITY\_ANONYMOUS\_MODIFICATIONS} & 308 & 33.9 (4) & 44.0 (14) & 0.0 (0) & 0.0 (0) \\ 
\rowcolor{gray!30} \emph{SECURITY\_LEAKED\_STACK\_TRACES} & 309 & 2762.8 (50) & 55.7 (5) & 0.0 (0) & 0.0 (0) \\ 
\emph{SECURITY\_HIDDEN\_ACCESSIBLE\_ENDPOINT} & 310 & 274.3 (11) & 119.7 (17) & 170.0 (5) & 0.0 (0) \\ 
\rowcolor{gray!30} \emph{SECURITY\_DECLARED\_AUTH\_NOT\_ENFORCED} & 311 & 481.3 (14) & 0.0 (0) & 114.9 (7) & 0.0 (0) \\ 
\emph{SECURITY\_CALL\_TIMEOUT} & 312 & 0.0 (0) & 0.0 (0) & 0.0 (0) & 0.0 (0) \\ 
\rowcolor{gray!30} \emph{SECURITY\_COMMAND\_INJECTION} & 313 & 0.0 (0) & 0.0 (0) & 0.0 (0) & 2.6 (3) \\ 
\emph{SECURITY\_NOSQL\_INJECTION} & 314 & 37.2 (9) & 0.0 (0) & 0.0 (0) & 0.0 (0) \\ 
\rowcolor{gray!30} \emph{SECURITY\_EXCESSIVE\_DATA\_EXPOSURE} & 315 & 0.0 (0) & 0.0 (0) & 0.0 (0) & 0.0 (0) \\ 
\emph{SECURITY\_MISSING\_HEADERS} & 316 & 679.0 (42) & 0.0 (0) & 0.0 (0) & 0.0 (0) \\ 
\rowcolor{gray!30} \emph{SECURITY\_LDAP\_INJECTION} & 317 & 0.0 (0) & 0.0 (0) & 0.0 (0) & 4.1 (5) \\ 
\midrule 
Total: 52 &  & 18355.8 [29] & 2943.0 [26] & 6678.1 [13] & 2155.5 [6] \\ 
\bottomrule 
\end{tabular} 

}
\end{table}

{
When it comes to detected faults, Table~\ref{tab:second-faults}
shows a summary of what found.
As there are \noracles distinct fault categories and \nsuts APIs, results are aggregated per fault category.
In these self-reported faults,
\cats seems to provide the best results.
Up to 29 different classes of faults are automatically detected on WFD, in the order of thousands.

Not all fault types have the same severity, and some of those detected faults might turn into false positive when manually checked.
Still, this data can provide valuable insight on what the current state-of-the-art fuzzers claim to be able to automatically detect.
It is a valuable starting point for possible follow up manual analyses, especially when analyzing specific fault categories in depth.
By using a common format like WFC, enhanced with visual, HTML-based web reports, it will make easier to identify key differences and enable more efficient manual inspections in fuzzer comparisons.
}

%%%%%%%%%%%%%%%%%%%%%%%%%%%%%%%%%%%%%%%%%%%%%%%%%%%%%%%%%%%%%%%%%%%%%%%%%%%%
\section{Discussion}
\label{sec:discussion}

In this paper, we have discussed three major issues related to the use and comparisons of fuzzers for REST APIs: \emph{authentication},  \emph{fault categorization} and \emph{case study selection}.
We have provided working solutions in the form of WFC, using general schema definitions and open-source library support.
These techniques are implemented as part of
{
 the most popular state-of-the-art tools, i.e.,  \cats, \evo, \schemathesis and \wuppie.
However, these technique could be applied to any other REST API fuzzer.
}

To enable sound empirical comparisons, we presented WFD, a collection of \nsuts REST APIs.
All the scaffolding needed to run experiments using WFC {is stored in a separated repository called WFE}.
This corpus of APIs and scaffolding support build on top of the existing EMB corpus.

To show the usefulness of WFC, {WFD and WFE}, we carried out
{two sets of empirical studies with a total of 8 existing fuzzers}, representing the current state-of-the-art in fuzzing REST APIs.
There have been potential issues on how comparisons have been carried out in the literature.
We have discussed these issues and provided working solutions to address them.
In this section, to improve the quality and rigour of future studies in software testing research, we summarize these discussions.

%\begin{description}
%
%\item[API Selection.]
{\bf API Selection.}
When comparing with an existing tool $X$, ideally should use a superset of the APIs used in the original study of $X$. Excluding APIs  due to potential selection bias might rise concerns about simply masking worse performance of a new proposed technique.
To address bias concerns regarding API selection, new APIs should be added in each new study from different research teams.
WFD has been created to address this issue, by collecting all Java open-source APIs used in previous studies, plus several new ones.
%

% \item[Tool Usability.]
{\bf Tool Usability.}
Comparing a novel technique with existing tools is important for scientific research.
However, to be used for comparisons, a tool should have some basic usability features, as we list in Figure~\ref{fig:usability}.
If not, it should be fair to exclude a tool from empirical comparisons.
Developers of new fuzzers should aim to support these basic features, as several existing fuzzers already do.
%

%\item[Authentication.]
{\bf Authentication.}
Real-world applications use authentication.
Not supporting authentication in a fuzzer would make it perform poorly on any realistic case study in which real-world APIs are used.
Authentication is a critical aspect for applications in industry.
However, there is no common mechanism or standard to share authentication information among fuzzers.
Each tool has its own custom mechanism, if any, which makes setting up experiments much more complex than it could be.
To address this issue, the WFC authentication format has been presented in this paper.
We invite the authors of other fuzzers to support WFC as well, to enable future comparisons on WFD.
{
As this is a community driven effort, currently handled by the authors of the four most popular REST API fuzzers in industry, feedback and involvement from other fuzzer authors is welcome.
}

{\bf Fault Comparisons.}
Comparing fuzzers only on faults of type HTTP status 500 would ignore all other kinds of faults that these fuzzers can find, especially critical ones related to security failures.
The presented WFC provides a candidate proposed standardized way to report these faults.
However, as these values are ``self-reported'', care needs to be taken when using such data to compare fuzzers.
Still, they  can provide useful extra information.

%\item[Test Generation.]
{\bf Test Generation.}
Test engineers in industry would use the tests generated by the fuzzers.
It does not matter what a fuzzer can find during its fuzzing session, if then no generated test given to the engineers can reproduce it.
As such, comparisons among tools must be carried out based on the generated tests.
None of the existing studies in the literature has done this, which led to keep hidden some important research challenges (e.g., how to best select the final test suites, how to deal with flaky assertions,  how to deal with mutable state and test dependencies).
A tool could provide tests in multiple different formats and programming languages, based on the preferences of users.
To simplify comparisons, it would be best if there was a common, widely accepted format for the generated test suites, which each fuzzer in the literature would support (regardless of the programming language the fuzzer is implemented with).
In this paper, we argue for Python for the test suite outputs for black-box testing (of course, a fuzzer could support other formats as well).
%\end{description}

%%%%%%%%%%%%%%%%%%%%%%%%%%%%%%%%%%%%%%%%%%%%%%%%%%%%%%%%%%%%%%%%%%%%%%%%%%%%
\section{Threats To Validity}
\label{sec:threats}

When comparing tools, there is always a chance of misunderstanding their documentation (if any is provided), or misusing them by mistake.
This could impact the final results, and the conclusions drawn from them.
As such, it is important to make available all used scripts in the experiments, as we do in this study.

To take into account the randomness of the employed fuzzers, each experiment was repeated 10 times.
Results were analyzed using  statistical methods, following common guidelines in the research literature~\cite{Hitchhiker14}.
In particular, we used the Friedman Test, the $\hat{A}_{12}$ Vargha-Delaney effect size, and the Mann-Whitney-Wilcoxon U-test.

As far as we know, with WFD this is the largest study of fuzzing open-source REST APIs in the literature to date~\cite{golmohammadi2023testing}.
Still, results on \nsuts open-source APIs might not generalize to other APIs, especially the ones developed in industry.
Selecting and setting-up APIs for experimentation takes a considerable amount of time.
One of the arguments we made in this paper is that, ideally, to reduce the impact of these issues such process should be community-driven.

Our analyses and discussions have been focused on testing REST APIs.
However, several of these highlights would likely apply to other research domains, e.g., when discussing how to select artifacts for experimentation.

%%%%%%%%%%%%%%%%%%%%%%%%%%%%%%%%%%%%%%%%%%%%%%%%%%%%%%%%%%%%%%%%%%%%%%%%%%%%
\section{Conclusions}
\label{sec:conclusions}

In this paper, we have presented Web Fuzzing Commons (WFC),  Web Fuzzing Dataset (WFD) { and Web Fuzzing Evaluations (WFE)}.
WFC provides the necessary schemas and tool support to enable specifying authentication information and report detected faults in a reliable manner.
WFD is a selection of \nsuts REST APIs, including authentication information in WFC format.
{All the scaffolding needed to run and analyze experiments on WFD is stored in WFE}.

To show the use of WFC, {WFD and WFE},
we have provided the largest study in the literature in terms of employed open-source APIs.
We compared {eight} state-of-the-art fuzzers:
\aratrl,
\cats,
\emrest,
\evo,
\llamaresttest,
\restler,
\schemathesis
and
\wuppie.

Our goal in this study was not to identify which tool or technique performs best, but rather to point out some concerns regarding the carrying out of empirical studies done in the literature.
To address these concerns, we have provided a list of actionable guidelines.
With this work, we aim at improving future studies on this topic.

In future work, we will extend the classification of faults in WFC each time new oracles are presented in the literature.
When new empirical studies will use open-source Java/Kotlin APIs not present in WFD, we will see to add them to WFD.

Technical improvements in our experiment scaffolding {in WFE} will aim at supporting better analyses via mutation testing.
{
Specific guidelines for fair comparisons will need to be established when fuzzers can use different computational resources, like available GPUs besides the CPUs (e.g., to run LLMs locally).
}

Our
WFC,\footnote{\url{https://github.com/WebFuzzing/Commons}}
WFD\footnote{\url{https://github.com/WebFuzzing/Dataset}}
{
and WFE\footnote{\url{https://github.com/WebFuzzing/Evaluations}}
}
are open-source on GitHub, with each release automatically uploaded to Zenodo (e.g., \cite{wfc010,zenodo400wfd}).
All employed fuzzers in our empirical study are open-source.

{
Due to the importance of the addressed problem,
this a community driven process, currently stewarded by the authors of the four most popular REST API fuzzers in industry.
}
In a research field prone to fragmentation, we have established a democratic process among fuzzer developers to steer future changes.

%%%%%%%%%%%%%%%%%%%%%%%%%%%%%%%%%%%%%%%%%%%%%%%%%%%%%%%%%%%%%%%%%%%%%%%%%%%%

\appendix

%-----------------
% \subsection{Criticism}
\section{Tool Exclusion Example}
\label{sec:resttestgen}

Our tool selection criteria are formally defined in Section~\ref{sec:selection-criteria}.
As tool selection is a critical component in any empirical study, let us clarify possible misunderstanding in our guidelines.
We are aware of no study in the literature in which all the existing REST API fuzzers have been compared.
To the best of our knowledge, the largest number of compared tools appeared in the work of Kim~\etal~\cite{Kim2022Rest}, where 10 REST API fuzzers have been compared.
Asking researchers to compare with 40 or more fuzzers each time they present a novel technique is not viable.
Not only it would take a considerable amount of time, but, also, showing results with descriptive figures/tables would be challenging, especially considering the page restrictions of conference publications.
Therefore, in the literature, it is a common practice to compare with around 4-5 existing tools.
The research question is how to select a reasonable, representative subset, especially if there are 40 to select from.

However, in academic peer-reviewing processes, it is not uncommon that peers criticize not using a specific tool in the comparisons.
For example, a previous version of this article was rejected because not including RestTestGen.
It is not in our intention to single out a tool like RestTestGen,
but, let us consider it as an example of why it would not be selected when following the criteria expressed in Section~\ref{sec:usability}.
These discussions can be useful to the research community when dealing with similar situations.

Considering criterion (1) about results in independent studies, for example according to the results in Table 3 in~\cite{Kim2022Rest}, RestTestGen achieved the 3rd line coverage (behind \evo and \schemathesis) coverage, the 4th branch coverage (behind \evo, \schemathesis and bBOXRT), and 6th unique error discovery (behind Tcases, \evo, \restler, \schemathesis and bBOXRT).
It was not among the top performing fuzzers in that independent comparisons.
Tools evolve and get updates, so following studies might show different results when used in new research studies.
In the independent study of Sartaj~\etal~\cite{sartaj2025rest}, several metrics and usage context were considered.
No fuzzer was best in all metrics.
However, for most of the scenarios \evo gave the best results, where in some scenarios it was RESTest that gave the best results.
Considering this selection criterion, there would be at least another four fuzzers to consider before selecting RestTestGen.

Considering criterion (2) about usage in industry, RestTestGen seems not having any official release (e.g., built jar file)
on GitHub.\footnote{\url{https://github.com/SeUniVr/RestTestGen/releases}}
It seems that users would need to clone its Git repository, and build it locally.
Without download statistics provide by GitHub releases, this makes bit harder to quantify how used it is in practice.
When looking at number of GitHub ``stars'' as indirect proxy for popularity and usage, with 62 stars at the time of writing, there would be at least 9 other fuzzers that we are aware of that are more popular than RestTestGen (
Dredd,
\schemathesis,
\restler,
CATS,
\evo,
OpenAPI-Fuzzer,
fuzz-lightyear,
RESTest
and
WuppieFuzz
).
Based on criterion (2) for fuzzer selection, there would be at least 9 other fuzzers to consider before looking at RestTestGen.
However, four of those seem no longer maintained.

Considering selection criterion (3), it does not seem there has been any proposed improvement of RestTestGen within the last year (i.e., 2025) in which a new version was studied that showed better results than the
state-of-the-art.\footnote{\url{https://github.com/SeUniVr/RestTestGen/blob/main/CITE.md}}
During 2025, seven new fuzzers have been proposed in the literature.
These are shown in Table~\ref{tab:2025}.
Only in one of them RestTestGen was used in the empirical comparisons (i.e., in the study for \apirl~\cite{foley2025apirl}).

Considering selection criterion (4), this is indeed subjective.
Once enough fuzzers are selected based on the previous three formal criteria (e.g., at least one per criterion, avoiding duplicates), there is no harm in adding more fuzzers for comparisons, if time and space constraint allow.

For example, it seems like RestTestGen is one of the very few fuzzers that can generate executable test cases (e.g., in JUnit format).
If a study involves using and comparing the generated tests, RestTestGen could be a good option to use.
But this depends on the type of study.

However,
considering the usability requirements in Figure~\ref{fig:usability}, technically, RestTestGen would not satisfy Requirement~3, regarding command-line input parameters.
Based on its documentation and example setups, RestTestGen has no support for command-line options.
Options (like where the OpenAPI schema is located) need to be written in configuration files, put on specific locations with specific names.
For comparison experiments, this is not a major limitation.
An experiment setup using for example Bash scripts could simply use commands such as \texttt{touch} and redirected \texttt{>>} to file \texttt{cat} commands to setup those files, in different folders (\texttt{mkdir}) to avoid clashes when running experiments in parallel.
Still, also for practitioners in industry, to avoid all these custom scaffolding made only for RestTestGen, it is our opinion it would make things easier for the community to support as well the option of specifying these parameters on the command-line (as most fuzzers already do).

%%%%%%%%%%%%%%%%%%%%%%%%%%%%%%%%%%%%%%%%%%%%%%%%%%%%%%%%%%%%%%%%%%%%%%%%%%%%%%%%%%
\section{Additional Data}

{
Table~\ref{tab:second-2xx},
Table~\ref{tab:second-500},
and Table~\ref{tab:second-coverage}
provide extra data for the second set of experiments carried out in this paper.
}

\begin{table}[!t]
\centering
\caption{ \label{tab:second-2xx}
{
Average results, out of 10 runs, for 2xx endpoint coverage percentage over the 4 compared fuzzers, on all the \nsuts APIs.
On each API, tool are compared by rank, where rank 1 is the best.
Rank values are presented in `()' parentheses after the average values.
In case of ties, ranks are averaged.
The best fuzzers on each API are highlighted in bold.
}
}
\vspace{-1.5\baselineskip}
 \resizebox{0.72\textwidth}{!}{
\begin{tabular}{ l  r r r r    }\\ 
\toprule 
SUT  &  \cats &  \evomaster &  \schemathesis &  \wuppiefuzz \\ 
\midrule 
\emph{adoptme} & {\bf 100.0 (2.0)} & {\bf 100.0 (2.0)} & {\bf 100.0 (2.0)} & 99.3 (4.0) \\ 
 \emph{arimaa} & {\bf 70.3 (1.0)} & 55.0 (3.0) & 66.8 (2.0) & 25.3 (4.0) \\ 
 \emph{bibliothek} & {\bf 12.5 (2.5)} & {\bf 12.5 (2.5)} & {\bf 12.5 (2.5)} & {\bf 12.5 (2.5)} \\ 
 \emph{blogapi} & 18.9 (4.0) & 20.9 (3.0) & {\bf 42.8 (1.0)} & 23.8 (2.0) \\ 
 \emph{catwatch} & {\bf 39.1 (2.0)} & {\bf 39.1 (2.0)} & {\bf 39.1 (2.0)} & 26.1 (4.0) \\ 
 \emph{commafeed} & 50.7 (2.0) & 39.5 (3.0) & {\bf 59.1 (1.0)} & 38.9 (4.0) \\ 
 \emph{cwa-verification} & {\bf 100.0 (1.0)} & 32.0 (3.0) & 60.0 (2.0) & 20.0 (4.0) \\ 
 \emph{digitalbanking} & 86.0 (4.0) & 88.9 (3.0) & 89.5 (2.0) & {\bf 96.0 (1.0)} \\ 
 \emph{erc20-rest-service} & {\bf 7.7 (2.5)} & {\bf 7.7 (2.5)} & {\bf 7.7 (2.5)} & {\bf 7.7 (2.5)} \\ 
 \emph{familie-ba-sak} & {\bf 2.2 (2.5)} & {\bf 2.2 (2.5)} & {\bf 2.2 (2.5)} & {\bf 2.2 (2.5)} \\ 
 \emph{features-service} & {\bf 100.0 (1.0)} & 87.2 (4.0) & 99.4 (2.0) & 96.7 (3.0) \\ 
 \emph{flight-search-api} & {\bf 79.2 (1.0)} & 55.8 (3.0) & 73.0 (2.0) & 47.5 (4.0) \\ 
 \emph{gdpr-kv} & {\bf 85.7 (1.0)} & 71.4 (3.0) & 71.4 (3.0) & 71.4 (3.0) \\ 
 \emph{genome-nexus} & {\bf 100.0 (1.5)} & 73.0 (3.0) & {\bf 100.0 (1.5)} & 30.4 (4.0) \\ 
 \emph{gestaohospital} & {\bf 93.5 (1.0)} & 52.5 (3.0) & 86.0 (2.0) & 26.0 (4.0) \\ 
 \emph{gravitee} & 36.2 (2.0) & 6.0 (3.0) & {\bf 42.5 (1.0)} & 0.0 (4.0) \\ 
 \emph{http-patch-spring} & {\bf 100.0 (2.0)} & {\bf 100.0 (2.0)} & {\bf 100.0 (2.0)} & 48.3 (4.0) \\ 
 \emph{jasper} & 62.6 (2.0) & 46.4 (3.0) & {\bf 70.2 (1.0)} & 0.0 (4.0) \\ 
 \emph{joinus} & {\bf 85.9 (1.0)} & 63.8 (3.0) & 59.5 (4.0) & 66.8 (2.0) \\ 
 \emph{kafka-rest} & {\bf 72.0 (1.0)} & 28.2 (2.0) & 10.0 (3.0) & 0.0 (4.0) \\ 
 \emph{komga} & {\bf 55.0 (1.0)} & 4.0 (4.0) & 46.0 (2.0) & 27.5 (3.0) \\ 
 \emph{languagetool} & {\bf 100.0 (1.5)} & 70.0 (3.0) & {\bf 100.0 (1.5)} & 0.0 (4.0) \\ 
 \emph{lovemining} & {\bf 77.8 (1.0)} & 72.2 (2.0) & 66.7 (3.0) & 22.2 (4.0) \\ 
 \emph{management-cassandra} & 65.2 (3.0) & {\bf 79.7 (1.0)} & 76.7 (2.0) & 19.0 (4.0) \\ 
 \emph{market} & 66.9 (3.0) & 73.8 (2.0) & {\bf 76.9 (1.0)} & 53.8 (4.0) \\ 
 \emph{microcks} & 46.1 (2.0) & {\bf 49.3 (1.0)} & 17.2 (4.0) & 20.3 (3.0) \\ 
 \emph{movies-xml} & 87.1 (2.0) & 42.9 (3.0) & {\bf 100.0 (1.0)} & 0.0 (4.0) \\ 
 \emph{notebook-manager} & {\bf 100.0 (2.0)} & {\bf 100.0 (2.0)} & {\bf 100.0 (2.0)} & 78.0 (4.0) \\ 
 \emph{ocvn} & 84.9 (2.0) & 84.1 (3.0) & {\bf 85.3 (1.0)} & 0.0 (4.0) \\ 
 \emph{ohsome-api} & {\bf 100.0 (1.0)} & 14.6 (3.0) & 99.5 (2.0) & 0.0 (4.0) \\ 
 \emph{pay-publicapi} & {\bf 0.0 (2.5)} & {\bf 0.0 (2.5)} & {\bf 0.0 (2.5)} & {\bf 0.0 (2.5)} \\ 
 \emph{person-controller} & {\bf 58.3 (1.5)} & 50.0 (3.0) & {\bf 58.3 (1.5)} & 25.0 (4.0) \\ 
 \emph{petclinic-rest} & {\bf 86.0 (1.5)} & 72.3 (3.0) & {\bf 86.0 (1.5)} & 0.0 (4.0) \\ 
 \emph{pn-paper-tracker} & 50.0 (3.5) & 62.5 (2.0) & {\bf 75.0 (1.0)} & 50.0 (3.5) \\ 
 \emph{proxyprint} & 57.0 (3.0) & {\bf 74.3 (1.0)} & 69.9 (2.0) & 1.7 (4.0) \\ 
 \emph{quartz-manager} & {\bf 72.7 (1.0)} & 63.6 (2.5) & 63.6 (2.5) & 55.5 (4.0) \\ 
 \emph{redis-sample} & {\bf 100.0 (2.0)} & 90.0 (4.0) & {\bf 100.0 (2.0)} & {\bf 100.0 (2.0)} \\ 
 \emph{reservations-api} & {\bf 82.9 (1.0)} & 57.1 (3.5) & 57.1 (3.5) & 67.1 (2.0) \\ 
 \emph{rest-ncs} & {\bf 100.0 (2.0)} & {\bf 100.0 (2.0)} & {\bf 100.0 (2.0)} & 96.7 (4.0) \\ 
 \emph{rest-news} & {\bf 100.0 (1.0)} & 85.7 (3.0) & 98.6 (2.0) & 32.9 (4.0) \\ 
 \emph{rest-scs} & {\bf 100.0 (2.5)} & {\bf 100.0 (2.5)} & {\bf 100.0 (2.5)} & {\bf 100.0 (2.5)} \\ 
 \emph{restcountries} & {\bf 100.0 (1.0)} & 98.2 (2.0) & 82.3 (3.0) & 0.0 (4.0) \\ 
 \emph{scout-api} & {\bf 90.6 (1.0)} & 86.9 (2.0) & 73.1 (3.0) & 0.0 (4.0) \\ 
 \emph{session-service} & {\bf 100.0 (1.0)} & 50.0 (3.0) & 87.5 (2.0) & 12.5 (4.0) \\ 
 \emph{spring-actuator-demo} & {\bf 100.0 (2.5)} & {\bf 100.0 (2.5)} & {\bf 100.0 (2.5)} & {\bf 100.0 (2.5)} \\ 
 \emph{spring-batch-rest} & {\bf 100.0 (1.5)} & 60.0 (3.0) & {\bf 100.0 (1.5)} & 0.0 (4.0) \\ 
 \emph{spring-ecommerce} & {\bf 44.4 (1.0)} & 40.7 (3.0) & 40.7 (3.0) & 40.7 (3.0) \\ 
 \emph{spring-rest-example} & {\bf 44.4 (2.5)} & {\bf 44.4 (2.5)} & {\bf 44.4 (2.5)} & {\bf 44.4 (2.5)} \\ 
 \emph{swagger-petstore} & 73.7 (3.0) & {\bf 89.5 (1.5)} & {\bf 89.5 (1.5)} & 0.0 (4.0) \\ 
 \emph{testing-system} & {\bf 100.0 (1.5)} & 68.2 (3.0) & {\bf 100.0 (1.5)} & 9.1 (4.0) \\ 
 \emph{tiltaksgjennomforing} & {\bf 8.9 (2.0)} & {\bf 8.9 (2.0)} & {\bf 8.9 (2.0)} & 0.0 (4.0) \\ 
 \emph{traccar} & 67.9 (3.0) & 74.6 (2.0) & {\bf 77.5 (1.0)} & 0.0 (4.0) \\ 
 \emph{tracking-system} & 67.8 (2.0) & {\bf 69.6 (1.0)} & 50.6 (3.0) & 0.0 (4.0) \\ 
 \emph{user-management} & {\bf 84.3 (1.0)} & 73.8 (3.0) & 81.4 (2.0) & 39.0 (4.0) \\ 
 \emph{webgoat} & 79.9 (3.0) & 83.3 (2.0) & {\bf 85.0 (1.0)} & 0.0 (4.0) \\ 
 \emph{youtube-mock} & {\bf 100.0 (1.5)} & 0.0 (3.5) & {\bf 100.0 (1.5)} & 0.0 (3.5) \\ 
 \midrule 
Average  & 72.4 (1.9) & 58.5 (2.6) & 69.5 (2.0) & 31.0 (3.5) \\ 
 Median  & 81.4 (2.0) & 63.7 (3.0) & 75.8 (2.0) & 23.0 (4.0) \\ 
 \midrule 
Friedman Test & \multicolumn{4}{r}{$\chi^2$ = 69.543, $p$-value = $< 0.001$ } \\ 
\bottomrule 
\end{tabular} 

}
\end{table}

\begin{table}[!t]
\centering
\caption{ \label{tab:second-500}
{
Average results, out of 10 runs, for the number of endpoints in which at least on test case leading to an HTTP 500 status was generated,  over the 4 compared fuzzers on all the \nsuts APIs.
On each API, tool are compared by rank, where rank 1 is the best.
Rank values are presented in `()' parentheses after the average values.
In case of ties, ranks are averaged.
The best fuzzers on each API are highlighted in bold.
}
}
\vspace{-1.5\baselineskip}
 \resizebox{0.72\textwidth}{!}{
\begin{tabular}{ l  r r r r    }\\ 
\toprule 
SUT  &  \cats &  \evomaster &  \schemathesis &  \wuppiefuzz \\ 
\midrule 
\emph{adoptme} & {\bf 14.0 (1.5)} & 0.5 (4.0) & {\bf 14.0 (1.5)} & 2.0 (3.0) \\ 
 \emph{arimaa} & {\bf 38.0 (1.0)} & 6.8 (3.0) & 36.6 (2.0) & 5.6 (4.0) \\ 
 \emph{bibliothek} & {\bf 8.0 (1.5)} & 0.0 (3.5) & {\bf 8.0 (1.5)} & 0.0 (3.5) \\ 
 \emph{blogapi} & {\bf 53.0 (1.0)} & 9.3 (4.0) & 52.8 (2.0) & 25.3 (3.0) \\ 
 \emph{catwatch} & {\bf 13.0 (1.5)} & 9.9 (3.0) & {\bf 13.0 (1.5)} & 7.9 (4.0) \\ 
 \emph{commafeed} & 27.4 (2.0) & 16.4 (4.0) & {\bf 29.0 (1.0)} & 26.0 (3.0) \\ 
 \emph{cwa-verification} & {\bf 5.0 (1.5)} & 0.0 (4.0) & {\bf 5.0 (1.5)} & 4.0 (3.0) \\ 
 \emph{digitalbanking} & 49.0 (2.0) & 15.0 (3.0) & {\bf 55.0 (1.0)} & 14.1 (4.0) \\ 
 \emph{erc20-rest-service} & {\bf 13.0 (1.5)} & 12.0 (3.5) & {\bf 13.0 (1.5)} & 12.0 (3.5) \\ 
 \emph{familie-ba-sak} & {\bf 150.0 (1.0)} & 15.0 (4.0) & 149.3 (2.0) & 28.0 (3.0) \\ 
 \emph{features-service} & {\bf 17.0 (1.0)} & 14.0 (3.0) & 14.0 (3.0) & 14.0 (3.0) \\ 
 \emph{flight-search-api} & {\bf 0.0 (2.5)} & {\bf 0.0 (2.5)} & {\bf 0.0 (2.5)} & {\bf 0.0 (2.5)} \\ 
 \emph{gdpr-kv} & {\bf 7.0 (1.5)} & 1.0 (4.0) & {\bf 7.0 (1.5)} & 1.2 (3.0) \\ 
 \emph{genome-nexus} & {\bf 0.0 (2.5)} & {\bf 0.0 (2.5)} & {\bf 0.0 (2.5)} & {\bf 0.0 (2.5)} \\ 
 \emph{gestaohospital} & {\bf 20.0 (1.5)} & 2.0 (3.5) & {\bf 20.0 (1.5)} & 2.0 (3.5) \\ 
 \emph{gravitee} & {\bf 290.2 (1.0)} & 187.9 (2.0) & 153.6 (3.0) & 0.0 (4.0) \\ 
 \emph{http-patch-spring} & {\bf 6.0 (1.5)} & 0.0 (3.5) & {\bf 6.0 (1.5)} & 0.0 (3.5) \\ 
 \emph{jasper} & {\bf 61.0 (1.5)} & 43.9 (3.0) & {\bf 61.0 (1.5)} & 0.0 (4.0) \\ 
 \emph{joinus} & {\bf 29.0 (1.0)} & 7.7 (4.0) & 19.7 (2.0) & 11.4 (3.0) \\ 
 \emph{kafka-rest} & 17.9 (2.0) & 2.4 (3.0) & {\bf 47.0 (1.0)} & 0.0 (4.0) \\ 
 \emph{komga} & {\bf 170.0 (1.0)} & 0.0 (4.0) & 28.0 (2.0) & 0.6 (3.0) \\ 
 \emph{languagetool} & {\bf 1.0 (2.0)} & {\bf 1.0 (2.0)} & {\bf 1.0 (2.0)} & 0.0 (4.0) \\ 
 \emph{lovemining} & {\bf 18.0 (1.0)} & 1.0 (3.0) & 17.7 (2.0) & 0.8 (4.0) \\ 
 \emph{management-cassandra} & 20.1 (3.0) & 25.9 (2.0) & {\bf 30.7 (1.0)} & 2.0 (4.0) \\ 
 \emph{market} & {\bf 13.0 (1.5)} & 5.0 (3.5) & {\bf 13.0 (1.5)} & 5.0 (3.5) \\ 
 \emph{microcks} & {\bf 53.2 (1.0)} & 15.1 (3.0) & 25.3 (2.0) & 6.4 (4.0) \\ 
 \emph{movies-xml} & {\bf 7.0 (1.5)} & 0.0 (3.5) & {\bf 7.0 (1.5)} & 0.0 (3.5) \\ 
 \emph{notebook-manager} & {\bf 5.0 (1.5)} & 3.0 (3.5) & {\bf 5.0 (1.5)} & 3.0 (3.5) \\ 
 \emph{ocvn} & 127.0 (2.0) & {\bf 197.3 (1.0)} & 65.3 (3.0) & 0.0 (4.0) \\ 
 \emph{ohsome-api} & {\bf 133.8 (1.0)} & 0.0 (3.5) & 120.2 (2.0) & 0.0 (3.5) \\ 
 \emph{pay-publicapi} & {\bf 16.0 (2.5)} & {\bf 16.0 (2.5)} & {\bf 16.0 (2.5)} & {\bf 16.0 (2.5)} \\ 
 \emph{person-controller} & {\bf 12.0 (1.5)} & 10.0 (3.5) & {\bf 12.0 (1.5)} & 10.0 (3.5) \\ 
 \emph{petclinic-rest} & 41.0 (2.0) & 11.9 (3.0) & {\bf 43.0 (1.0)} & 0.0 (4.0) \\ 
 \emph{pn-paper-tracker} & 3.0 (3.0) & {\bf 4.0 (1.5)} & {\bf 4.0 (1.5)} & 1.0 (4.0) \\ 
 \emph{proxyprint} & {\bf 50.0 (1.0)} & 39.2 (3.0) & 46.8 (2.0) & 6.1 (4.0) \\ 
 \emph{quartz-manager} & {\bf 10.0 (1.0)} & 4.0 (4.0) & 8.0 (2.0) & 6.0 (3.0) \\ 
 \emph{redis-sample} & {\bf 4.0 (1.5)} & 0.9 (4.0) & {\bf 4.0 (1.5)} & 1.0 (3.0) \\ 
 \emph{reservations-api} & {\bf 7.0 (1.5)} & 1.0 (4.0) & {\bf 7.0 (1.5)} & 3.9 (3.0) \\ 
 \emph{rest-ncs} & 1.5 (2.0) & 0.0 (3.5) & {\bf 4.0 (1.0)} & 0.0 (3.5) \\ 
 \emph{rest-news} & {\bf 3.0 (1.0)} & 1.0 (3.0) & 2.0 (2.0) & 0.0 (4.0) \\ 
 \emph{rest-scs} & 1.5 (2.0) & 0.0 (4.0) & {\bf 2.0 (1.0)} & 0.6 (3.0) \\ 
 \emph{restcountries} & 0.9 (3.0) & {\bf 1.0 (1.5)} & {\bf 1.0 (1.5)} & 0.0 (4.0) \\ 
 \emph{scout-api} & 33.6 (3.0) & {\bf 35.0 (1.0)} & 34.6 (2.0) & 0.0 (4.0) \\ 
 \emph{session-service} & {\bf 8.0 (1.5)} & 2.6 (4.0) & {\bf 8.0 (1.5)} & 7.0 (3.0) \\ 
 \emph{spring-actuator-demo} & {\bf 2.0 (1.0)} & 0.0 (3.5) & 1.0 (2.0) & 0.0 (3.5) \\ 
 \emph{spring-batch-rest} & {\bf 5.0 (1.5)} & 1.0 (3.0) & {\bf 5.0 (1.5)} & 0.0 (4.0) \\ 
 \emph{spring-ecommerce} & 26.0 (2.0) & 14.0 (3.5) & {\bf 26.9 (1.0)} & 14.0 (3.5) \\ 
 \emph{spring-rest-example} & {\bf 9.0 (1.5)} & 1.0 (3.5) & {\bf 9.0 (1.5)} & 1.0 (3.5) \\ 
 \emph{swagger-petstore} & 8.0 (3.0) & {\bf 12.1 (1.0)} & 12.0 (2.0) & 0.0 (4.0) \\ 
 \emph{testing-system} & 9.0 (2.0) & 0.0 (3.5) & {\bf 10.6 (1.0)} & 0.0 (3.5) \\ 
 \emph{tiltaksgjennomforing} & {\bf 77.0 (1.5)} & 2.0 (3.0) & {\bf 77.0 (1.5)} & 0.0 (4.0) \\ 
 \emph{traccar} & {\bf 0.0 (2.5)} & {\bf 0.0 (2.5)} & {\bf 0.0 (2.5)} & {\bf 0.0 (2.5)} \\ 
 \emph{tracking-system} & {\bf 15.0 (1.0)} & 0.0 (3.5) & 4.7 (2.0) & 0.0 (3.5) \\ 
 \emph{user-management} & {\bf 21.0 (1.5)} & 1.0 (4.0) & {\bf 21.0 (1.5)} & 2.4 (3.0) \\ 
 \emph{webgoat} & 187.0 (2.0) & 35.0 (3.0) & {\bf 202.9 (1.0)} & 0.0 (4.0) \\ 
 \emph{youtube-mock} & {\bf 1.0 (1.0)} & 0.0 (3.5) & 0.1 (2.0) & 0.0 (3.5) \\ 
 \midrule 
Average  & 34.3 (1.7) & 14.0 (3.1) & 28.2 (1.7) & 4.3 (3.5) \\ 
 Median  & 13.0 (1.5) & 2.2 (3.5) & 13.0 (1.5) & 0.9 (3.5) \\ 
 \midrule 
Friedman Test & \multicolumn{4}{r}{$\chi^2$ = 107.165, $p$-value = $< 0.001$ } \\ 
\bottomrule 
\end{tabular} 

}
\end{table}

\begin{table}[!t]
\centering
\caption{ \label{tab:second-coverage}
{
Average results, out of 10 runs, for code coverage percentage over the 4 compared fuzzers, on all the \nsuts APIs.
On each API, tool are compared by rank, where rank 1 is the best.
Rank values are presented in `()' parentheses after the average values.
In case of ties, ranks are averaged.
The best fuzzers on each API are highlighted in bold.
}
}
\vspace{-1.5\baselineskip}
 \resizebox{0.72\textwidth}{!}{
\begin{tabular}{ l  r r r r r    }\\ 
\toprule 
SUT  &  \base &  \cats &  \evomaster &  \schemathesis &  \wuppiefuzz \\ 
\midrule 
\emph{adoptme} & 8.4 (5.0) & {\bf 83.5 (1.0)} & 82.5 (3.0) & 83.5 (2.0) & 70.5 (4.0) \\ 
 \emph{arimaa} & 6.3 (5.0) & {\bf 27.2 (1.0)} & 24.8 (3.0) & 26.2 (2.0) & 19.2 (4.0) \\ 
 \emph{bibliothek} & 36.1 (5.0) & {\bf 40.5 (2.5)} & {\bf 40.5 (2.5)} & {\bf 40.5 (2.5)} & {\bf 40.5 (2.5)} \\ 
 \emph{blogapi} & 8.1 (5.0) & 31.1 (3.0) & 30.1 (4.0) & {\bf 50.9 (1.0)} & 37.1 (2.0) \\ 
 \emph{catwatch} & 9.8 (5.0) & {\bf 41.1 (1.0)} & 35.9 (3.0) & 36.3 (2.0) & 20.6 (4.0) \\ 
 \emph{commafeed} & 10.9 (5.0) & 14.3 (2.0) & 14.2 (3.0) & {\bf 15.0 (1.0)} & 13.8 (4.0) \\ 
 \emph{cwa-verification} & 21.9 (5.0) & {\bf 66.4 (1.0)} & 45.5 (3.0) & 50.6 (2.0) & 43.0 (4.0) \\ 
 \emph{digitalbanking} & 15.0 (5.0) & 56.5 (2.0) & 55.5 (4.0) & 55.9 (3.0) & {\bf 58.8 (1.0)} \\ 
 \emph{erc20-rest-service} & 7.9 (5.0) & {\bf 43.5 (1.0)} & 26.7 (2.5) & 26.7 (2.5) & 25.1 (4.0) \\ 
 \emph{familie-ba-sak} & 16.0 (5.0) & {\bf 16.7 (2.0)} & 16.7 (4.0) & {\bf 16.7 (2.0)} & {\bf 16.7 (2.0)} \\ 
 \emph{features-service} & 21.0 (5.0) & {\bf 78.8 (1.0)} & 64.9 (4.0) & 77.9 (2.0) & 76.8 (3.0) \\ 
 \emph{flight-search-api} & 11.5 (5.0) & {\bf 71.9 (1.0)} & 45.7 (3.0) & 58.3 (2.0) & 41.1 (4.0) \\ 
 \emph{gdpr-kv} & 3.5 (5.0) & {\bf 14.9 (1.0)} & 14.7 (3.0) & 14.8 (2.0) & 12.9 (4.0) \\ 
 \emph{genome-nexus} & 12.8 (5.0) & 25.9 (3.0) & {\bf 31.2 (1.0)} & 27.1 (2.0) & 22.6 (4.0) \\ 
 \emph{gestaohospital} & 19.9 (5.0) & {\bf 66.6 (1.0)} & 56.5 (3.0) & 65.4 (2.0) & 38.1 (4.0) \\ 
 \emph{gravitee} & 6.1 (5.0) & 13.2 (2.0) & 9.1 (3.0) & {\bf 13.4 (1.0)} & 6.1 (4.0) \\ 
 \emph{http-patch-spring} & 5.3 (5.0) & 79.1 (2.0) & 78.5 (3.0) & {\bf 81.9 (1.0)} & 52.1 (4.0) \\ 
 \emph{jasper} & 13.0 (5.0) & 37.6 (2.0) & 30.6 (3.0) & {\bf 40.9 (1.0)} & 13.0 (4.0) \\ 
 \emph{joinus} & 2.9 (5.0) & {\bf 28.3 (1.0)} & 15.1 (3.0) & 10.5 (4.0) & 21.0 (2.0) \\ 
 \emph{kafka-rest} & 5.2 (5.0) & {\bf 39.8 (1.0)} & 26.6 (3.0) & 27.0 (2.0) & 5.2 (4.0) \\ 
 \emph{komga} & 15.8 (5.0) & {\bf 33.0 (1.0)} & 17.8 (4.0) & 30.7 (2.0) & 21.7 (3.0) \\ 
 \emph{languagetool} & 1.7 (5.0) & 13.8 (2.0) & 5.5 (3.0) & {\bf 35.4 (1.0)} & 1.7 (4.0) \\ 
 \emph{lovemining} & 13.7 (5.0) & 62.4 (2.0) & {\bf 65.8 (1.0)} & 61.8 (3.0) & 29.4 (4.0) \\ 
 \emph{management-cassandra} & 6.7 (5.0) & 39.9 (3.0) & {\bf 47.2 (1.0)} & 44.0 (2.0) & 18.3 (4.0) \\ 
 \emph{market} & 7.0 (5.0) & 26.3 (3.0) & 27.5 (2.0) & {\bf 35.5 (1.0)} & 21.9 (4.0) \\ 
 \emph{microcks} & 3.1 (5.0) & 12.2 (2.0) & {\bf 12.4 (1.0)} & 6.4 (4.0) & 7.0 (3.0) \\ 
 \emph{movies-xml} & 11.1 (5.0) & {\bf 79.1 (1.0)} & 26.9 (3.0) & 79.0 (2.0) & 11.4 (4.0) \\ 
 \emph{notebook-manager} & 29.5 (5.0) & 83.2 (3.0) & 83.8 (2.0) & {\bf 84.8 (1.0)} & 76.4 (4.0) \\ 
 \emph{ocvn} & 10.1 (5.0) & 33.4 (3.0) & 34.1 (2.0) & {\bf 34.2 (1.0)} & 10.4 (4.0) \\ 
 \emph{ohsome-api} & 7.1 (5.0) & 48.8 (2.0) & 36.5 (3.0) & {\bf 49.7 (1.0)} & 10.0 (4.0) \\ 
 \emph{pay-publicapi} & 10.2 (5.0) & {\bf 16.3 (1.5)} & 15.8 (3.0) & {\bf 16.3 (1.5)} & 15.7 (4.0) \\ 
 \emph{person-controller} & 12.8 (5.0) & {\bf 73.5 (1.5)} & 72.5 (3.0) & {\bf 73.5 (1.5)} & 20.9 (4.0) \\ 
 \emph{petclinic-rest} & 3.5 (4.5) & 32.9 (2.0) & 31.5 (3.0) & {\bf 33.2 (1.0)} & 3.5 (4.5) \\ 
 \emph{pn-paper-tracker} & 6.0 (5.0) & 7.7 (3.5) & 11.2 (2.0) & {\bf 11.8 (1.0)} & 7.7 (3.5) \\ 
 \emph{proxyprint} & 4.3 (5.0) & 43.0 (3.0) & 47.9 (2.0) & {\bf 51.6 (1.0)} & 4.5 (4.0) \\ 
 \emph{quartz-manager} & 34.4 (5.0) & {\bf 53.5 (1.0)} & 50.1 (3.0) & 50.6 (2.0) & 48.7 (4.0) \\ 
 \emph{redis-sample} & 77.6 (5.0) & {\bf 93.5 (1.0)} & 86.0 (4.0) & 92.7 (2.5) & 92.7 (2.5) \\ 
 \emph{reservations-api} & 24.4 (5.0) & {\bf 76.1 (1.0)} & 55.9 (4.0) & 60.8 (3.0) & 63.6 (2.0) \\ 
 \emph{rest-ncs} & 5.1 (5.0) & 92.0 (2.0) & 74.6 (3.0) & {\bf 94.5 (1.0)} & 53.8 (4.0) \\ 
 \emph{rest-news} & 13.9 (5.0) & {\bf 90.4 (1.0)} & 83.4 (3.0) & 90.3 (2.0) & 45.8 (4.0) \\ 
 \emph{rest-scs} & 4.7 (5.0) & 62.0 (3.0) & {\bf 65.1 (1.0)} & 64.7 (2.0) & 61.4 (4.0) \\ 
 \emph{restcountries} & 3.5 (4.5) & 77.3 (2.0) & {\bf 77.5 (1.0)} & 74.5 (3.0) & 3.5 (4.5) \\ 
 \emph{scout-api} & 12.0 (4.5) & {\bf 39.5 (1.0)} & 39.4 (2.0) & 39.1 (3.0) & 12.0 (4.5) \\ 
 \emph{session-service} & 21.4 (5.0) & 81.8 (2.0) & 57.9 (3.0) & {\bf 85.7 (1.0)} & 25.8 (4.0) \\ 
 \emph{spring-actuator-demo} & 71.0 (5.0) & {\bf 87.1 (2.5)} & {\bf 87.1 (2.5)} & {\bf 87.1 (2.5)} & {\bf 87.1 (2.5)} \\ 
 \emph{spring-batch-rest} & 15.0 (4.5) & {\bf 57.5 (1.0)} & 34.3 (3.0) & 57.2 (2.0) & 15.0 (4.5) \\ 
 \emph{spring-ecommerce} & 15.7 (5.0) & {\bf 45.8 (1.0)} & 39.6 (3.0) & 40.0 (2.0) & 32.8 (4.0) \\ 
 \emph{spring-rest-example} & 11.1 (5.0) & {\bf 48.9 (2.0)} & {\bf 48.9 (2.0)} & {\bf 48.9 (2.0)} & 48.6 (4.0) \\ 
 \emph{swagger-petstore} & 29.5 (4.5) & 65.9 (3.0) & 73.4 (2.0) & {\bf 73.9 (1.0)} & 29.5 (4.5) \\ 
 \emph{testing-system} & 10.1 (5.0) & 76.6 (2.0) & 66.2 (3.0) & {\bf 78.0 (1.0)} & 27.5 (4.0) \\ 
 \emph{tiltaksgjennomforing} & 7.2 (4.5) & {\bf 9.0 (2.0)} & {\bf 9.0 (2.0)} & {\bf 9.0 (2.0)} & 7.2 (4.5) \\ 
 \emph{traccar} & 5.4 (5.0) & 7.1 (3.0) & 7.2 (2.0) & {\bf 7.3 (1.0)} & 5.4 (4.0) \\ 
 \emph{tracking-system} & 19.5 (5.0) & {\bf 40.1 (1.0)} & 39.9 (2.0) & 39.3 (3.0) & 20.6 (4.0) \\ 
 \emph{user-management} & 6.5 (5.0) & {\bf 74.0 (1.0)} & 63.5 (3.0) & 70.3 (2.0) & 46.4 (4.0) \\ 
 \emph{webgoat} & 19.6 (5.0) & {\bf 58.2 (1.0)} & 52.4 (3.0) & 56.6 (2.0) & 20.8 (4.0) \\ 
 \emph{youtube-mock} & 36.9 (4.5) & {\bf 80.8 (1.0)} & 46.0 (3.0) & 80.0 (2.0) & 36.9 (4.5) \\ 
 \midrule 
Average  & 14.8 (4.9) & 50.0 (1.8) & 43.6 (2.7) & 49.4 (1.9) & 30.0 (3.7) \\ 
 Median  & 11.0 (5.0) & 47.3 (2.0) & 40.2 (3.0) & 50.1 (2.0) & 21.8 (4.0) \\ 
 \midrule 
Friedman Test & \multicolumn{5}{r}{$\chi^2$ = 166.394, $p$-value = $< 0.001$ } \\ 
\bottomrule 
\end{tabular} 

}
\end{table}

%%%%%%%%%%%%%%%%%%%%%%%%%%%%%%%%%%%%%%%%%%%%%%%%%%%%%%%%%%%%%%%%%%%%%%%%%%%%

\bibliographystyle{ACM-Reference-Format} % this requires ACM-Reference-Format.bst in same folder
%\bibliographystyle{acm}  % this is ancient from 80s, which does not support URL and DOI

%%https://arxiv.org/help/submit_tex#latex
%
% IMPORTANT: For final version arXiv, use generated bbl.
%            In such case, do not use compile.sh to build final pdf, but rather call pdflatex directly, eg
%            pdflatex arxiv


%\bibliography{papers}

%If needed
%\input{appendix.tex}

\end{document}